%% file: A_computational_spectral_approach_to_interest_rate_models.tex
\documentclass[a4paper,13pt]{article}
\usepackage[colorinlistoftodos]{todonotes}
\input{macro}

\usepackage[title,titletoc,toc]{appendix}

\title{A computational spectral approach to interest rate models}

\date{\today}

\author{
\textsc{Luca Di Persio} 
            \qquad
 \textsc{Gregorio Pellegrini}
        \mbox{}\\ %
        Department of Computer Science, 
       University of Verona \\
        Strada Le Grazie 15, 37134 Verona, Italy\\
\texttt{luca.dipersio@univr.it} \\  \texttt{gregorio.pellegrini@univr.it}\\ $\ $\\
\textsc{Michele Bonollo} 
        \mbox{}\\ %
       Iason Ltd, Milan, Italy and IMT Lucca,\\
       Piazza S.Francesco 19, 55100 Lucca, Italy\\
\texttt{michele.bonollo@imtlucca.it}
}

\begin{document}

\maketitle


\begin{abstract}

The Polynomial Chaos Expansion (PCE) technique recovers a finite second order random variable exploiting suitable linear combinations of orthogonal polynomials which are functions of a given stochastic quantity $\xi$, hence acting as a kind of random basis.
The PCE methodology has been developed as a mathematically rigorous Uncertainty Quantification (UQ) method which aims at providing
reliable numerical estimates for some uncertain physical quantities defining the dynamic of certain engineering models and their related simulations. \newline

In the present paper we exploit the PCE approach to analyze some equity and interest rate models considering, without loss of generality, the one dimensional case. In particular we will take into account  those models which are based on the Geometric Brownian Motion (gBm), e.g. the Vasicek model, the CIR model, etc.
We also provide several numerical applications and results which are discussed for a set of volatility values. The latter allows us to test the PCE technique on a quite large set of different scenarios, hence providing a rather complete and  detailed investigation on PCE-approximation's features and properties, such as the convergence of statistics, distribution and quantiles. 

Moreover we give results concerning both an efficiency and an accuracy study of our approach by comparing {\it our outputs} with the ones  obtained adopting the Monte Carlo approach in its standard form as well as in its enhanced version.

\smallskip $\ $\\
\end{abstract}

\section*{Introduction}

In this paper we present a Polynomial Chaos Expansion (PCE) approach to  the solution of the following It\^o stochastic differential equation (SDE)
\begin{equation}
dX_t = r(t,X_t)dt + \sigma(t,X_t)dW_t \label{sec:eq1}
\end{equation}
 where  $X_0$ is the initial value of the unknown stochastic process $X_t$, the  term $r(t,X_t)$, resp. $\sigma(t,X_t)$, represents the {\it drift}, resp.  the {\it volatility} of the process, while $W_t$ is a standard  Brownian motion and it is considered up to a certain finite time $T>0$ to which we will refer, taking into account the financial setting, as the maturity time,  of, e.g., some underlying. 
According to \cite[Section 4.5]{Platen}, if the coefficients  $r(t,x)$ and $\sigma(t,x)$ are smooth enough, then we have the existence and uniqueness of the solution to \eqref{sec:eq1}. \newline


From a numerical point of view, between the most  popular, and simple,  method to numerically approximate the solutions for a large class of SDEs, there are the so-called Monte Carlo (MC) methods, which rely on  pseudo-random sampling of independent increments of the Brownian Motion in \eqref{sec:eq1}, see, e.g, 
\cite[Chapter 9]{Platen},  
\cite[Chapter 3]{Glasserman} and references therein.

The PCE technique is based on a radically different approach,  namely it expresses the solution by means of polynomial basis of the probability space where the solution to \eqref{sec:eq1}, at certain time $T$, is defined, see, e.g., \cite[Chapter 3]{Spectral} for further details.

Strictly speaking PCE recovers a random variable in terms of a linear combination of functionals whose entries are known random variables called {\it germs} or { \it basic variables}. Several method are available to compute these coefficients. In particular we choose the  Non-Intrusive Spectral Projection (NISP) method which employs a set of deterministic realizations, see, e.g.,  \cite[Chapter 3]{Spectral} for further details. 
Moreover the theory developed by Doss in \cite{Doss}, allows us to apply NISP avoiding unfeasible numerical problems of NISP approach to get numerical solution to \eqref{sec:eq1}.
\newline

Originally the {\it Polynomial Chaos} idea was introduced by Norbert Wiener in his 1938 paper, see \cite{Norbert}, where he applies his generalized harmonic analysis, see, e.g., \cite{a1}. 
Then P.D. Spanos and R.G. Ghanem in \cite[Section 2.4, Subsection 3.3.6]{Spanos}, combined the PCE method with a finite element one, in order to compute the solution of PDEs in presence of uncertain parameters.  Afterwards latter approach has been  applied in several frameworks, e.g. to the simulation of probabilistic chemical reactions, see \cite{Villegas} , to stochastic optimal trajectories generation, see \cite{OptimalTrajectory} , to sensitivity analysis, see \cite{Crestaux}, in order to refine specific numerical methods, see \cite{Xiu}, and also to a rather large set of problems arising in  engineering and computational fluid dynamics (CFD), see, e.g., \cite[Chapter 6]{Spectral}, \cite{Oladyshkin,Peccati} , and references therein.


Our analysis has been focused on three well known one dimensional  SDEs whose solutions are related, respectively, to the  Geometric Brownian Motion equity model, see, e.g., \cite[Chapter 11, section 3]{Hull}, to the Vasicek model, see \cite{Vasicek},  and to the Cox,  Ingersoll , Ross (CIR) interest rate model, see \cite{Cox}. We also refer to \cite[Chapter 3, Section 3.2.3]{Brigo}, for more details and references about the aforementioned models. \newline 
To get an exhaustive comparison with already obtained results, e.g. by using  Monte Carlo approaches, the PCE-approximation for each of the cited financial models, has been implemented  for a set of three different volatility values, namely $\sigma = \{15\%,25\%,30\%\}$. Moreover we have analyzed the convergence of both the mean and the variance, to analytical values, as well as the distribution and two quantiles of the PCE-approximation, providing a wide test beach for PCE machinery.


In order to show the advantages of the PCE approach, the results of our PCE-implementation have been compared {\it versus} the ones obtained using both the standard Monte Carlo (MC) and the {\it{quasi}}-Monte Carlo (QMC) techniques. The former is based on pseudo-random sampling and its convergence properties basically rely on the \emph{Law of large numbers} and the \emph{Central limit theorem}. The latter uses low-discrepancy sequences to simulate the process and it is theoretically based on the Koksma-Hlawka-inequality, see \cite{Hickenell}, which provides error bounds for computations involving QMC methods, see, e.g., \cite[Chapter 5]{Glasserman}. \newline

We would like to underline that  the first two cases, namely the gBm-model and the Vasicek model, are meaningful since they have an analytical solution to the related SDE, therefore we provide a rigorous study of the convergence property of the PCE method applied to them. In the  CIR case, since we do not have an analytical solution, the application of the PCE technique allows us also to derive some considerations on the convergence properties of the related approximating solutions.\newline

The paper is organized as follows: in Section \ref{sec:PCE} the PCE approach is described in a general setting, while Section \ref{sec:NISP} deals with the NISP approach. Section \ref{sec:PCEeSDE} points out how to apply PCE in order to decompose the solution of the considered SDE, by using usual numerical method for solving (\ref{sec:eq1}), and how to employ Doss Theory in the PCE-approximation setting.
In Section \ref{sec:secGBM}, \ref{sec:secVAS}, \ref{sec:secCIR}, we consider numerical applications of PCE technique, from the point of view of the Doss Theory,  to approximate, respectively,  the solutions to the  gBm equity model, to the Vasicek model and to the CIR model. 

In section \ref{sec:Comparison}, we give an overview of the PCE results in term of accuracy and computational time cost for each one of the  considered models.




\section{Polynomial Chaos Expansion} \label{sec:PCE}

Let $(\Omega,\Sigma,\PP)$ be a probability space, where $\Omega$ is the  set of elementary events, $\Sigma$ is a $\sigma$-algebra of subsets of $\Omega$ and $\PP$ is a probability measure on $\Sigma$. 

Let us consider the Hilbert space of scalar real-valued random variables $L^2(\Omega,\Sigma,\PP)$, whose generic element is a random variable $X$ defined on $(\Omega,\Sigma,\PP)$, and such that
\begin{equation*}
\EE[X^2] = \int_\Omega \big(X(\omega)\big)^2d\PP(\omega) < +\infty\:.
\end{equation*}

Notice that, as for Lebesgue spaces, the elements $X \in L^2(\Omega,\Sigma,\PP)$ are equivalent classes of random variables. Note that $L^2(\Omega,\Sigma,\PP)$  is a Hilbert space endowed with the following scalar product
\begin{equation*}
\EE[XY] = \left<X,Y\right>_\PP = \int_\Omega X(\omega)Y(\omega)d\PP(\omega) \;,
\end{equation*}
being 
\begin{equation*}
\norm{X}^2_{L^2(\Omega,\Sigma,\PP)}=\EE[X^2] = \int_\Omega \big(X(\omega)\big)^2d\PP(\omega) \;,
\end{equation*}
the usual norm.
To shorten the notation $\normp{X}^2:= \norm{X}^2_{L^2(\Omega,\Sigma,\PP)}$, and the related convergence will be always referred as \emph{mean square convergence} or \emph{strong convergence}. \newline

Among elements in $L^2(\Omega,\Sigma,\PP)$ there is the class of \emph{basic random variables}, which is used to decompose, as entries of functionals, the quantity of interest $Y$ such as a random variable of interest, the solution of the SDE at time $T$. 
We notice that not all the functions $\xi:\ \Omega \to D$ can be used to perform such a decomposition since they have to satisfy, see, e.g.,  \cite[Section 3]{Ernst}, at least the following two properties
\begin{itemize}
\item $\xi$ has finite raw moments of all orders
\item the distribution function $F_\xi(x) := \Prob{\xi \le x}$ of the basic random variables is continuous,  $f_\xi$ being  its probability density function (pdf).
\end{itemize}

Since the increments of the Brownian motion  are independent and normally distributed, from now on let us consider as basic random variable $\xi \sim \mathcal{N}(0,1/2)$. The latter choice is motivate in Remark 1 of Sec. \ref{sec:flow}

The elements in $L^2(\Omega,\Sigma,\PP)$ can be gathered in two groups: in the first one we have the {\it basic random variables}, let us indicate them by $\xi$,  ruling the decomposition, for simplicity let us call it $Set_\xi$, while the second set is composed by {\it generic elements}, let us say $Y$,  which we want to decompose using elements of the first set, which is referred as $Set_Y$.


Let us denote by $\sigma(\xi)$ the $\sigma$-algebra generated by the basic random variable $\xi$, hence $\sigma(\xi) \subset \Sigma$. If we want  to polynomially decompose the random variable $Y$ in terms of $\xi$, then $Y$  has to be,  at least,  measurable with respect to the $\sigma$-algebra $\sigma(\xi)$. 
Exploiting the  Doob-Dynkin Lemma, see, e.g.,  \cite[ Lemma 1.13]{Kallenberg}, we have that $Y$ is $\sigma(\xi)$-measurable, by detecting a Borel measurable function $g: \R \to \R$,  such that $Y = g (\xi)$.  In what follows, without  loss of generality, we restrict ourselves to consider the decomposition in $L^2(\Omega,\sigma(\xi),\PP)$, moreover the basic random variable $\xi$ determines the class of orthogonal polynomials $\{\Psi_i(\xi)\}_{i \in \N}$ it is always indicated as the \emph{generalized polynomial chaos} (gPC) basis.  We underline that their orthogonality properties is detected by means of the measure induced by $\xi$ in the image space $(D,\mathcal{B}(D))$, where $\mathcal{B}(D)$ denotes the Borel $\sigma$-algebra of $D$, in particular, for each $i,j \in \N$, we have 
\begin{equation}
\sprod{\Psi_i}{\Psi_j} = \intPxi{\Psi_i}{\Psi_j} =  \intPxiR{\Psi_i}{\Psi_j} \:. \; \label{sec:eq2}
\end{equation}

\begin{figure}[!h]
\centering
\includegraphics[scale=0.45]{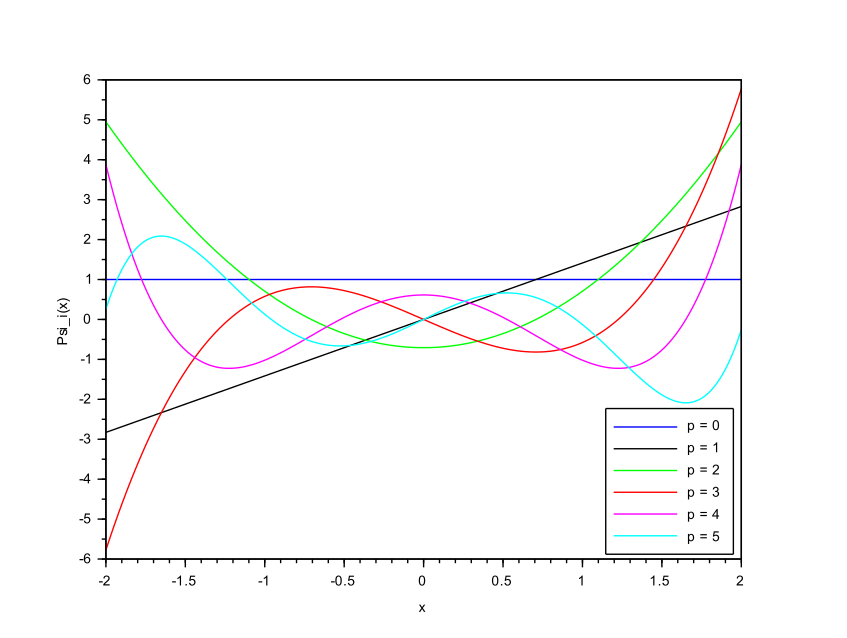}
\caption{The  Hermite polynomials up to degree $5$}  \label{fig:fig1}
\end{figure}

Since $\xi \sim \mathcal{N}(0,1/2)$, the related set $\{\Psi_i(x)\}_{i \in \N}$ is represented by the  family of  Hermite polynomials defined on the whole real line, namely $D =\R$, and

\begin{equation}\label{Hermite polynomials}
\begin{cases}
\Psi_0(x) &= 1 \\
\Psi_1(x) &= 2x \\
\Psi_2(x) &= 4x^2-2\\
&\vdots
\end{cases}
\end{equation}\label{sec:eq3}
Figure \ref{fig:fig1} provides the graph of the first six orthonormal polynomials, achieved by scaling each  $\Psi_i$ in \eqref{Hermite polynomials} by  its norm in $L^2(\Omega,\sigma(\xi),\PP)$, namely, $\forall i \in \N$,  $\Psi_i$  is divided by $\normp{\Psi_i}:=\sqrt{2^i i!} $. 

Latter polynomials constitute a maximal system in $ L^2(\Omega,\sigma(\xi),\PP)$,  
therefore every finite second order random variable $Y$ can be approximated as follows
\begin{equation}
Y^{(p)} = \sum_{i=0}^p c_i \Psi_i(\xi)\;, \label{sec:eq4}
\end{equation}
for suitable coefficients $c_i$ which depend on the random variable $Y$. We refer to eq. \eqref{sec:eq4} as the \emph{truncated PCE}, at degree $p$, of $Y$. Exploiting previous definitions, taking $i \in \{0,\dots,p\}$, and considering the orthogonality property of the polynomials $\{\Psi_i(\xi)\}_{i \in \N}$, we have
\begin{equation}
c_i = \frac{1}{\normp{\Psi_i}^2} \sprod{Y}{\Psi_i} = \frac{1}{\normp{\Psi_i}^2} \sprod{g}{\Psi_i} \label{sec:eq5}\;,
\end{equation}
 and, since $Y=g(\xi)$, we also obtain
\begin{equation}
\sprod{Y}{\Psi}  = \sprod{g}{\Psi} = \int_\Omega g(\xi(\omega))\Psi_i(\xi(\omega))dP(\omega) = \int_{\R}g(x)\Psi_i(x)f_\xi(x)dx \label{sec:eq6}\;,
\end{equation}

moreover $Y^{(p)}$ converges in mean square sense to $Y$, see, e.g.,  \cite[Section 3.1]{Ernst}. The convergence rate of the PCE-approximation \eqref{sec:eq4} in $L^2(\Omega,\sigma(\xi),\PP)$ norm is strictly linked to the magnitude of the coefficients of the decomposition. Indeed by the Parseval's identity, we have  
\begin{equation*}
\normp{Y}^2 = \sum_{i=0}^{+\infty}{c_i^2}\normp{\Psi_i}^2 \; ,
\end{equation*}
furthermore, using the orthogonality property of Hermite polynomials in $L^2(\Omega,\sigma(\xi),\PP)$, the norm of \eqref{sec:eq4} is given by
\begin{equation*}
\normp{Y^{(p)}}^2 = \sum_{i=0}^{p}{c_i^2}\normp{\Psi_i}^2 \; ,
\end{equation*}
then, exploiting the fundamental properties of the orthogonal projection in Hilbert Space, see, e.g., \cite[Theorem 4.11]{Rudin}, we can estimate the mean square error as
\begin{equation}
\normp{Y-Y^{(N)}}^2= \normp{Y}^2-\normp{Y^{(N)}}^2 = \sum_{i=N+1}^{+\infty}{c_i^2}\normp{\Psi_i}^2 \;, \label{sec:eq7}
\end{equation}
thus the coefficients rules the convergence rate.

We would like to underline that the PCE of $Y^{(p)}$ approximates the $Y$-statistics using the $c_i$ coefficients appearing in eq. \eqref{sec:eq4}, e.g. the first two centered moments are determined by 
\begin{equation} \label{sec:eq8}
 \Expect{Y^{(p)}} =c_0  \;,
\end{equation}
\begin{equation} \label{sec:eq9}
\Var{Y^{(p)}}  =\sum_{i=1}^p c_i^2 \normp{\Psi_i}^2 \:. 
\end{equation}

\section{Non-Intrusive Spectral Projection} \label{sec:NISP}

Let us consider the case of a  functional $\mathcal{M}$ which governs the dynamic of a quantity  we want to study. We assume that $\mathcal{M}$ is characterized by a random output $Y$ depending on the input $\xi$, which is itself a random variable. The goal is to describe the behavior of $Y$ in dependence of the variations of $\xi$, and with respect to the action of the functional $\mathcal{M}$. In what follows, without loss of generality,  we consider the $1-dimensional$ case, 
indeed the results we shall obtain can be easily generalized to greater dimension. 
In our setting  $\mathcal{M}$ will be the solution functional to an SDE describing a specific {\it interest rate model}, see later Sec. \ref{sec:secGBM}, \ref{sec:secVAS}, \ref{sec:secCIR} for further details, and we also allow the {\it output} $Y$ to depend on a set of parameters, let us call such a set $\Theta \in \R^n$, with $n\ge2$,  characterizing the interest rate dynamics, namely 
\begin{equation*}
Y  = \mathcal{M}(\xi,\Theta) \:.
\end{equation*}
In order to shorten the notation the parameters $\Theta$ are omitted in the $\mathcal{M}$ definition.
Our aim is to exploit the so-called Non-Intrusive Spectral Projection (NISP) method to obtain the  truncated PCE of the output $Y$, taking into account that the basic random variable $\xi$ inherits all the assumptions made in the previous section. Therefore we have the following {\it spectral projection}
\begin{equation*}
Y^{(p)} = \sum_{i=0}^p c_i \Psi_i(\xi)\;,
\end{equation*}
where the coefficients are defined as in (\ref{sec:eq5}). \newline

We recall that, see, e.g. \cite[Section 3.3]{Spectral}, the NISP approach computes the scalar product in (\ref{sec:eq5})  by Gaussian Quadrature formulae, in particular,  $\forall i \in \{0,\dots,p\}$, we have the following Gauss-Hermite type result
\begin{equation}
c_i \approx  \frac{1}{\normp{\Psi_i}^2}\sum_{j=1}^N {\mathcal{M}}(\xi_j)\Psi_i(\xi_j)w_j \;, \label{sec:eq10}
\end{equation}
where $\{\xi_j\}_{j=1}^N$, $\{w_j\}_{j=1}^N$ are the quadrature nodes and the weights of the one-dimensional Gaussian integration rule.  Moreover (\ref{sec:eq10}) requires the evaluation of the process $\mathcal{M}$ at a well defined set of realizations for the input random variable $\xi$.
By the very definition of NISP, $Y^{(p)}$ achieves spectral convergence with respect to the $Y= \mathcal{M}(\xi)$ for increasing degree $p$, see, e.g., \cite[Appendix B]{Spectral}.

We note that  the number of quadrature points $N$ are not linked a priori to the degree $p$ of the truncation. Instead, the precision of the coefficients $c_i$ is related with $N$, thus a reasonable and standard choice that has used in literature, and which we will use in what follows, is to take $N=p$, since the polynomials involved are at least of degree $p$. 
Moreover, in what follows, we refer to $\{\xi_j  \}_{j=1}^N$   as quadrature nodes as well as a set of independent realizations of the basic random variable.

\subsection{Flowchart of the NISP computation } \label{sec:flow}
In order to better explain the NISP approach, let us describe it exploiting the following flowchart
\begin{center}
{ \footnotesize
\begin{tikzpicture}[
  font=\sffamily,
  every matrix/.style={ampersand replacement=\&,column sep=.7cm,row sep=1cm},
  source/.style={draw,thick,rounded corners,fill=yellow!20,inner sep=.2cm},
 source2/.style={draw,thick,fill=blue!30,inner sep=.2cm},
  process/.style={draw,thick,shape=ellipse,fill=blue!20,inner sep=.2cm},
  sink/.style={source,fill=green!20},
  datastore/.style={draw,very thick,shape=datastore,inner sep=.1cm},
  dots/.style={gray,scale=2},
  to/.style={->,>=stealth',shorten >=1pt,semithick,font=\sffamily\footnotesize},
  every node/.style={align=center}]
\tikzstyle{line} = [draw, -latex']
\matrix{
\& \node[source] (l1) {Basic random variable \\ $\xi \sim \mathcal{N}(0,1/2)$};\\

\&\node[source] (l3) {Gaussian quadrature formula};\\
\node[source] (l4a) {Sampling $\{\xi_j\}_{j=1}^N$ \\ quadrature nodes}; \&
\&  \node[source] (l4b) {weights $\{w_j\}_{j=1}^N$};\\
\& \node[source2](l5a) {Simulation of the process $\mathcal{M}(\xi)$ \\ at $\{\xi_j\}_{j=1}^N$}; \\
\&\node[source2] (l6) {PCE coefficients computations \\ equation (\ref{sec:eq10})};\\
\&\node[source2](l7) {Post-processing analysis of \\ the truncated PCE $Y^{(p)}$}; \\
\node[source2] (l8a) {Statistics: \\ mean, variance}; \& 
\node[source2] (l8b) {Sampling };\& 
\node[source2] (l8c)  {Quantiles}; \\
  };

\draw[to] (l1) -- (l3);
 \draw[|-,-|,->, thick,] (l3.south) |-+(0,-1em)-| (l4a.north);
 \draw[|-,-|,->, thick,] (l3.south) |-+(0,-1em)-| (l4b.north);
 \draw[|-,-|,->, thick,] (l4a.south) |-+(0,-1em)-| (l5a.north);
  \path [line] (l4b) |- (l6);
\draw[to] (l5a) -- (l6);
\draw[to] (l6) -- (l7);
 \draw[|-,-|,->, thick,] (l7.south) |-+(0,-1.5em)-| (l8a.north);
 \draw[|-,-|,->, thick,] (l7.south) |-+(0,-1.5em)-| (l8b.north);
 \draw[|-,-|,->, thick,] (l7.south) |-+(0,-1.5em)-| (l8c.north);

\end{tikzpicture}

}
\end{center}

We would like to underline that the set of orthogonal polynomials, which constitute a basis in $ L^2(\Omega,\sigma(\xi),\PP)$, is computed off-line once  the basic random variable $\xi$ has been given and independently on the particular model we want to decompose. Analogously, the same happens for the set of $N=p$ realizations of $\xi$ and weights needed to apply the Gaussian quadrature method.

A different situation happens about simulation of the process and related computations. Indeed, we have to deal with data of the type $\{\mathcal{M}(\xi_j)\}_{j=1}^N$, which are sensible to changes of the model parameters. This is why such a difference has been highlighted in our flowchart using yellow, resp. blue, rectangles, which represent  the off-line computations, resp. the model correlated computations. \newline

\begin{remark}. Let us focus our attention of the basic random variable. Other possibilities are available, e.g. we can take $\xi \sim \mathcal{U}(0,1)$, $\xi \sim \exp(1)$, etc., nevertheless  the {\it Gaussian choice} appears as the most natural one, since the Wiener increments are also distributed as Gaussian variables. In particular in our computations we consider  $\xi \sim \mathcal{N}(0,1/2)$, where the specification of the variance value is due to software restrictions. In particular we have used the Scilab toolbox which  provides a powerful environment to implement PCE decomposition, also giving to the user a very flexible and efficient tool for the concrete computations involved by the PCE method itself, particularly by exploiting
The Scilab toolbox, the Hermite polynomials require such input random variable in order to satisfy (\ref{sec:eq5}), see \cite{NISP1,NISP2} for further details. \newline
\end{remark}
\begin{remark}.
The sampling of size $L$ for the PCE of $ Y^{(p)}$ is achieved by detecting a unique sample $\{\xi_l\}_{l=1}^L$ of size $L$, of the basic random variable $\xi$,  then we use eq. \eqref{sec:eq10}, to each realization, namely  
\begin{equation*}
Y^{(p)}_l = \sum_{i=0}^p c_i \Psi_i(\xi_l) \;, l=1,\dots,L\;,
\end{equation*} 
therefore the collection
\begin{equation*}
\left\{Y^{(p)}_l\right\}_{l=1}^L \;, 
\end{equation*} 
is the required PCE sampling.
\end{remark}

\section{Polynomial Chaos Expansion and Stochastic Differential Equations} \label{sec:PCEeSDE}

Let us consider a stochastic process $\{X_t\}_{t\ge 0}$ which satisfies the following SDE 
\begin{equation}
dX_t = r(X_t)dt + \sigma(X_t)dW_t  \label{sec:eq11}\:,
\end{equation}
where $W_t:=\{W_t\}_{t \in [0,T]}$ is a $\R$-valued Brownian motion on the filtered probability space $(\Omega,\Sigma,\Sigma_t,\PP)$, $\{\Sigma_t\}_{t \in [0,T]}$ being  the filtration generated by $W_t$.

Our aim is to write the PCE approximation for the process $X_T$, solution to \eqref{sec:eq11}, at a given, positive and finite time $T$, namely the random variable representing the process at that time. Without loss of generality, in what follows we consider the one dimensional version of the eq. \eqref{sec:eq11}, hence $X_T$ is a scalar, real-valued random variable. In particular we will focus our attention on three specific models which can be described by the eq. \eqref{sec:eq11}, namely  the geometric Brownian motion (gBm), the Vasicek model and the CIR interest rate model. 
Since in each of the aforementioned models the random variable $X_T$ has finite variance, there exists a suitable probability space $(\Omega,\Sigma_T,\PP)$, where $X_T$ is well defined and such that $X_T \in L^2(\Omega,\Sigma_T,\PP)$. Therefore  the PCE method applies
provided the definition of the the functional $\mathcal{M}$ such that $X_T = \mathcal{M}(\xi)$, for a suitable choice of the basic variable 
$\xi$ as seen in Sec. \ref{sec:flow}.

\subsection{Usual Numerical Methods for SDE and PCE} \label{sec:NumPCE}

Numerical methods used to approximate solutions to eq. (\ref{sec:eq1}), usually exploit the independent increments of the Wiener process. For instance, by considering the Euler–Maruyama scheme, see, e.g., \cite[Section 10.2]{Platen}, for each $k = \{0,\dots,L-1\}$, we have
\begin{equation*}
X_{t_{k+1}} =X_{t_k} +r(t_k,X_{t_k}) \Delta t_k+ \sigma(t_k,X_{t_k}) \sqrt{\Delta t_k} \mathcal{N}(0,1) \;,
\end{equation*}
where $\{t_k\}_{k =0}^L$, $L \in \N$ is a set of increasing time steps $0=t_0 < t_1,\dots,t_{L-1}<t_L= T$, and  $X_0$ is the value of the process at starting time $t=0$, while  $\Delta t_k = t_{k+1}-t_k$. To shorten the notation let us define $X_k := X_{t_k}$, for each time step, 
%
then the solution of \eqref{sec:eq11} at time $T$  reads as follows
\begin{align}
X_T = X_L =  X_0 \prod_{k=0}^{L-1}  \left(1+r(t_k,X_{t_k}) \Delta t_k+ \sigma(t_k,X_{t_k}) \sqrt{\Delta t_k}\sqrt{2}\xi_k \right)\;,
\end{align}
or equivalently
\begin{equation*}
X_T = \mathcal{M}(\XI) \qquad \XI = (\xi_0,\dots,\xi_{L-1}) \;,
\end{equation*}
where  $\{\xi_k\}_{k=0}^{L-1}$ is the aforementioned set of i.i.d. Gaussian random variables with zero mean, and with variance equal to $1/2$, namely they are independent $\mathcal{N}(0,1/2)$. Therefore  we are in position to apply the  PCE decomposition, see, e.g.,  \cite[Secton 3.2]{Ernst} for further details concerning multivariate decomposition.
We would like to underline that, even if we consider the Euler-Maruyama scheme, the previous approach can be applied considering  any other method which is based on independent increments of the Brownian motion, such as Milstein method, see, e.g., \cite[Section 10.2, Section 10.3]{Platen}.

Let us recall that 
to get reliable results $L$ has to be great enough, usually $L=100$, which implies a high computational cost since the complexity of the NISP method increases dramatically when more than $20$ inputs are involved. The latter problem is referred as the \emph{curse of dimensionality}, see \cite{Spectral}.

\subsection{Solution functional of a SDE} \label{sec:secDoss}

The theory developed in \cite{Doss} gives conditions to express the solution of an autonomous SDEs as a functional of the Brownian motion. In what follows let us consider an autonomous stochastic differential equation \eqref{sec:eq11},  by assuming that the drift and the volatility are Lipschitz real valued functions, the solution process $\{X_t\}_{t \ge 0}$ of \eqref{sec:eq11} can be expressed as
\begin{equation}
X_t = H(D_t,W_t),\;\; \; t \in [0,T]\;, \label{sec:eq13}
\end{equation}
where $W_t$ denotes the standard Wiener process at time $t$ and $H:\  \R \times \R \to \R$ is a suitable function such that
\begin{equation}
\begin{cases}
\frac{\partial H(x,y)}{\partial y} = \sigma(H(x,y)) \\
H(x,0) = x \label{sec:eq14}
\end{cases}
\;,
\end{equation}
for every fixed $x \in \R$. \newline

Moreover $\{D_t\}_{t \ge 0}$ in \eqref{sec:eq13} is a continuous process, adapted to the filtration of the Brownian motion, and such that for nearly all $\omega \in \Omega$
{\footnotesize
\begin{equation}
\begin{cases}
\dot{D}(\omega) = \exp\left\{- \int_0^{W_t(\omega)}\sigma'(H(D(\omega),W_t(\omega)))\right\} \biggl( \mu \left(H(D(\omega),W_t(\omega))\right) -\frac{1}{2}\sigma(H(D(\omega),W_t(\omega)))\sigma'(H(D(\omega),W_t(\omega))) \biggl)\hspace{0.15cm} ,\; t>0 \\
D(\omega) = X_0  \hspace{13.5 cm},\; t=0\label{sec:eq15}
\end{cases}
\;,
\end{equation}}
thus $\{D_t\}_{t \ge 0}$ is determined path-wise by means of the aforementioned deterministic ODE. Moreover the dependence on time $t$ of $D(\omega)$ is skipped to shorten the notation. For further details on properties of these functions, see, e.g., \cite{Doss}. \newline

Latter approach is suitable for the PCE-approximation method, since it defines the solution of the SDE \eqref{sec:eq11} as a function of the Wiener process, see \eqref{sec:eq13}. 

In particular we can developed the following  general {\it road map},  essentially based on Doss analysis, to apply the PCE-approximation:
\begin{itemize}
\item Given the SDE \eqref{sec:eq11}, the solution functional $H$ is detected by solving \eqref{sec:eq14}.
\item By the very definition of the Weiner process\begin{equation*}
W_t \sim \mathcal{N}(0,t)  \;,  t \in [0,T],
\end{equation*}
therefore $W_t = \sqrt{2t} \xi$, where the {\it basic random variable} $\xi$ is $\mathcal{N}(0,1/2)$, see Sec. \ref{sec:NISP}. Exploiting such feature and integrating the differential equation \eqref{sec:eq15}, where the Brownian Motion is expressed as above, the stochastic process $\{D_t\}_{t \ge 0}$ is path-wise determined  in terms of $\xi$.
\item Eventually, by means of \eqref{sec:eq13}, the solution process is computed as $X_T = H(D_T,W_T)$ for $T>0$. The aforementioned steps allows to express $X_T$ in terms if the {\it basic random variable} $\xi$, providing a NISP friendly definition of $X_T$, see Sec. \ref{sec:NISP}.
\end{itemize}

It is worth to mention that previous steps are discussed in full detail in what follows,  for each numerical application  considered, namely for each  functional solution $H(D_t,W_t)$ related to the specific dynamics of the equity, resp. of the interest rate models, we have taken into account.

\section{PCE approximation of the Geometric Brownian motion} \label{sec:secGBM}

In what follows we analyze the geometric Brownian motion defined as the solution of the following stochastic differential equation
\begin{equation}
dS_t= rS_tdt+\sigma S_t dW_t \label{sec:eq16}\;,
\end{equation}
where $r,\sigma \in \R^+$ and $W_t$ is the usual Wiener process. In particular let us focus our attention on $S_T$, for $T>0$. \newline


The PCE method is based on the definition of a suitable process $\mathcal{M}$. Let us consider (\ref{sec:eq16}) and apply 
the approach described in Sec. \ref{sec:secDoss}, where the solution at time $T$ is defined as
\begin{equation}
S_T = H(D_T,W_T)\;. \label{sec:eq17}
\end{equation}
Since $\sigma(y) = \sigma y$, by means of \eqref{sec:eq14}, $H: \R \times \R \to \R$ is given by
\begin{equation}
H(x,y) = x e^{\sigma y}\;. \label{sec:eq18}
\end{equation}
Furthermore the random variable $D_T$ is determined by integrating \eqref{sec:eq15} up to time $T$. Thus for almost all $\omega \in \Omega$, we have 
{\small
\begin{equation}
\begin{cases}
\dot{D}(\omega) = \exp\big\{- \sigma W_t (\omega)\big\} \Big( rH(D(\omega),W_t(\omega))-\frac{\sigma^2}{2}H(D(\omega),W_t(\omega)) \Big) \hspace{1.95cm} t>0\\
D(\omega) = S_0  \hspace{10 cm} t=0\label{sec:eq19}
\end{cases}
\:,
\end{equation}}
where $\dot{D}(\omega)$ denotes the derivatives with respect to time of $D_t(\omega)$, for every fixed sample event $\omega \in \Omega$. We would like to underlying that the dependence on time, $t$, in previous equation is omitted to shorten the notation.
Furthermore
\begin{equation}
W_t \sim \mathcal{N}(0,t) \label{sec:eq20} \;,  t \in [0,T],
\end{equation}
which allows to set $W_t = \sqrt{2t} \xi$, and by plugging it into \eqref{sec:eq19}, the solution at time $D_T$ can be expressed as a functional depending on the input random variable $\xi$, namely $D_T = \mathcal{M}_1(\xi,\Theta)$. 

We note that if the gBm-model is used  to describe the behaviour of the interest rate $S_t$, e.g. considering interest rate for equity markets, the related output at fixed time $T$, i.e. $S_T$, depends not only on $\xi$ but on other parameters, gathered by $\Theta = (r,\sigma) \in \R^2$, which characterize its dynamics. 

The solution at time $T$ of \eqref{sec:eq19} is numerically computed by means of an adaptive Runge-Kutta method of order 4 (RK4), where the absolute tolerance, resp. relative tolerance, is set as $1e$-$7$, resp. $1e$-$5$; see, e.g.,  \cite[Section II.1, Scetion II.4]{Hairer}, for further details. 

Thus, by means of \eqref{sec:eq17}, we obtain
\begin{equation}
 \mathcal{M}(\xi,\Theta):= S_T = D_T e^{\sigma W_T} = \mathcal{M}_1(\xi,\Theta)e^{\sigma \sqrt{2T} \xi } \label{sec:eq21} \:.
\end{equation}
To shorten the notation, in what follows the dependence on $\Theta$ is omitted.  \newline


Turning back to the analysis of $S_T$, we apply the NISP method,  exploiting eq. (\ref{sec:eq10}), in order to obtain the truncated following PCE 
\begin{equation*}
S_T^{(p)} = \sum_{j=0}^p c_i \Psi_i(\xi) \;,
\end{equation*}
in particular let us set $\{S_{T,j}\}_{j=1}^N = \left\{ \mathcal{M}(\xi_j)\right\}_{j=1}^N$, where $N=p$, as the evaluation of the process (\ref{sec:eq21}) at $\{\xi_j\}_{j=1}^N$,  defined by the Gaussian quadrature formula as stated in  (\ref{sec:eq10}). 
Latter realizations are determined in two steps:
\begin{itemize}
\item each Gaussian quadrature nodes, belonging to $\{\xi_j\}_{j=1}^N$, can be interpreted as the image of a suitable $\omega_j \in \Omega$ through $\xi$. Therefore by integrating \eqref{sec:eq19} for each path in $\{\omega_1,\dots,\omega_N\} \subset \Omega$ we achieve a set of realization of $D_T$, let us say $\{\mathcal{M}_1(\xi_j)\}_{j=1}^N$.
 It is worth to mention that  we are integrating $N$ independent ODEs.
\item By means of \eqref{sec:eq21} we get the required set $\{S_{T,j}\}_{j=1}^N$ by point-wise multiplication of  $\{\mathcal{M}_1(\xi_j)\}_{j=1}^N$ by $\{e^{\sigma \sqrt{2\cdot T}\xi_j}\}_{j=1}^N$
\end{itemize}

\subsection{Numerical application: overview of the computations} \label{sec:Ngbm}
In what follows we give a PCE-approximation of the solution to the  SDE (\ref{sec:eq16}), for some particular values of the parameters involved, see Table \ref{tab:tab1}. \newline
\begin{table}[!h]
\centering
\begin{tabular}{c|cccc}
\centering Parameters & \centering  r & \centering  $S_0$  & \centering T \tabularnewline
\hline
\centering Values & \centering 3\% & \centering 100 & \centering 1 \tabularnewline
\end{tabular}
\caption{Parameters of gBm } \label{tab:tab1}
\end{table} 

In order to make the discussion as complete as possible, our method is tested on a set of volatility values
\begin{equation*} 
\sigma =\Big \{ 15\%,25\%,30\% \Big \}\:.
\end{equation*}
Since eq. \eqref{sec:eq16} has an analytical solution, such a solution will constitute our benchmark, in particular, at time $T$, we have 
\begin{equation}
S_T= S_0e^{\left(r-\frac{1}{2}\sigma^2 \right)T +\sigma W_T} \label{sec:eq22}\;,
\end{equation}
whose mean and variance are respectively
\begin{align*}
\Expect{S_T} &= S_0 e^{rT} \;,\\ 
\Var{S_T} &=  S_0^2 e^{2rT}\left(e^{\sigma^2T}-1\right)\:.
\end{align*}
Then let us compute the absolute errors of the mean and the variance, namely
\begin{align*}
\epsilon^{(p)}_{MEAN} &= \abs{\Expect{S_T} -\Expect{S^{(p)}_T}}\;,\\
\epsilon^{(p)}_{VAR} &= \abs{\Var{S_T}-\Var{S^{(p)}_T} }\;,
\end{align*}
the absolute value of the relative error being also considered. Due to  the logarithmic scale, we  look at  their absolute values, highlighting the related order, rather than its numerical value. Therefore
\begin{align*}
RE^{(p)}_{MEAN} &= \frac{\epsilon^{(p)}_{MEAN}}{\Expect{S_T}}\;,\\
RE^{(p)}_{VAR} &= \frac{\epsilon^{(p)}_{VAR}}{\Var{S_T}}\:,
\end{align*}
We note that  such data have been computed for an increasing set of degrees
$
p = \{1,2,\dots,15\}
$\:.
$\ $\newline

 For the aforementioned set of degree, and for each value of the volatility $\sigma$, we will compute two quantiles $\hat{Q}_\gamma$, where $\gamma = 99\%$ and $\gamma=99.9\%$, of the PCE approximation $S_T^{(p)}$, also writing their analytical values.
Recalling that the standard statistics for $Q_{\gamma}$ is the $\gamma$-th sample quantile, namely the $(K+1)$-th realization of the sampling of $S_T^{(p)}$, sort in ascending order, such that
$
K \le [\gamma M]
$, 
where $M$ is the size of the sampling, and $[\cdot]$ denotes the integer part of the real number within the brackets. In our analysis we employ a  Latin Hypercube Sampling (LHS) of $S_T^{(p)}$, see, e.g., \cite{Tang, Owen} and \cite{Iman} for further references, of size $M=5000$. Thus we  have $K_{99\%} = 4951$, while  $K_{99.9\%} = 4996$, see, e.g., \cite{David} for further details, and we compute the absolute error of $\hat{Q}_{\gamma}$ versus the analytical values ${Q}_\gamma$.

In particular we are aiming at computing the absolute error of $\hat{Q}_{\gamma}$ with respect to the analytical values, which are
\begin{equation*}
Q_{\gamma} = \exp\left\{\left(r-\frac{\sigma^2}{2}\right)T+\sigma \sqrt{T}Z_{\gamma}\right\}\;,
\end{equation*}
where $Z_{\gamma}$ represents the quantile of the normal random variable of zero mean and unitary variance. 
Thus the error are defined as
\begin{equation*}
\epsilon_{\gamma} = \abs{\hat{Q}_{\gamma}-{Q}_{\gamma}}\;,
\end{equation*}

As last comparison let us estimate ${Q}_\gamma$, where $\gamma = 99\%$ and $\gamma=99.9\%$, by means of the aforementioned sample quantile applied, where a standard Monte Carlo sampling of the analytical solution $S_T$ to \eqref{sec:eq22}.

The accuracy of such computation is represented by the standard error of the estimated average, namely for a set $\{Q_\gamma^{MC}(l)\}_{l=1}^L$ of $L=200$ independent estimates of the quantile, whose arithmetic average is \linebreak $\bar{Q}_\gamma^{MC}= (1/L)\cdot \sum_{l=1}^L Q_\gamma^{MC}(l)$, we detect its standard error 
\begin{equation*}
SE_{Q^{MC}_\gamma} = \frac{\hat{\sigma}}{\sqrt{L}}\;,
\end{equation*}
where the estimated variance is
\begin{equation*}
\hat{\sigma}^2 = \frac{1}{L-1}\sum_{l=1}^L \left(Q_\gamma^{MC}(l)-\bar{Q}_\gamma^{MC}\right)^2\;,
\end{equation*}

\begin{remark}.
We would like to underline that the choice of the basic estimator for quantiles, i.e. the sample quantile, 
is due to focus our attention to the efficacy of the method, instead of taking care of the estimates accuracy as well as providing a fair comparison between the data achieved. Nevertheless the dedicated literature provides other techniques, which are often more accurate, such as the \emph{Two-phase quantile estimator} presented in \cite{Chen},   the \emph{L-estimator}, or the d \emph{Harrel Davis} (HD) estimators, see, e.g., \cite{Mausser} for further details.
\end{remark}

\subsection{$\boldsymbol{\sigma=15\%}$}

In this section we set $\sigma=15\%$, while other parameters are displayed in Table \ref{tab:tab1}.

\begin{figure}[!h]
\centering
\includegraphics[scale=0.33]{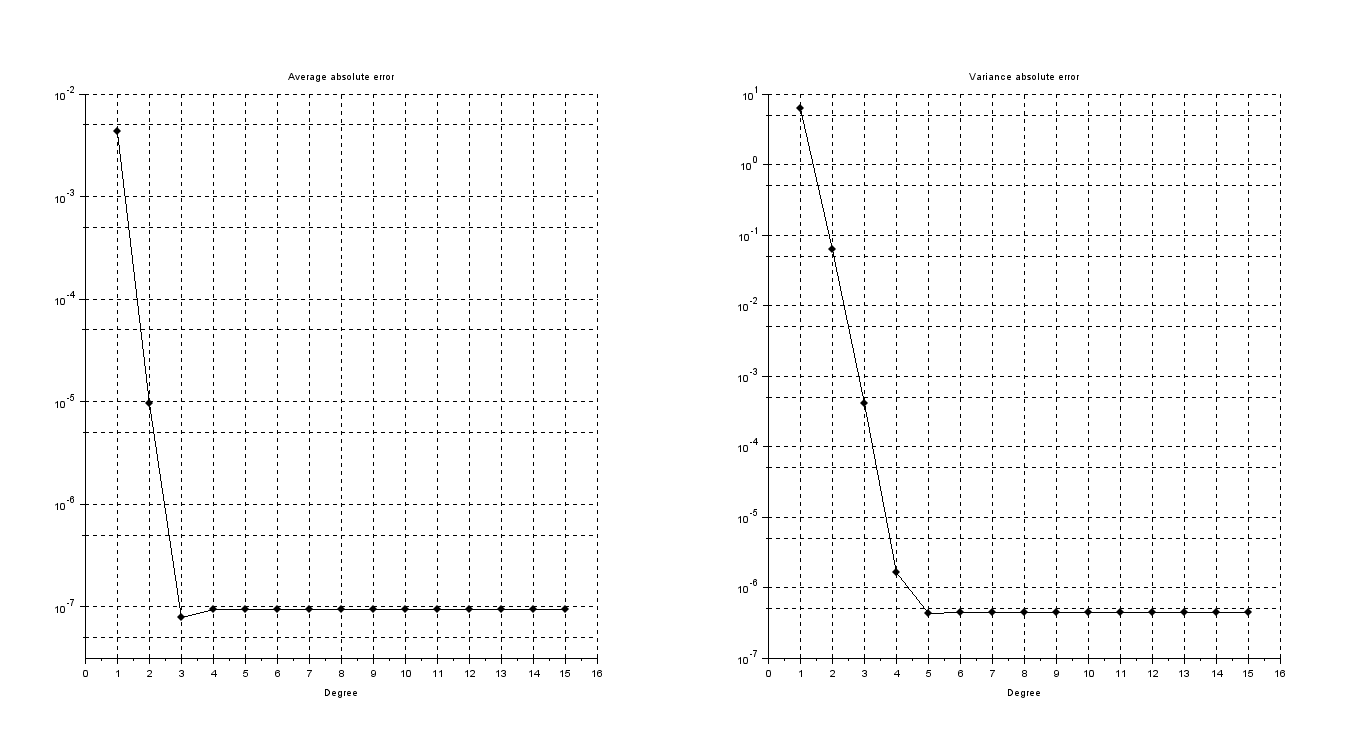}
\caption{Semilogy scale plot of the absolute error of the mean (left) and the variance (right) computed via PCE for gBm at time $T=1$, whose parameters are $r = 3 \%$, $\sigma =15 \%$ and starting value $S_0= 100$,  for a set of degrees $p = \{1,2,\dots,15\}$.}  \label{fig:fig2}
\end{figure}

\begin{figure}[!h]
\centering
\includegraphics[scale=0.33]{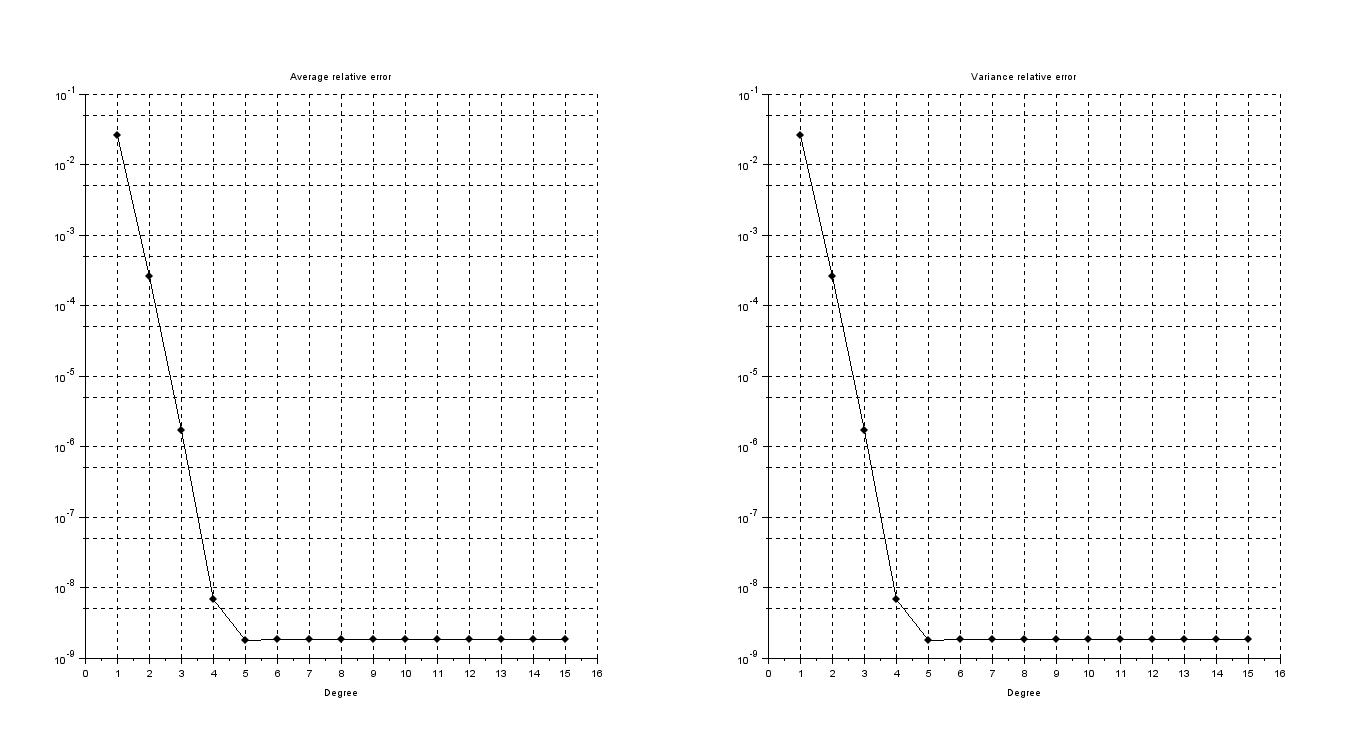}
\caption{Semilogy scale plot of the absolute value of the relative error of the mean (left) and the variance (right) computed via PCE for gBm at time $T=1$, whose parameters are $r = 3 \%$, $\sigma = 15 \%$ and starting value $S_0= 100$, for a set of degrees $p = \{1,2,\dots,15\}$.}  \label{fig:fig3}
\end{figure}

First let us display in Figure \ref{fig:fig2}, Figure  \ref{fig:fig3} and Table \ref{tab:tab2}, the absolute and relative error of the average, resp. variance, of the PCE-approximation of the gBm. 

\begin{table}[!h]
\centering
\begin{tabular}{ccccc}
 Degree of PCE &  Average Error & Variance Error & Average relative error & Variance relative error  \tabularnewline
\hline
 1 & 4.3211e-03 & 6.2663e+00 & 4.1934e-05 & 2.5934e-02 \tabularnewline 
 2 & 9.6288e-06 & 6.2400e-02 & 9.3443e-08 & 2.5825e-04 \tabularnewline 
 3 & 7.8038e-08 & 4.1041e-04 & 7.5732e-10 & 1.6986e-06 \tabularnewline 
 4 & 9.3644e-08 & 1.6273e-06 & 9.0877e-10 & 6.7351e-09 \tabularnewline 
 5 & 9.3664e-08 & 4.3078e-07 & 9.0896e-10 & 1.7829e-09 \tabularnewline 
 6 & 9.3664e-08 & 4.3922e-07 & 9.0896e-10 & 1.8178e-09 \tabularnewline 
 7 & 9.3664e-08 & 4.3925e-07 & 9.0896e-10 & 1.8179e-09 \tabularnewline 
 8 & 9.3664e-08 & 4.3925e-07 & 9.0896e-10 & 1.8179e-09 \tabularnewline 
 9 & 9.3664e-08 & 4.3925e-07 & 9.0896e-10 & 1.8179e-09 \tabularnewline 
 10 & 9.3664e-08 & 4.3925e-07 & 9.0896e-10 & 1.8179e-09 \tabularnewline 
 11 & 9.3664e-08 & 4.3925e-07 & 9.0896e-10 & 1.8179e-09 \tabularnewline 
 12 & 9.3663e-08 & 4.3925e-07 & 9.0895e-10 & 1.8179e-09 \tabularnewline 
 13 & 9.3664e-08 & 4.3925e-07 & 9.0895e-10 & 1.8179e-09 \tabularnewline 
 14 & 9.3664e-08 & 4.3925e-07 & 9.0895e-10 & 1.8179e-09 \tabularnewline 
 15 & 9.3664e-08 & 4.3925e-07 & 9.0895e-10 & 1.8179e-09 \tabularnewline 
\end{tabular}
\caption{Absolute and relative errors of average and variance for PCE approximation of gBm at time $T=1$, whose parameters are $r = 3 \%$, $\sigma = 15 \%$ and starting value $S_0= 100$} \label{tab:tab2}
\end{table}
\newpage 

\begin{figure}[!h]
\centering
\includegraphics[scale=0.32]{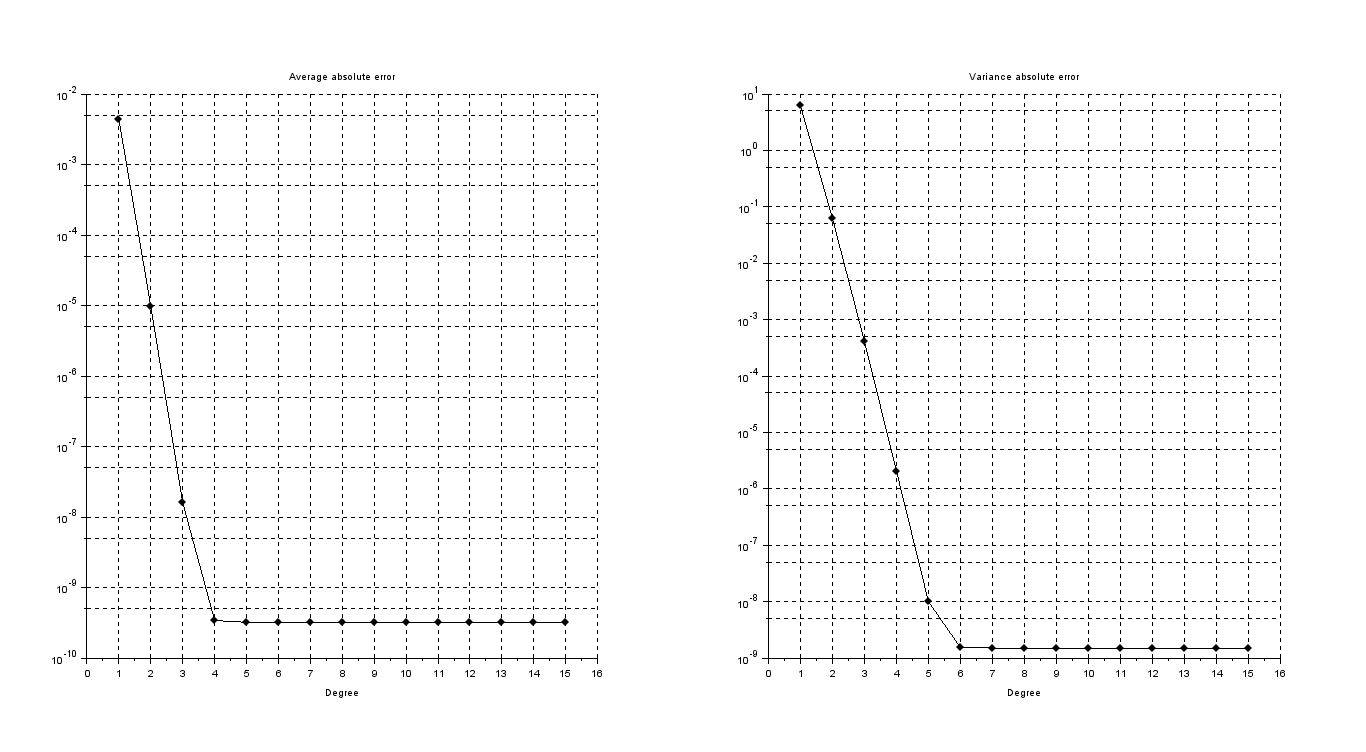}
\caption{Semilogy scale plot for the absolute error of the mean (left) and variance (right) computed via PCE of gBm at time $T=1$ ( whose parameters are $r = 3 \%$, $\sigma = 15 \%$ and starting value $S_0= 100$) with higher precision at computing $\{S_{T,j}\}_{j=1}^N$}  \label{fig:fig4}
\end{figure}

\begin{figure}[!h]
\centering
\includegraphics[scale=0.32]{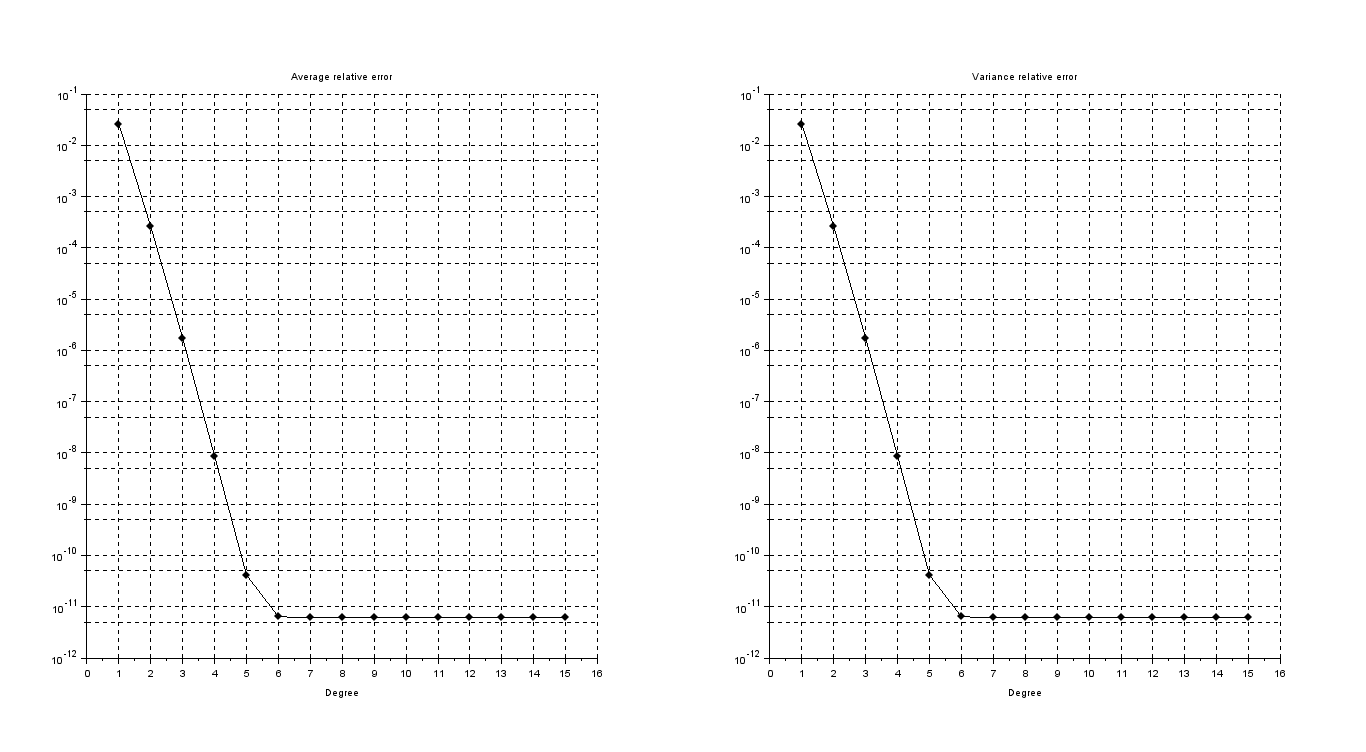}
\caption{Semilogy scale plot for the absolute value of the relative error of the mean (left) and variance (right) computed via PCE of gBm at time $T=1$ ( whose parameters are $r = 3 \%$, $\sigma = 15 \%$ and starting value $S_0= 100$) with higher precision at computing $\{S_{T,j}\}_{j=1}^N$}  \label{fig:fig5}
\end{figure}

Both average and variance error are stationary, despite an expected spectral convergence for increasing degree $p$. This can be motivated by the presence of an error, which is not due to the PCE method, that corrupts the expected spectral converge of $S_T^{(p)}$. 
There are only two possible causes: the analytical approximation of the solution $S_T$ with (\ref{sec:eq17}) and the error coming form numerical computations of $\left\{ \mathcal{M}(\xi_j)\right\}_{j=1}^N$, see eq. \eqref{sec:eq21}, used in (\ref{sec:eq10}). 
Let us focus our attention on the latter: as  displayed in  Figure \ref{fig:fig4},  Figure \ref{fig:fig5} and Table \ref{tab:tab3}, by increasing the precision used to compute $\left\{ \mathcal{M}(\xi_j)\right\}_{j=1}^N$ the errors decrease. This is achieved by setting the absolute tolerance, resp. relative tolerance, of RK4 method, used to compute \eqref{sec:eq19}, at $1e$-$15$, resp. $1e$-$10$.

 Hence PCE errors are influenced by such approximation required to compute the coefficients. In Fig. \ref{fig:fig4} we can see the spectral convergence of the error for the PCE approximation up to 4-th degree for mean, resp. up to the 6-th degree for variance.  Note that, although the increased accuracy, the error displays again a stationary behavior, but actually no improvement on numerical methods can be implemented, which allows us to conclude that this is the only source of error. \newline

\begin{table}[!h]
\centering
\begin{tabular}{ccccc}
 Degree of PCE &  Average Error & Variance Error & Average relative error& Variance relative error  \tabularnewline
\hline
 1 & 4.3212e-03 & 6.2663e+00 & 4.1935e-05 & 2.5934e-02 \tabularnewline 
 2 & 9.7228e-06 & 6.2400e-02 & 9.4355e-08 & 2.5826e-04 \tabularnewline 
 3 & 1.5946e-08 & 4.1085e-04 & 1.5475e-10 & 1.7004e-06 \tabularnewline 
 4 & 3.4008e-10 & 2.0681e-06 & 3.3003e-12 & 8.5592e-09 \tabularnewline 
 5 & 3.2061e-10 & 9.9706e-09 & 3.1114e-12 & 4.1265e-11 \tabularnewline 
 6 & 3.2060e-10 & 1.5312e-09 & 3.1112e-12 & 6.3373e-12 \tabularnewline 
 7 & 3.2060e-10 & 1.5023e-09 & 3.1112e-12 & 6.2175e-12 \tabularnewline 
 8 & 3.2057e-10 & 1.5025e-09 & 3.1109e-12 & 6.2186e-12 \tabularnewline 
 9 & 3.2057e-10 & 1.5022e-09 & 3.1109e-12 & 6.2173e-12 \tabularnewline 
 10 & 3.2057e-10 & 1.5020e-09 & 3.1109e-12 & 6.2165e-12 \tabularnewline 
 11 & 3.2048e-10 & 1.5026e-09 & 3.1101e-12 & 6.2187e-12 \tabularnewline 
 12 & 3.2111e-10 & 1.5039e-09 & 3.1162e-12 & 6.2241e-12 \tabularnewline 
 13 & 3.2078e-10 & 1.5025e-09 & 3.1130e-12 & 6.2183e-12 \tabularnewline 
 14 & 3.2074e-10 & 1.5029e-09 & 3.1126e-12 & 6.2202e-12 \tabularnewline 
 15 & 3.2071e-10 & 1.5019e-09 & 3.1123e-12 & 6.2160e-12 \tabularnewline 
\end{tabular}
\caption{Absolute error of the average and the variance of PCE approximation of gBm at time $T=1$ ( whose parameters are $r = 3 \%$, $\sigma = 15 \%$ and starting value $S_0= 100$) for higher precision at computing $\{S_{T,j}\}_{j=1}^N$} \label{tab:tab3}
\end{table}

\begin{figure}[!h]
\centering
\includegraphics[scale=0.33]{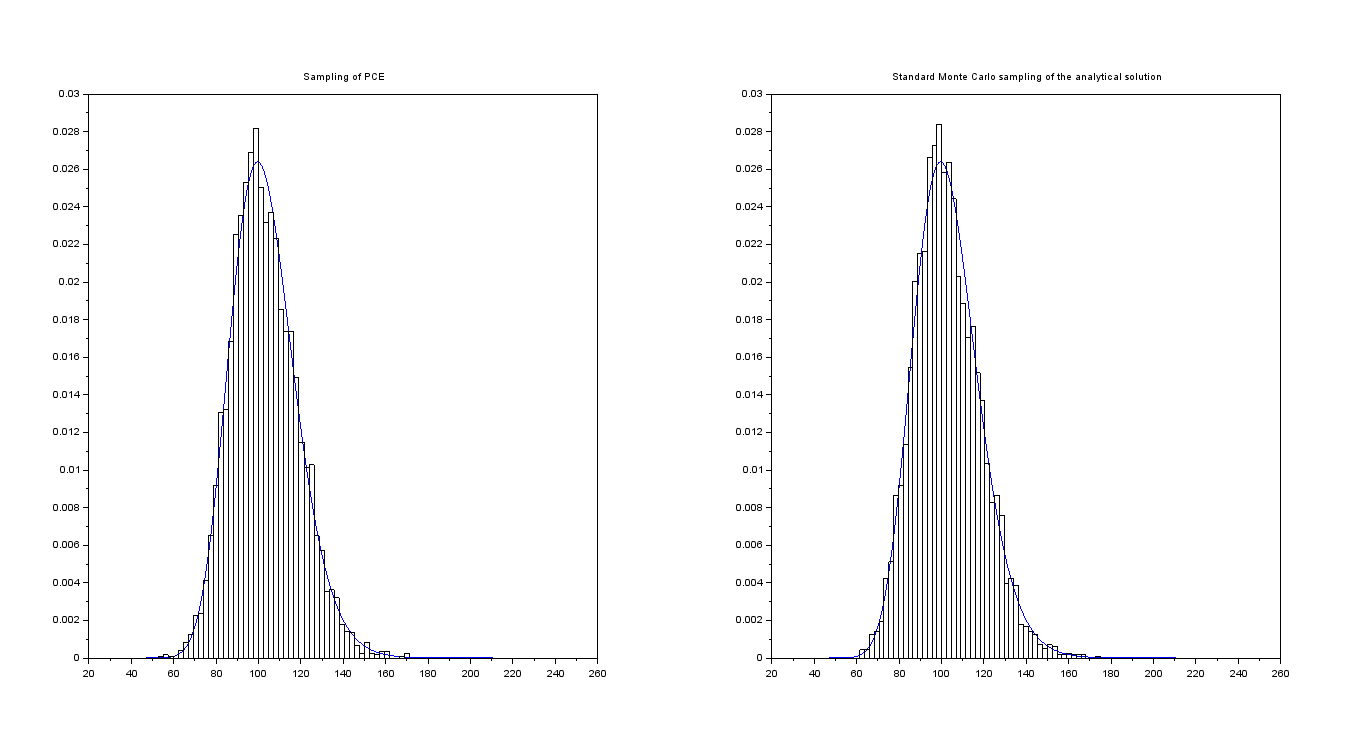}
\caption{Probability density function of gBm at time $T=1$, whose parameters are $r = 3 \%$, $\sigma = 15 \%$ and starting value $S_0 = 100$, (blue curve) and histogram of a Monte Carlo sampling (size = $5000$) of the $S_T^{(p)}$ for $p=15$ (left). The right plot displays the analytical probability density function of gBm, for the same parameters values, and a standard Monte Carlo sampling of size $5000$ of the gBm-analytical solution.}  \label{fig:fig6}
\end{figure}

\newpage
Coming back to  Table \ref{tab:tab1}, let us compare a sampling of size $5000$ of $S_T^{(p)}$ for $p=15$, achieved with standard Monte Carlo technique, with the probability density function of the analytical solution to eq. \eqref{sec:eq16} at time $T$, whose distribution is lognormal, namely
\begin{equation*}
 X =  \mathcal{N}\left(\log(S_0)+\left(r-\frac{\sigma^2}{2}\right)T\ ,\  \sigma^2 T \right) \:,\: S_T = e^X\:,
\end{equation*}
computed  values are shown in Figure \ref{fig:fig6} were the standard Monte Carlo sampling of $S_T$ is shown.

%
Let us determine the $99\%$ and $99.9\%$ quantiles of PCE-approximation, by means of the sample quantile based on Latin Hypercube Sampling (LHS) technique. These values  are compared with the analytical ones, see Sec. \ref{sec:Ngbm} for further details.

\begin{table}[!h]
\centering
\begin{tabular}{cccc}
 Degree of PCE &  $\epsilon_{99\%}$ & $\epsilon_{99.9\%}$   \tabularnewline
\hline
 1 & 5.6531e+00 & 1.1050e+01 \tabularnewline 
 2 & 2.7440e-01 & 6.0752e-01 \tabularnewline 
 3 & 8.1900e-02 & 6.6493e-01 \tabularnewline 
 4 & 8.3246e-02 & 7.5604e-01 \tabularnewline 
 5 & 8.1763e-02 & 7.5891e-01 \tabularnewline 
 6 & 8.1672e-02 & 7.5879e-01 \tabularnewline 
 7 & 8.1672e-02 & 7.5878e-01 \tabularnewline 
 8 & 8.1673e-02 & 7.5878e-01 \tabularnewline 
 9 & 8.1673e-02 & 7.5878e-01 \tabularnewline 
 10 & 8.1673e-02 & 7.5878e-01 \tabularnewline 
 11 & 8.1673e-02 & 7.5878e-01 \tabularnewline 
 12 & 8.1673e-02 & 7.5878e-01 \tabularnewline 
 13 & 8.1673e-02 & 7.5878e-01 \tabularnewline 
 14 & 8.1673e-02 & 7.5878e-01 \tabularnewline 
 15 & 8.1673e-02 & 7.5878e-01 \tabularnewline 
\end{tabular}
\caption{Absolute errors of the two quantiles $\hat{Q}_{99\%}$  and $\hat{Q}_{99.9\%}$ of the PCE approximation of gBm at time $T=1$,  whose parameters are $r = 3 \%$, $\sigma = 15 \%$ and starting value $S_0= 100$.} \label{tab:tab4}
\end{table}

\begin{figure}[!h]
\centering
\includegraphics[scale=0.5]{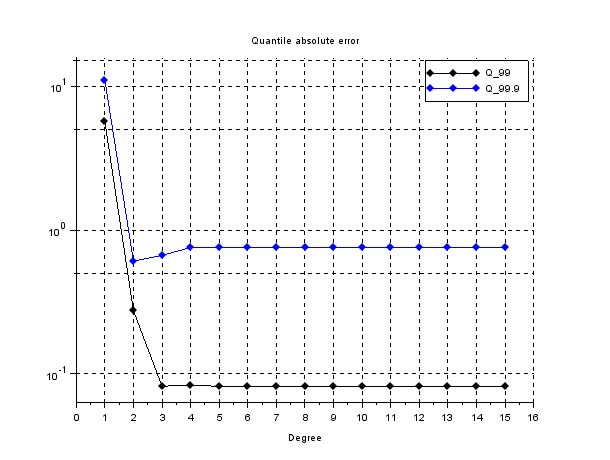}
\caption{Absolute errors of the two quantiles $\hat{Q}_{99\%}$  and $\hat{Q}_{99.9\%}$ of the PCE approximation of gBm at time $T=1$,  whose parameters are $r = 3 \%$, $\sigma = 15 \%$ and starting value $S_0= 100$.} \label{fig:fig7} 
\end{figure}

Eventually let us compute the two aforementioned quantiles by means of a standard Monte Carlo sampling of the analytical solution $S_T$, see \eqref{sec:eq22}. As discussed in Sec. \ref{sec:Ngbm}, the standard errors of the quantiles are shown
\begin{align*}
&SE_{Q_{99\%}^{MC}} = 0.0817117         \\ &SE_{Q_{99.9\%}^{MC}} =    0.2478969
\end{align*}

\subsection{$\boldsymbol{\sigma=25\%}$}

In this section we set $\sigma=25\%$, while the other parameters takes the values of Table \ref{tab:tab1}. \newline

\begin{figure}[!h]
\centering
\includegraphics[scale=0.33]{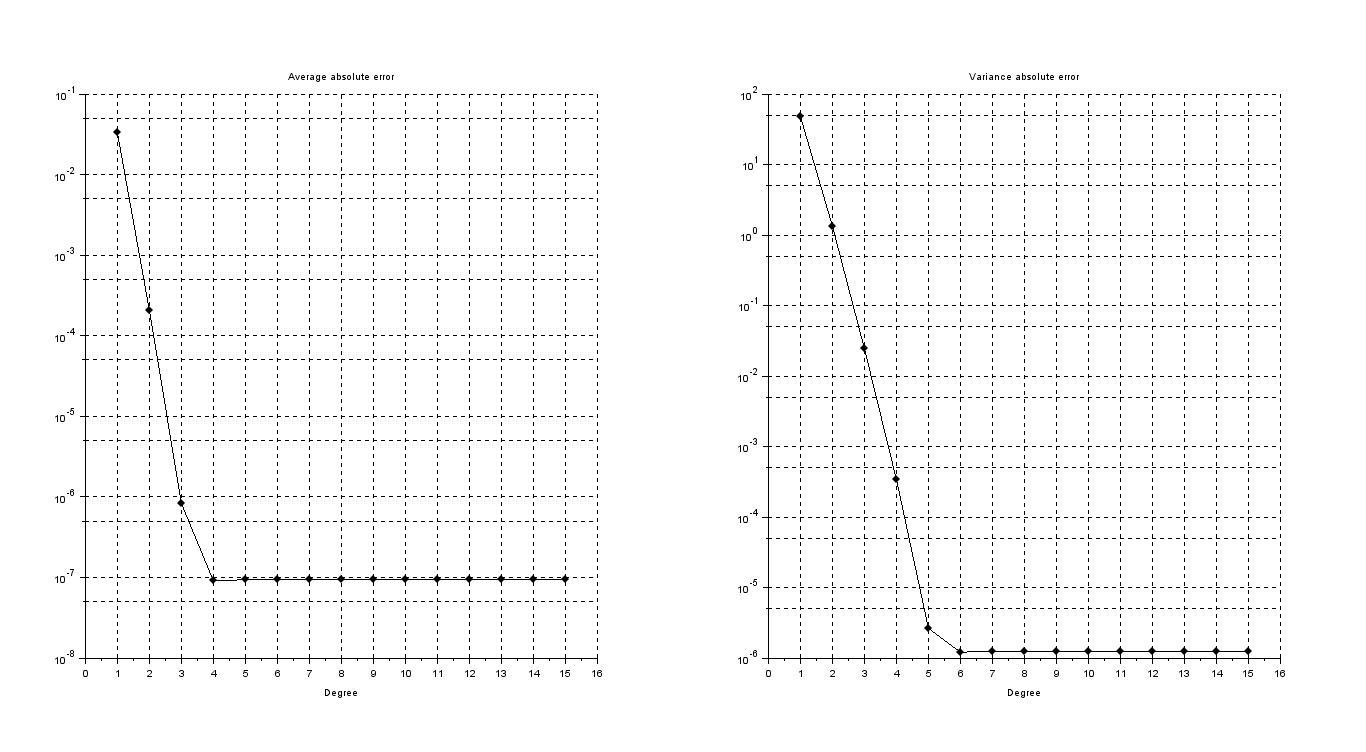}
\caption{Semilogy scale plot of the absolute error of the mean (left) and the variance (right) computed via PCE for gBm at time $T=1$, whose parameters are $r = 3 \%$, $\sigma = 25 \%$ and starting value $S_0= 100$,  for a set of degrees $p = \{1,2,\dots,15\}$.}  \label{fig:fig8}
\end{figure}

\begin{figure}[!h]
\centering
\includegraphics[scale = 0.33]{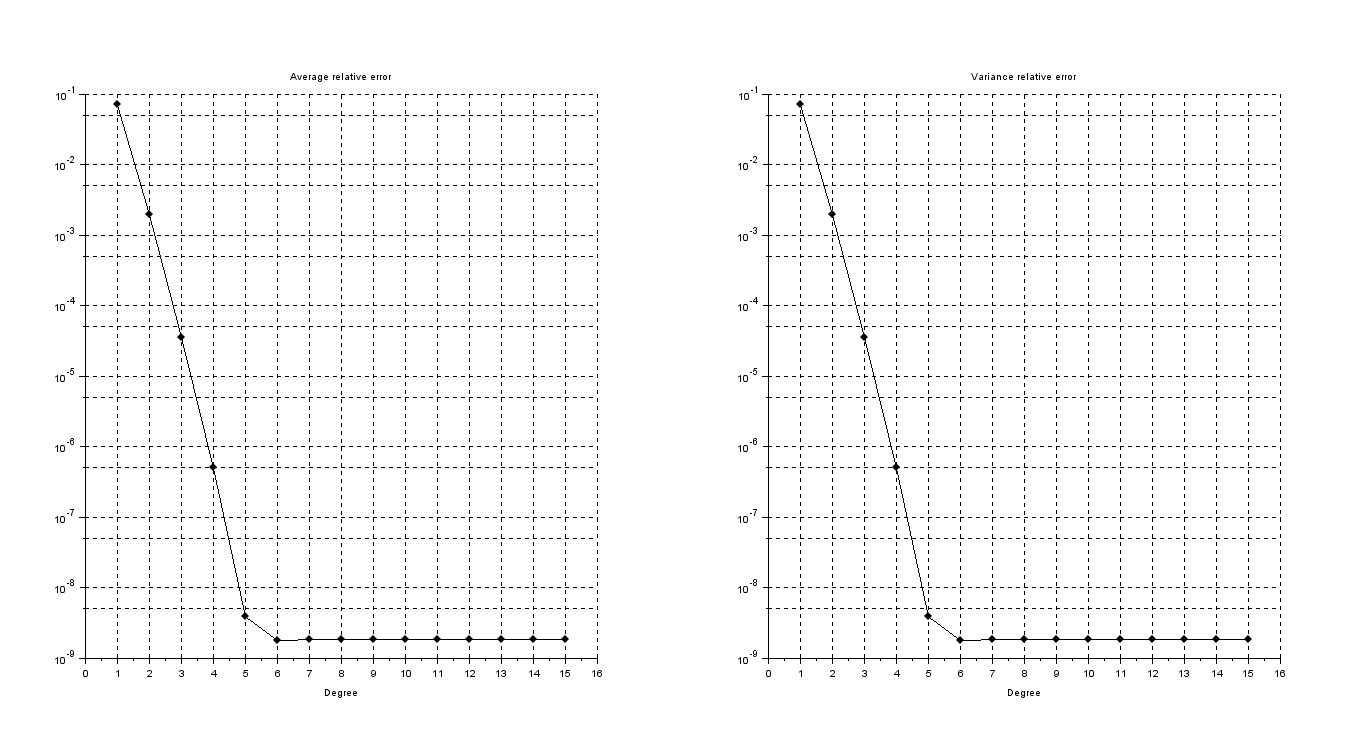}
\caption{Semilogy scale plot of the absolute value of the relative error of the mean (left) and the variance (right) computed via PCE for gBm at time $T=1$, whose parameters are $r = 3 \%$, $\sigma = 25 \%$ and starting value $S_0= 100$, for a set of degrees $p = \{1,2,\dots,15\}$.}  \label{fig:fig9}
\end{figure}

First let us display in Figure \ref{fig:fig8}, Figure  \ref{fig:fig9} and Table \ref{tab:tab5} the absolute and relative error of the average, resp. variance, of the PCE-approximation of gBm. 

\begin{table}[!h]
\centering
\begin{tabular}{ccccc}
 Degree of PCE &  Average Error & Variance Error & Average relative error & Variance relative error  \tabularnewline
\hline
 1 & 3.2989e-02 & 4.8289e+01 & 3.2014e-04 & 7.0513e-02 \tabularnewline 
 2 & 2.0607e-04 & 1.3341e+00 & 1.9998e-06 & 1.9481e-03 \tabularnewline 
 3 & 8.2671e-07 & 2.4397e-02 & 8.0228e-09 & 3.5625e-05 \tabularnewline 
 4 & 9.0548e-08 & 3.3971e-04 & 8.7872e-10 & 4.9606e-07 \tabularnewline 
 5 & 9.3735e-08 & 2.6355e-06 & 9.0965e-10 & 3.8484e-09 \tabularnewline 
 6 & 9.3744e-08 & 1.2087e-06 & 9.0973e-10 & 1.7649e-09 \tabularnewline 
 7 & 9.3744e-08 & 1.2457e-06 & 9.0973e-10 & 1.8190e-09 \tabularnewline 
 8 & 9.3744e-08 & 1.2460e-06 & 9.0973e-10 & 1.8195e-09 \tabularnewline 
 9 & 9.3744e-08 & 1.2460e-06 & 9.0973e-10 & 1.8195e-09 \tabularnewline 
 10 & 9.3744e-08 & 1.2460e-06 & 9.0973e-10 & 1.8195e-09 \tabularnewline 
 11 & 9.3744e-08 & 1.2460e-06 & 9.0974e-10 & 1.8195e-09 \tabularnewline 
 12 & 9.3744e-08 & 1.2460e-06 & 9.0973e-10 & 1.8195e-09 \tabularnewline 
 13 & 9.3744e-08 & 1.2460e-06 & 9.0973e-10 & 1.8195e-09 \tabularnewline 
 14 & 9.3744e-08 & 1.2460e-06 & 9.0973e-10 & 1.8195e-09 \tabularnewline 
 15 & 9.3744e-08 & 1.2460e-06 & 9.0973e-10 & 1.8195e-09 \tabularnewline 
\end{tabular}
\caption{Absolute and relative errors of average and variance for PCE approximation of gBm  at time $T=1$, whose parameters are $r = 3 \%$, $\sigma = 25 \%$ and starting value $S_0= 100$} \label{tab:tab5}
\end{table}
\newpage 
As in the case where $\sigma = 15\%$, both average and variance error are stationary, despite an expected spectral convergence for increasing degree $p$. This can be motivated by the presence of an error, not due to PCE method, that corrupts the expected spectral converge of $S_T^{(p)}$. 
As before, there are only two possible causes: the approximation of the solution of (\ref{sec:eq16}) with (\ref{sec:eq17}) and the error coming form numerical computations of $\left\{ \mathcal{M}(\xi_j)\right\}_{j=1}^N$, by means of \eqref{sec:eq21}, used in (\ref{sec:eq10}). 
Focusing the attention on the latter: as  displayed in  Table \ref{tab:tab6}, by increasing the precision used to compute $\left\{ \mathcal{M}(\xi_j)\right\}_{j=1}^N$ the errors decrease, namely we set the absolute and relative tolerance in RK4 method as $1e$-$15$ and $1e$-$10$. Hence PCE errors are influenced by such a approximation required to compute the coefficients. In Tab.  \ref{tab:tab6} we can see the spectral convergence of the error for the PCE approximation up to 4-th degree for mean, resp up to the 6-th degree for variance.  Note that, although the increased accuracy, the error displays again a stationary behavior, but actually no improvement on numerical methods can be implemented, hence we can once again conclude  that this is the only source of error. \newline

\begin{table}[!h]
\centering
\begin{tabular}{ccccc}
 Degree of PCE &  Average Error & Variance Error & Average relative error& Variance relative error  \tabularnewline
\hline
 1 & 3.2989e-02 & 4.8289e+01 & 3.2015e-04 & 7.0513e-02 \tabularnewline 
 2 & 2.0617e-04 & 1.3341e+00 & 2.0007e-06 & 1.9481e-03 \tabularnewline 
 3 & 9.2078e-07 & 2.4398e-02 & 8.9356e-09 & 3.5627e-05 \tabularnewline 
 4 & 3.5168e-09 & 3.4097e-04 & 3.4129e-11 & 4.9789e-07 \tabularnewline 
 5 & 3.2968e-10 & 3.8858e-06 & 3.1993e-12 & 5.6741e-09 \tabularnewline 
 6 & 3.2060e-10 & 4.1608e-08 & 3.1112e-12 & 6.0758e-11 \tabularnewline 
 7 & 3.2058e-10 & 4.5733e-09 & 3.1111e-12 & 6.6780e-12 \tabularnewline 
 8 & 3.2057e-10 & 4.2648e-09 & 3.1109e-12 & 6.2276e-12 \tabularnewline 
 9 & 3.2058e-10 & 4.2643e-09 & 3.1111e-12 & 6.2268e-12 \tabularnewline 
 10 & 3.2055e-10 & 4.2613e-09 & 3.1108e-12 & 6.2225e-12 \tabularnewline 
 11 & 3.2048e-10 & 4.2659e-09 & 3.1101e-12 & 6.2291e-12 \tabularnewline 
 12 & 3.2108e-10 & 4.2641e-09 & 3.1159e-12 & 6.2265e-12 \tabularnewline 
 13 & 3.2077e-10 & 4.2605e-09 & 3.1129e-12 & 6.2213e-12 \tabularnewline 
 14 & 3.2074e-10 & 4.2667e-09 & 3.1126e-12 & 6.2303e-12 \tabularnewline 
 15 & 3.2070e-10 & 4.2598e-09 & 3.1122e-12 & 6.2203e-12 \tabularnewline 
\end{tabular}
\caption{Absolute error of the average and the variance of PCE approximation of gBm at time $T=1$, whose parameters are $r = 3 \%$, $\sigma = 25 \%$ and starting value $S_0= 100$, for higher precision at computing $\{S_{T,j}\}_{j=1}^N$} \label{tab:tab6}
\end{table}

\begin{figure}[!h]
\centering
\includegraphics[scale=0.33]{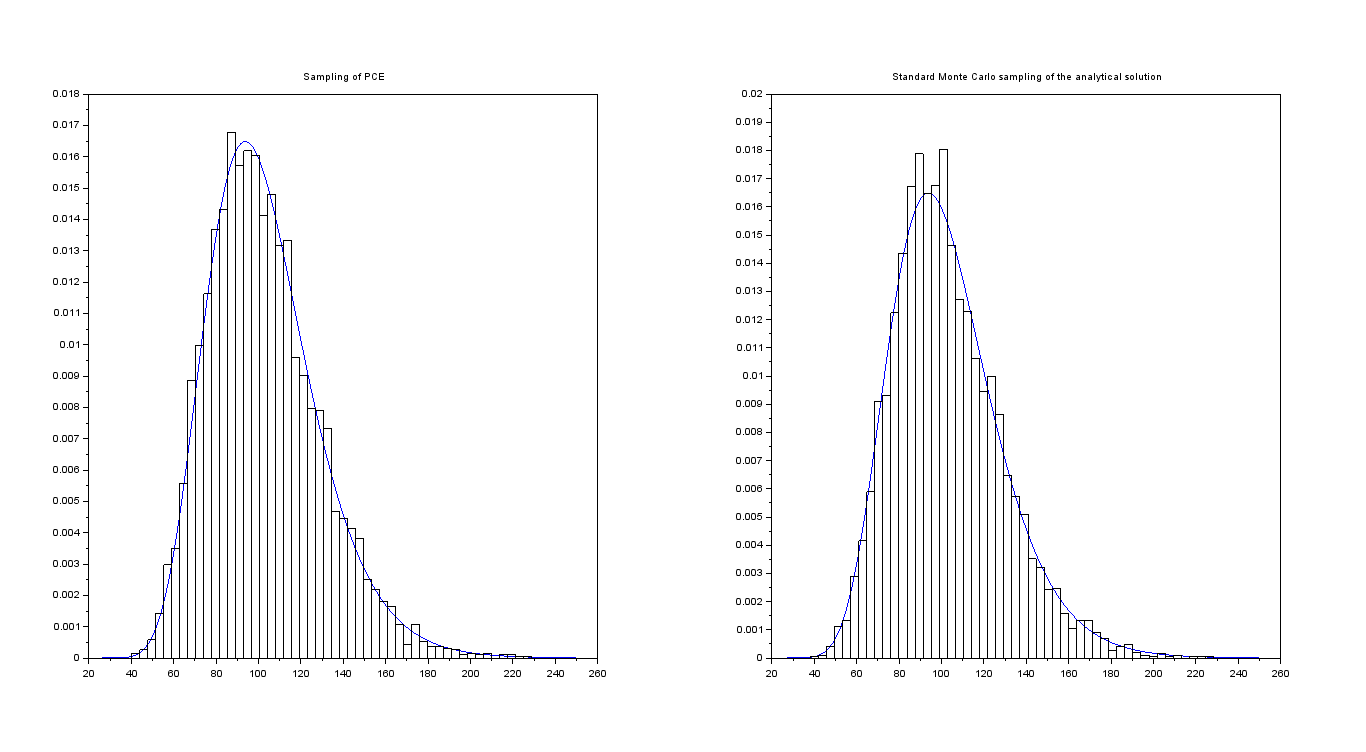}
\caption{Probability density function of gBm at time $T=1$, whose parameters are $r = 3 \%$, $\sigma = 25 \%$ and starting value $S_0 = 100$, (blue curve) and histogram of a Monte Carlo sampling (size = $5000$) of the $S_T^{(p)}$ for $p=15$ (left). The right plot displays the analytical probability density function of gBm, for the same parameters values, and a standard Monte Carlo sampling of size $5000$ of the gBm-analytical solution.}  \label{fig:fig10}
\end{figure}

\newpage
Coming back to  Table \ref{tab:tab1}, let us compute a sampling of size $5000$ of $S_T^{(p)}$ for $p=15$, achieved with standard Monte Carlo technique, which is compared with the probability density function of the analytical solution to eq. \eqref{sec:eq16} at time $T$, whose distribution is lognormal, namely 
\begin{equation*}
 X =  \mathcal{N}\left(\log(S_0)+\left(r-\frac{\sigma^2}{2}\right)T\ ,\  \sigma^2 T \right) \:,\: S_T = e^X\:,
\end{equation*}
the computed  values are shown in Figure \ref{fig:fig10} were the standard Monte Carlo sampling of $S_T$ is shown.

\newpage

\begin{figure}[!h]
\centering
\includegraphics[scale=0.5]{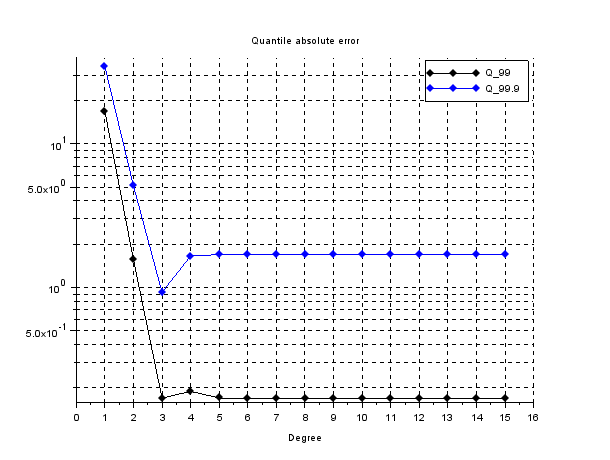}
\caption{Absolute errors of the two quantiles $\hat{Q}_{99\%}$  and $\hat{Q}_{99.9\%}$ of the PCE approximation of gBm at time $T=1$,  whose parameters are $r = 3 \%$, $\sigma = 25 \%$ and starting value $S_0= 100$.} \label{fig:fig11} 
\end{figure}

Then, for each degree $p$, we compute the two quantiles $\hat{Q}_{\alpha}$ for $ \gamma = 99\%$ and $\gamma=99.9\%$ by means of the sample quantile, see Sec. \ref{sec:Ngbm}, based on Latin Hypercube Sampling (LHS) technique.
 Obtained values are shown in Figure  \ref{fig:fig11} and in Table \ref{tab:tab7}.

\begin{table}[!h]
\centering
\begin{tabular}{cccc}
 Degree of PCE &  $\epsilon_{99\%}$ & $\epsilon_{99.9\%}$   \tabularnewline
\hline
 1 & 1.6863e+01 & 3.4495e+01 \tabularnewline 
 2 & 1.5613e+00 & 5.1013e+00 \tabularnewline 
 3 & 1.7022e-01 & 9.2564e-01 \tabularnewline 
 4 & 1.8926e-01 & 1.6542e+00 \tabularnewline 
 5 & 1.7039e-01 & 1.6940e+00 \tabularnewline 
 6 & 1.6839e-01 & 1.6917e+00 \tabularnewline 
 7 & 1.6839e-01 & 1.6911e+00 \tabularnewline 
 8 & 1.6840e-01 & 1.6911e+00 \tabularnewline 
 9 & 1.6840e-01 & 1.6911e+00 \tabularnewline 
 10 & 1.6840e-01 & 1.6911e+00 \tabularnewline 
 11 & 1.6840e-01 & 1.6911e+00 \tabularnewline 
 12 & 1.6840e-01 & 1.6911e+00 \tabularnewline 
 13 & 1.6840e-01 & 1.6911e+00 \tabularnewline 
 14 & 1.6840e-01 & 1.6911e+00 \tabularnewline 
 15 & 1.6840e-01 & 1.6911e+00 \tabularnewline 
\end{tabular}
\caption{Absolute errors of the two quantiles $\hat{Q}_{99\%}$  and $\hat{Q}_{99.9\%}$ of the PCE approximation of gBm at time $T=1$,  whose parameters are $r = 3 \%$, $\sigma = 25 \%$ and starting value $S_0= 100$.} \label{tab:tab7}
\end{table}

Eventually let us compute the two aforementioned quantiles by means of a standard Monte Carlo sampling of the analytical solution of \eqref{sec:eq16}. The standard error of the quantile estimators are
\begin{align*}
&SE_{Q_{99\%}^{MC}} = 0.1764685        \\ &SE_{Q_{99.9\%}^{MC}} =       0.5685217 
\end{align*}

\newpage

\subsection{$\boldsymbol{\sigma=30\%}$}

In this section we set $\sigma=30\%$, while the other parameters are taken form Table \ref{tab:tab1}.

\begin{figure}[!h]
\centering
\includegraphics[scale=0.33]{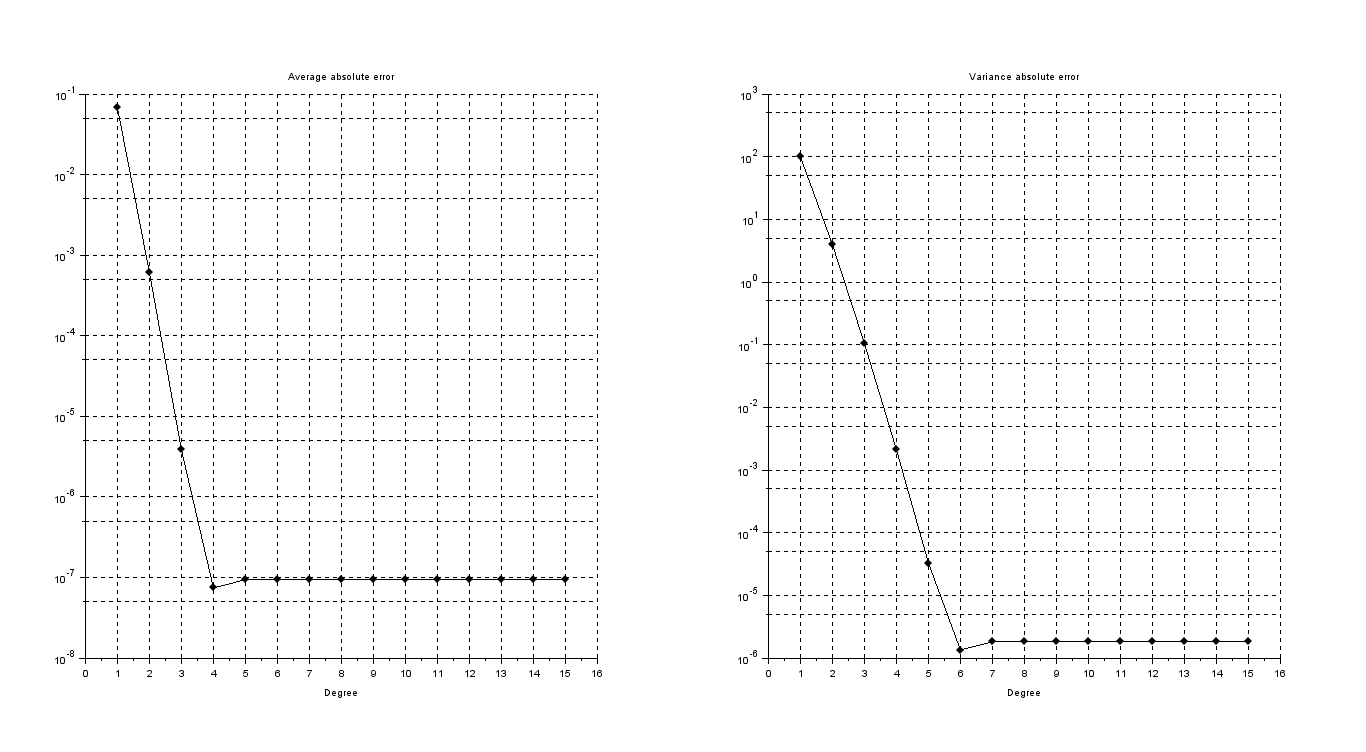}
\caption{Semilogy scale plot of the absolute error of the mean (left) and the variance (right) computed via PCE for gBm at time $T=1$, whose parameters are $r = 3 \%$, $\sigma = 30 \%$ and starting value $S_0= 100$,  for a set of degrees $p = \{1,2,\dots,15\}$.}  \label{fig:fig12}
\end{figure}

\begin{figure}[!h]
\centering
\includegraphics[scale=0.33]{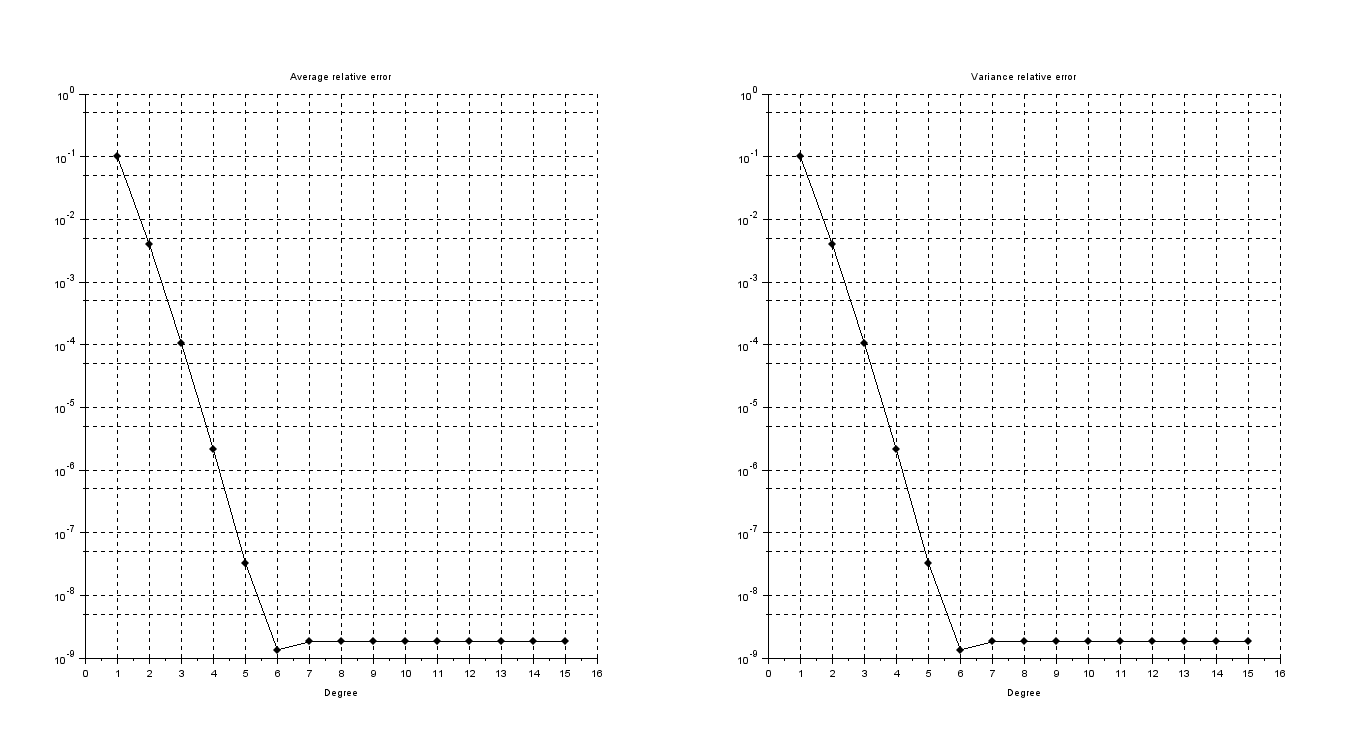}
\caption{Semilogy scale plot of the absolute value of the relative error of the mean (left) and the variance (right) computed via PCE for gBm at time $T=1$, whose parameters are $r = 3 \%$, $\sigma = 30 \%$ and starting value $S_0= 100$, for a set of degrees $p = \{1,2,\dots,15\}$.}  \label{fig:fig13}
\end{figure}

First let us display in Figure \ref{fig:fig12}, Figure  \ref{fig:fig13} and Table \ref{tab:tab8} the absolute and relative error of the average, resp. variance, of the PCE-approximation of the gBm. 

\begin{table}[!h]
\centering
\begin{tabular}{ccccc}
 Degree of PCE &  Average Error & Variance Error & Average relative error & Variance relative error  \tabularnewline
\hline
 1 & 6.7908e-02 & 1.0006e+02 & 6.5901e-04 & 1.0006e-01 \tabularnewline 
 2 & 6.1100e-04 & 3.9769e+00 & 5.9295e-06 & 3.9770e-03 \tabularnewline 
 3 & 3.8351e-06 & 1.0472e-01 & 3.7217e-08 & 1.0473e-04 \tabularnewline 
 4 & 7.4151e-08 & 2.1059e-03 & 7.1960e-10 & 2.1059e-06 \tabularnewline 
 5 & 9.3718e-08 & 3.2736e-05 & 9.0948e-10 & 3.2737e-08 \tabularnewline 
 6 & 9.3798e-08 & 1.3416e-06 & 9.1026e-10 & 1.3417e-09 \tabularnewline 
 7 & 9.3798e-08 & 1.8147e-06 & 9.1026e-10 & 1.8148e-09 \tabularnewline 
 8 & 9.3798e-08 & 1.8204e-06 & 9.1026e-10 & 1.8205e-09 \tabularnewline 
 9 & 9.3798e-08 & 1.8205e-06 & 9.1026e-10 & 1.8205e-09 \tabularnewline 
 10 & 9.3798e-08 & 1.8205e-06 & 9.1026e-10 & 1.8205e-09 \tabularnewline 
 11 & 9.3798e-08 & 1.8205e-06 & 9.1026e-10 & 1.8205e-09 \tabularnewline 
 12 & 9.3798e-08 & 1.8205e-06 & 9.1026e-10 & 1.8205e-09 \tabularnewline 
 13 & 9.3798e-08 & 1.8205e-06 & 9.1026e-10 & 1.8205e-09 \tabularnewline 
 14 & 9.3798e-08 & 1.8205e-06 & 9.1026e-10 & 1.8205e-09 \tabularnewline 
 15 & 9.3798e-08 & 1.8205e-06 & 9.1026e-10 & 1.8205e-09 \tabularnewline 
\end{tabular}
\caption{Absolute and relative errors of average and variance for PCE approximation of gBm  at time $T=1$, whose parameters are $r = 3 \%$, $\sigma = 30 \%$ and starting value $S_0= 100$} \label{tab:tab8}
\end{table}
\newpage 
As for the previous settings, i.e. $\sigma = 15\%$ and $\sigma = 25 \%$, both average and variance error are stationary. The motivations are the same: the analytical approximation of (\ref{sec:eq16}) with (\ref{sec:eq19}) and the error coming form numerical computations of $\left\{ \mathcal{M}(\xi_j)\right\}_{j=1}^N$, by means of \eqref{sec:eq21}, used in (\ref{sec:eq10}). 

As shown in Table \ref{tab:tab9} the errors decrease by increasing the accuracy to compute the $\left\{ \mathcal{M}(\xi_j)\right\}_{j=1}^N$. Moreover we can see the spectral convergence of the error for the PCE approximation up to 4-th degree for mean, resp up to the 6-th degree for variance.  Note that, although the increased accuracy, the error displays again a stationary behavior, but actually no improvement on numerical methods can be implemented, which allows to say that this is the only source of error. \newline

\begin{table}[!h]
\centering
\begin{tabular}{ccccc}
 Degree of PCE &  Average Error & Variance Error & Average relative error& Variance relative error  \tabularnewline
\hline
 1 & 6.7908e-02 & 1.0006e+02 & 6.5901e-04 & 1.0006e-01 \tabularnewline 
 2 & 6.1110e-04 & 3.9769e+00 & 5.9304e-06 & 3.9770e-03 \tabularnewline 
 3 & 3.9292e-06 & 1.0473e-01 & 3.8131e-08 & 1.0473e-04 \tabularnewline 
 4 & 1.9968e-08 & 2.1077e-03 & 1.9378e-10 & 2.1078e-06 \tabularnewline 
 5 & 4.0106e-10 & 3.4562e-05 & 3.8921e-12 & 3.4563e-08 \tabularnewline 
 6 & 3.2092e-10 & 4.8507e-07 & 3.1144e-12 & 4.8508e-10 \tabularnewline 
 7 & 3.2065e-10 & 1.1972e-08 & 3.1118e-12 & 1.1972e-11 \tabularnewline 
 8 & 3.2063e-10 & 6.2851e-09 & 3.1115e-12 & 6.2852e-12 \tabularnewline 
 9 & 3.2064e-10 & 6.2283e-09 & 3.1116e-12 & 6.2285e-12 \tabularnewline 
 10 & 3.2063e-10 & 6.2219e-09 & 3.1115e-12 & 6.2220e-12 \tabularnewline 
 11 & 3.2057e-10 & 6.2321e-09 & 3.1109e-12 & 6.2322e-12 \tabularnewline 
 12 & 3.2115e-10 & 6.2226e-09 & 3.1166e-12 & 6.2228e-12 \tabularnewline 
 13 & 3.2085e-10 & 6.2201e-09 & 3.1137e-12 & 6.2203e-12 \tabularnewline 
 14 & 3.2078e-10 & 6.2329e-09 & 3.1130e-12 & 6.2330e-12 \tabularnewline 
 15 & 3.2075e-10 & 6.2165e-09 & 3.1127e-12 & 6.2167e-12 \tabularnewline 
\end{tabular}
\caption{Absolute error of the average and the variance of PCE approximation of gBm at time $T=1$ ( whose parameters are $r = 3 \%$, $\sigma = 30 \%$ and starting value $S_0= 100$) for higher precision at computing $\{S_{T,j}\}_{j=1}^N$} \label{tab:tab9}
\end{table}
 \newpage

\begin{figure}[!h]
\centering
\includegraphics[scale=0.33]{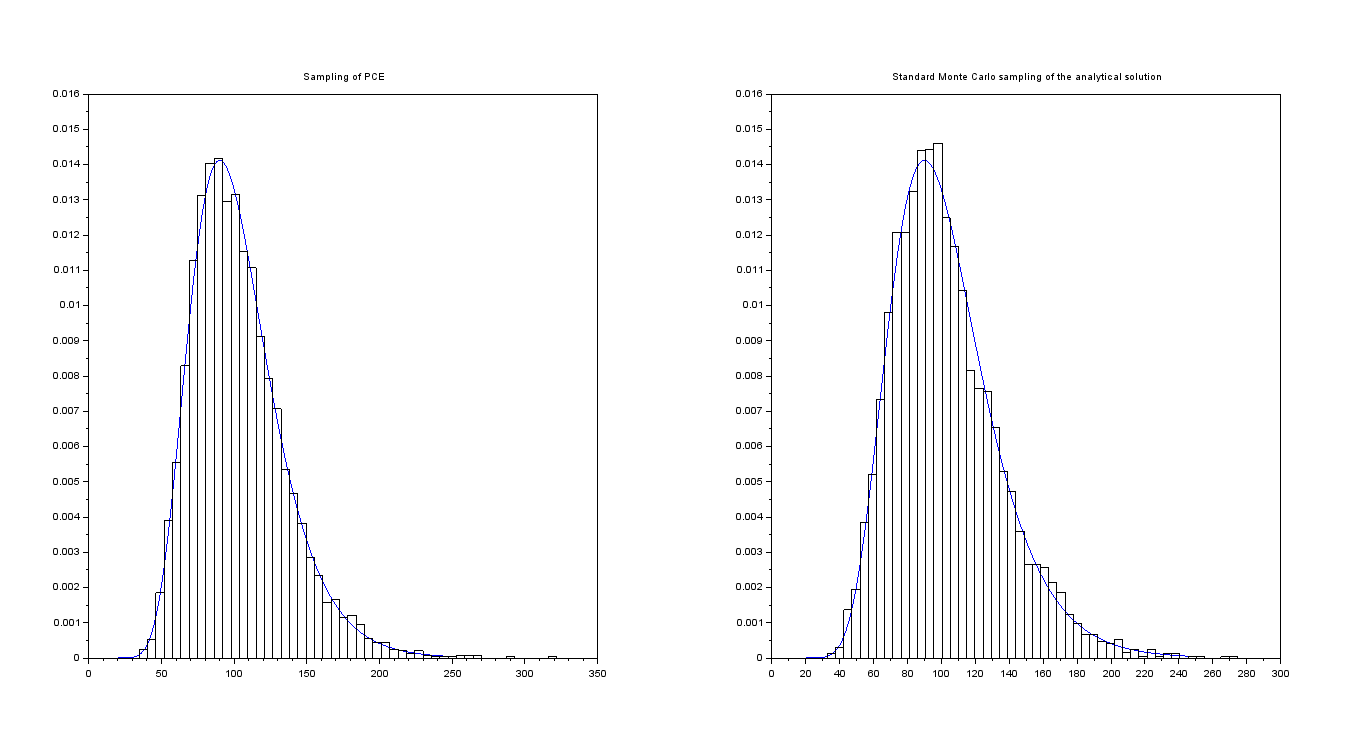}
\caption{Probability density function of gBm at time $T=1$, whose parameters are $r = 3 \%$, $\sigma = 30 \%$ and starting value $S_0 = 100$, (blue curve) and histogram of a Monte Carlo sampling (size = $5000$) of the $S_T^{(p)}$ for $p=15$ (left). The right plot displays the analytical probability density function of gBm, for the same parameters values, and a standard Monte Carlo sampling of size $5000$ of the gBm-analytical solution.}  \label{fig:fig14}
\end{figure}

Coming back to  Table \ref{tab:tab1}, a sampling of size $5000$ of $S_T^{(p)}$ for $p=15$, achieved with standard Monte Carlo sampling, is compared with the probability density function of the analytical solution to eq. \eqref{sec:eq16} at time $T$, whose distribution is lognormal, namely
\begin{equation*}
 X =  \mathcal{N}\left(\log(S_0)+\left(r-\frac{\sigma^2}{2}\right)T\ ,\  \sigma^2 T \right) \:,\: S_T = e^X\:,
\end{equation*}
the computed  values are shown in Figure \ref{fig:fig14} were the standard Monte Carlo sampling of $S_T$ is shown.

\begin{figure}[!h]
\centering
\includegraphics[scale=0.45]{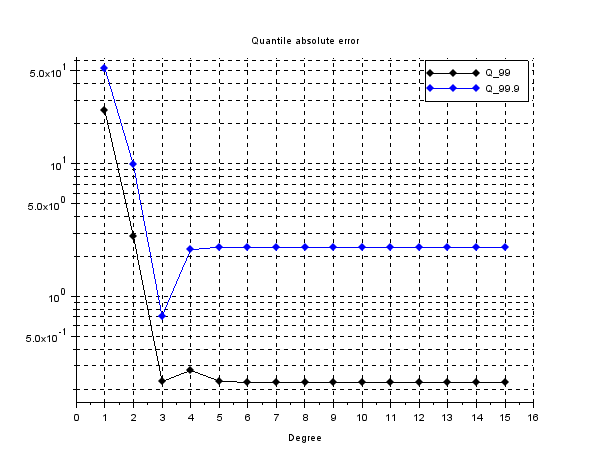}
\caption{Absolute errors of the two quantiles $\hat{Q}_{99\%}$  and $\hat{Q}_{99.9\%}$ of the PCE approximation of gBm at time $T=1$,  whose parameters are $r = 3 \%$, $\sigma = 30 \%$ and starting value $S_0= 100$.} \label{fig:fig15} 
\end{figure}

\newpage 
Then, for each degree $p$, we compute the two quantiles $\hat{Q}_{\alpha}$ for $ \gamma = 99\%$ and $\gamma=99.9\%$ by means of the sample quantile, see Sec. \ref{sec:Ngbm}, based on Latin Hypercube Sampling (LHS) technique.
 Obtained values are shown in Figure  \ref{fig:fig15} and in Table \ref{tab:tab10}.

\begin{table}[!h]
\centering
\begin{tabular}{cccc}
 Degree of PCE &  $\epsilon_{99\%}$ & $\epsilon_{99.9\%}$   \tabularnewline
\hline
 1 & 2.5084e+01 & 5.2332e+01 \tabularnewline 
 2 & 2.8318e+00 & 9.8090e+00 \tabularnewline 
 3 & 2.2776e-01 & 7.0824e-01 \tabularnewline 
 4 & 2.7654e-01 & 2.2441e+00 \tabularnewline 
 5 & 2.2992e-01 & 2.3467e+00 \tabularnewline 
 6 & 2.2387e-01 & 2.3402e+00 \tabularnewline 
 7 & 2.2386e-01 & 2.3380e+00 \tabularnewline 
 8 & 2.2393e-01 & 2.3378e+00 \tabularnewline 
 9 & 2.2393e-01 & 2.3378e+00 \tabularnewline 
 10 & 2.2393e-01 & 2.3378e+00 \tabularnewline 
 11 & 2.2393e-01 & 2.3378e+00 \tabularnewline 
 12 & 2.2393e-01 & 2.3378e+00 \tabularnewline 
 13 & 2.2393e-01 & 2.3378e+00 \tabularnewline 
 14 & 2.2393e-01 & 2.3378e+00 \tabularnewline 
 15 & 2.2393e-01 & 2.3378e+00 \tabularnewline 
\end{tabular}
\caption{Absolute errors of the two quantiles $\hat{Q}_{99\%}$  and $\hat{Q}_{99.9\%}$ of the PCE approximation of gBm at time $T=1$,  whose parameters are $r = 3 \%$, $\sigma = 30 \%$ and starting value $S_0= 100$.} \label{tab:tab10}
\end{table}

As described in Sec. \ref{sec:Ngbm}, let us compute the standard error of the $99\%$ and $99.9\%$ quantiles, by means of $L=100$ independent estimates. 

The  achieved values are
$ SE_{Q_{99\%}^{MC}} =0.2235454$, resp., $SE_{Q_{99.9\%}^{MC}} = 0.7501973$.


\newpage

\section{Vasicek interest rate model} \label{sec:secVAS}

In what follows we consider the Vasicek model, see \cite{Vasicek}, and \cite[Section 4.4.3]{Shreve} for further details, which is defined by
\begin{equation}
dR_t = \left(\alpha-\beta R_t \right) dt+\sigma dW_t \label{sec:eq23} \;,
\end{equation}
where $\alpha,\beta$ and $\sigma$ are positive, real-valued constants and $R_0$ is the starting value of the unknown process. The analytical solution to the SDE \eqref{sec:eq23}, at time $T>0$, reads as follows
\begin{equation}
R_T = e^{-\beta T} R_0+\frac{\alpha}{\beta}\left( 1-e^{-\beta T}\right)+\sigma e^{-\beta T} \int_{0}^Te^{\beta s} dW(s) \label{sec:eq24}\;,
\end{equation}
therefore  $R_T$ is normally distributed and its mean, resp. its variance, is given by
\begin{equation*}
\Expect{R_T} = R_0 e^{-\beta T}+\frac{\alpha}{\beta}(1-e^{-\beta T}) \;\;, {\it resp. by }\: \;
\Var{R_T} = \frac{\sigma^2}{2 \beta}(1-e^{-2 \beta T})\:.
\end{equation*}
Note that we have used the notation $\left\{R_t\right\}_{t \geq 0}$ in order to highlight that the Vasicek model \eqref{sec:eq23} is used to describe the evolution of interest rates. We also underline that such a model belongs to the class of the so-called {\it one factor short rate model} which includes, e.g., the  Rendleman–Bartter model, the CIR model, the  Hull–White model, the Ho–Lee model, etc., since $R_t$ depends only on one  market risk factor, see, e.g.,  \cite{Andrew,Jarrow} and \cite[Section 3.2.1]{Brigo} .

It is worth to mention that the Vasicek model solution allows negative values for the interest rate evaluated at given time $T$, hence its use in credit market framework is often criticized  by practitioners.

\subsection{PCE approximation} \label{sec:PCEvas}

Analogously to what we have done concerning the study of the gBm model, see  Sec. \ref{sec:secGBM}, we can rewrite the solution to (\ref{sec:eq23}) using a functional $H$ such that
\begin{equation}
R_T = H(D_T,W_T)\;, \label{sec:eq25}
\end{equation}
for every fixed  time $T$, which is often referred as {\it maturity time}.
In particular, exploiting the Doss approach described in Section \ref{sec:secDoss},  the scalar function function $H(x,y)$ is determined by solving the ODE \eqref{sec:eq22}, hence obtaining
\begin{equation*}
H(x,y) = \sigma y+x\:.
\end{equation*}
Then the process $\{D_t\}_{t \ge 0}$ is path-wise  determined by solving  
\begin{equation} \label{sec:eq26}
\begin{cases}
\dot{D}{(\omega)} = \alpha -\beta H(D(\omega),W_t(\omega)) \qquad \ \ \ t>0\\
D(\omega)= R_0 \qquad \qquad \qquad \qquad \qquad \quad  \ {t=0} 
\end{cases}\;,
\end{equation}
for almost all $\omega \in \Omega$, where $\dot{D}$ represents the derivative with respect to time. Moreover such time dependence of $D$ is omitted to shorten the notation.

As in Sec. \ref{sec:secGBM}, since the Wiener process is 
\begin{equation}
W_t \sim \mathcal{N}(0,t) \label{sec:eq27} \;,  t \in [0,T],
\end{equation}
then we set $W_t = \sqrt{2t} \xi$, and we consider the {\it basic random variable} $\xi$ to be  Gaussian, namely we take $\xi\sim\mathcal{N}(0,1/2)$ in order to satisfy the restrictions imposed by the software we have used to numerically implement the PCE method.

Proceeding as we made studying the gBm-model, the solution at time $T$ of \eqref{sec:eq26} is numerically computed by means of an adaptive Runge-Kutta method of order 4 (RK4), where the absolute tolerance is set as $1e$-$7$ and the relative one is $1e$-$5$, see, e.g.,  \cite[Section II.1, Scetion II.4]{Hairer}, for further details. Even if an analytical solution is available of \eqref{sec:eq26}, we use the numerical one in order to be coherent with the discussion of Sec. \ref{sec:secGBM}.

 Then by plugging \eqref{sec:eq27} into \eqref{sec:eq26}, and integrating it up to time $T$, the solution $D_T$ can be expressed as a functional depending on the input random variable $\xi$, namely $D_T = \mathcal{M}_1(\xi,\Theta)$, the latter input \linebreak $\Theta=(\alpha,\beta,\sigma) \in \R^3$ collects the parameters that characterizes the dynamics of the Vasicek interest rate model. Due to \eqref{sec:eq27}, the  eq. \eqref{sec:eq25} becomes
\begin{equation}
 \mathcal{M}(\xi,\Theta):= R_T= \sigma W_T +\mathcal{M}_1(\xi,\Theta)=\sigma \sqrt{2T}\xi +\mathcal{M}_1(\xi,\Theta)  \:. \label{sec:eq28}
\end{equation}
As made before, in what follows we shorten the notation by omitting the dependence on vector parameters $\Theta$. \newline

Applying the NISP procedure up to degree $p$, the PCE decomposition of $R_T$ is given by
\begin{equation*}
R_T^{(p)} = \sum_{i=0}^p c_i \Psi_i(\xi) \;,
\end{equation*}
where the coefficients are detecting using (\ref{sec:eq10}) and evaluating $R_T$ in \eqref{sec:eq28} on a set of realizations of the basic random variable $\xi$, namely
considering $N$ values $\{\xi_j\}_{j=1}^N$. Moreover, in what follows, we set $N=p$. The obtained {\it simulated} values for $R_T$ will be indicated as follows
\begin{equation*}
\{R_{T,j}\}_{j=1}^N = \{\mathcal{M}(\xi_j)\}_{j=1}^N \:.
\end{equation*}

The aforementioned realizations of the \eqref{sec:eq28} are determined in two steps:
\begin{itemize}
\item each Gaussian quadrature nodes, belonging to $\{\xi_j\}_{j=1}^N$, can be interpreted as the image of a suitable $\omega_j \in \Omega$ through $\xi$. Therefore by integrating \eqref{sec:eq26} for each path in $\{\omega_1,\dots,\omega_N\} \subset \Omega$ we achieve a set of realization of $D_T$, let us say $\{\mathcal{M}_1(\xi_j)\}_{j=1}^N$.
 It is worth to mentioned that  we are integrating $N$ independent ODEs.
\item By means of \eqref{sec:eq28} we get the required set $\{R_{T,j}\}_{j=1}^N$, where the first summand  $\{{\sigma \sqrt{2 T}\xi_j}\}_{j=1}^N$.
\end{itemize}
%

\subsection{Numerical applications: overview of the computations}  \label{sec:Nvas}
Let us implement the theoretical approach developed in previous section by a working example which consists in considering the 
 Vasicek model for a set of volatility
\begin{equation*}
\sigma = \{15\%,25\%,30\%\}\;,
\end{equation*}
while the other parameters as set according to Table \ref{tab:tab11}. \newline 

\begin{table}[!h]
\centering
\begin{tabular}{c|ccccc}
\centering Parameters & \centering $\alpha$ & \centering $\beta$   & \centering $R_0$ & \centering T \tabularnewline
\hline
\centering Values & \centering 0.1 &\centering 0.2 & \centering 110 & \centering1 \tabularnewline
\end{tabular}
\caption{Parameters and initial value for the Vasicek interest rate model.} \label{tab:tab11}
\end{table}

First we compute the absolute error as well as the relative error of the mean and variance of $R_T^{(p)}$, with respect to degree $ p=\{0,1,2,\dots,15\} $. In particular, for every given value $p$, the absolute errors for the mean and for the variance are as follows
\begin{align*}
\epsilon^{(p)}_{MEAN} &= \abs{\Expect{R_T} -\Expect{R^{(p)}_T}}\;,\\
\epsilon^{(p)}_{VAR} &= \abs{\Var{R_T}-\Var{R^{(p)}_T}}\;,
\end{align*}
while the relative errors are given by
\begin{align*}
RE^{(p)}_{MEAN} &= \frac{\epsilon^{(p)}_{MEAN}}{\Expect{R_T}}\;,\\
RE^{(p)}_{VAR} &= \frac{\epsilon^{(p)}_{VAR}}{\Var{R_T}}\:.
\end{align*}
Due to  the logarithmic scale employed, we  consider their absolute values, highlighting the related order, rather than their numerical value.  \newline

Then let us compute the two quantiles $\hat{Q}_\gamma$, where $\gamma = 99\%$ and $\gamma=99.9\%$, of the PCE approximation $R_T^{(p)}$, where $ p =0,1,\ldots,15$.
Proceeding as in Sec. \ref{sec:secGBM}, we consider the $\gamma$-th sample quantile, namely the $(K+1)$-th realization of the sampling of $R_T^{(p)}$ such that $K \le [\gamma M]$, where $M$ is the size of the sampling, and $[\cdot]$ denotes the integer part of the real number within the brackets. In what follows we always employ a  Latin Hypercube Sampling (LHS) technique of $R_T^{(p)}$, see \cite{Tang, Owen} and also \cite{Iman} for further references, of size $M=5000$. In particular we  have $K_{99\%} = 4951$, while  $K_{99.9\%} = 4996$, and we compute the absolute error of $\hat{Q}_{\gamma}$ versus the analytical values ${Q}_\gamma$. Therefore we evaluate
$
\epsilon_{\gamma} = \abs{\hat{Q}_\gamma-{Q}_\gamma}
$, 
where $Q_\gamma$ is the quantile of a normal random variable of mean $\Expect{R_T}$ and variance $\Var{R_T}$ and, indicating its cumulative density function by $F_{R_T}$, we have $
Q_\gamma = F_{R_T}^{-1}(\gamma)
$.  \newline

Furthermore the same quantiles are computed by means of standard Monte Carlo approach using the analytical solution $R_T$ to \eqref{sec:eq23}. The accuracy of the achieved values is expressed in terms of their standard error $SE_{Q_{\gamma}^{MC}}$, for $\gamma = 99\%, 99.9\%$. Therefore the sample quantiles are computed $L=200$ times, getting $\{Q_{\gamma}^{MC}(l)\}_{l=1}^L$ independent estimates. Then the standard error reads as
\begin{equation*}
SE_{Q_{\gamma}^{MC}} = \frac{\hat{\sigma}}{\sqrt{L}} \;,
\end{equation*}
where $\hat{\sigma}^2 = \frac{1}{L-1}\sum_{l=1}^L \left(Q_{\gamma}^{MC}(l)-\bar{Q}_{\gamma}^{MC}\right)^2$ and $\bar{Q}_{\gamma}^{MC}$ is the arithmetic average of $\{Q_{\gamma}^{MC}(l)\}_{l=1}^L$.

\begin{remark}.
We would like to note that the choice of the basic estimator for quantiles, i.e. the sample quantile, 
has been taken to focus our attention to the efficacy of the method instead of its  estimates accuracy as well as to provide a fair comparison between the data achieved. Nevertheless the dedicated literature provides more accurate techniques, such as the \emph{Two-phase quantile estimator}, presented in \cite{Chen}, the \emph{L-estimator} and \emph{Harrel Davis} (HD) estimators, see, e.g., \cite{Mausser} for further details.
\end{remark}

\subsection{$\boldsymbol{\sigma = 15\%}$}

In this section the value of the volatility is set to $\sigma = 15\%$, while the other parameters are chosen as in Table \ref{tab:tab11}. \newline

\begin{figure}[!h]
\centering
\includegraphics[scale=0.32]{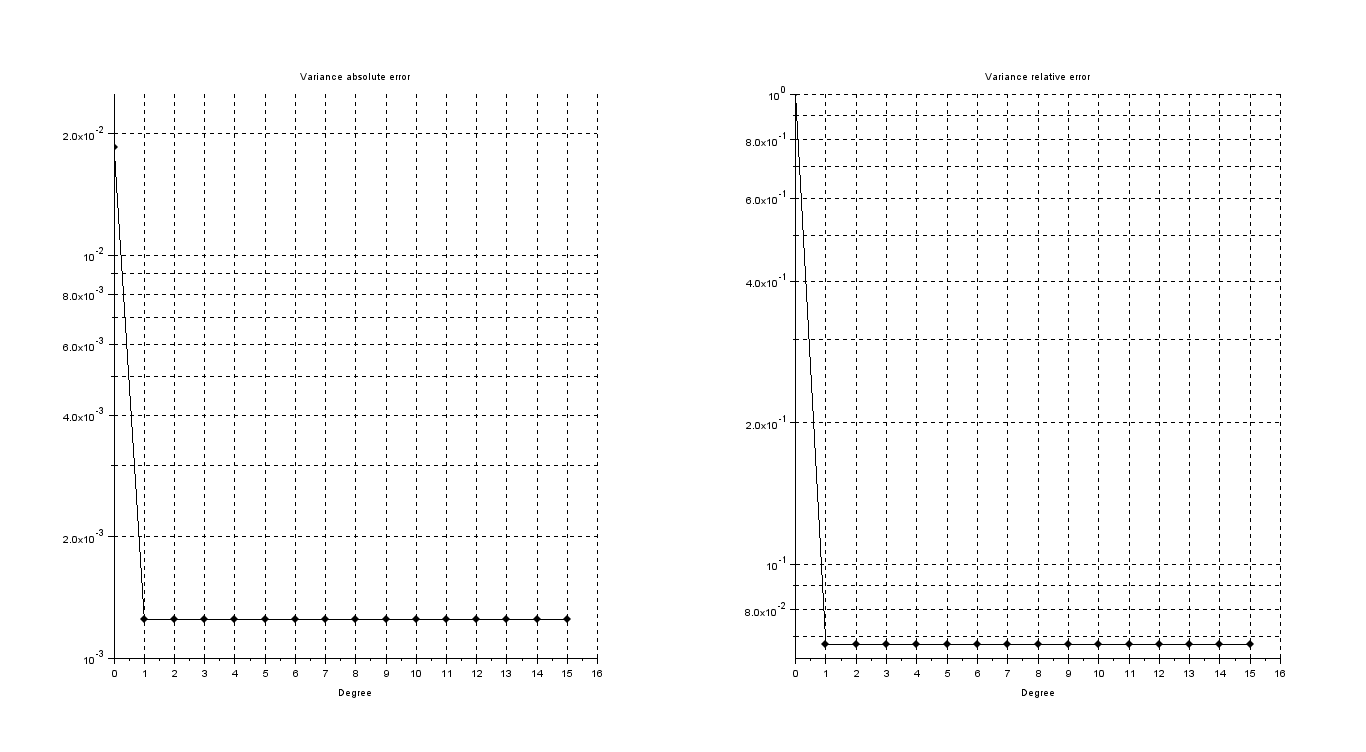}
\caption{Semilogy scale plot of the absolute error of the variance of $R_T^{(p)}$ evaluated at time $T=1$ (left) and the absolute  relative error of the variance (right) of $R_T^{(p)}$ at time $T=1$.  In both cases the parameters of the Vasicek interest rate model are $\alpha = 0.1$, $\beta = 0.2$, $\sigma = 0.15$.}  \label{fig:fig16}
\end{figure}

First let us display in Figure \ref{fig:fig16} the absolute and relative error of the variance, while in Table \ref{tab:tab12} are collected both the absolute and absolute value of relative error for mean and variance concerning the Vasicek interest rate model.

\begin{table}[!h]
\centering
\begin{tabular}{ccccc}
 Degree of PCE &  Average Error & Variance Error & Average relative error & Variance relative error   \tabularnewline
\hline

 0 & 8.2289e-08 & 1.8544e-02 & 9.1279e-10 & 1.0000e+00 \tabularnewline 
 1 & 8.2303e-08 & 1.2490e-03 & 9.1295e-10 & 6.7349e-02 \tabularnewline 
 2 & 8.2312e-08 & 1.2490e-03 & 9.1305e-10 & 6.7349e-02 \tabularnewline 
 3 & 8.2348e-08 & 1.2490e-03 & 9.1345e-10 & 6.7349e-02 \tabularnewline 
 4 & 8.2321e-08 & 1.2490e-03 & 9.1315e-10 & 6.7349e-02 \tabularnewline 
 5 & 8.2349e-08 & 1.2490e-03 & 9.1346e-10 & 6.7349e-02 \tabularnewline 
 6 & 8.2328e-08 & 1.2490e-03 & 9.1322e-10 & 6.7349e-02 \tabularnewline 
 7 & 8.2308e-08 & 1.2490e-03 & 9.1300e-10 & 6.7349e-02 \tabularnewline 
 8 & 8.2327e-08 & 1.2490e-03 & 9.1321e-10 & 6.7349e-02 \tabularnewline 
 9 & 8.2300e-08 & 1.2490e-03 & 9.1291e-10 & 6.7349e-02 \tabularnewline 
 10 & 8.2298e-08 & 1.2490e-03 & 9.1289e-10 & 6.7349e-02 \tabularnewline 
 11 & 8.2352e-08 & 1.2490e-03 & 9.1349e-10 & 6.7349e-02 \tabularnewline 
 12 & 8.2311e-08 & 1.2490e-03 & 9.1303e-10 & 6.7349e-02 \tabularnewline 
 13 & 8.2364e-08 & 1.2490e-03 & 9.1362e-10 & 6.7349e-02 \tabularnewline 
 14 & 8.2321e-08 & 1.2490e-03 & 9.1315e-10 & 6.7349e-02 \tabularnewline 
 15 & 8.2357e-08 & 1.2490e-03 & 9.1355e-10 & 6.7349e-02 \tabularnewline 
\end{tabular}
\caption{Absolute error of the average and the variance of $R_T^{(p)}$ at time $T=1$, with respect to the Vasicek model parameters $\alpha = 0.1$, $\beta = 0.2$, $\sigma = 15 \%$.} \label{tab:tab12}
\end{table}

\newpage

The absolute and relative errors of mean and variance are stationary, see Table \ref{tab:tab12}. This can be motivated by looking at the solution to (\ref{sec:eq28}). By integrating \eqref{sec:eq26}, upon plugging the definition of $H(x,y)$, we get

\begin{equation*}
RR_T = \sigma W_T +  e^{-\beta T} R(0)+\frac{\alpha}{\beta}\left( 1-e^{-\beta T}\right)-\sigma \beta e^{-\beta T} \int_0^T W_s e^{\beta s} ds \;,
\end{equation*}
we call such solution as $RR_T$, in order to distinguish it from the analytical solution of the Vasicek interest rate model \eqref{sec:eq23}. Then, exploiting \eqref{sec:eq27}, $W_s = \sqrt{2s}\xi$ and $W_T = \sqrt{2T}\xi$
\begin{equation}
RR_T = \sigma W_T + e^{-\beta T} R(0)+\frac{\alpha}{\beta}\left( 1-e^{-\beta T}\right)-\xi \cdot \left(\sigma \beta   e^{-\beta T}\int_0^T \sqrt{2s} e^{\beta s} ds\right)\;. \label{sec:eq29}
\end{equation}
Notice that the same random variable $\xi$ is used both for $W_T$ and $W_s$ definition, since the latter is involved by detection of $D_T$, which is computed separately from $W_T$, thus the former is not linked with the latter. Therefore we can gather $\xi$ and due to $\xi \sim \mathcal{N}(0,1/2)$, we conclude that $RR_T$ is normally distributed with
\begin{align}
\Expect{RR_T} &=  e^{-\beta T} R(0)+\frac{\alpha}{\beta}\left( 1-e^{-\beta T}\right)\;,\label{sec:eq30}\\
\Var{RR_T} &= \sigma^2 \left(\sqrt{T}-\beta e^{-\beta T} \int_0^T \sqrt{s} e^{\beta s} ds  \right)^2\;. \label{sec:eq31}
\end{align}
Let us consider the PCE-approximation $R_T^{(p)}$, for a fixed degree $p$, then by triangular inequality
\begin{equation*}
\abs{\Expect{R_T}-\Expect{R^{(p)}_T}} \le \abs{\Expect{RR_T}-\Expect{R^{(p)}_T}} +\abs{\Expect{RR_T}-\Expect{R_T}} \;,
\end{equation*}
The last summand on the right is null, by means of \eqref{sec:eq30}, therefore the error of the mean is driven by the first summand. Due to \eqref{sec:eq8} and \eqref{sec:eq10}, the numerical approximation of $\{R_{T,j}\}_{j=1}^N$ influences the accuracy of PCE-statistics, therefore such error corrupts the spectral convergence of the PCE method. Indeed in Tab. \ref{tab:tab12} it becomes stationary at a value close to absolute tolerance of the RK4 method, i.e. $1e$-$7$, which is used to compute the $\{R_{T,j}\}_{j=1}^N$ . \newline

\begin{figure}[!h]
\centering
\includegraphics[scale=0.43]{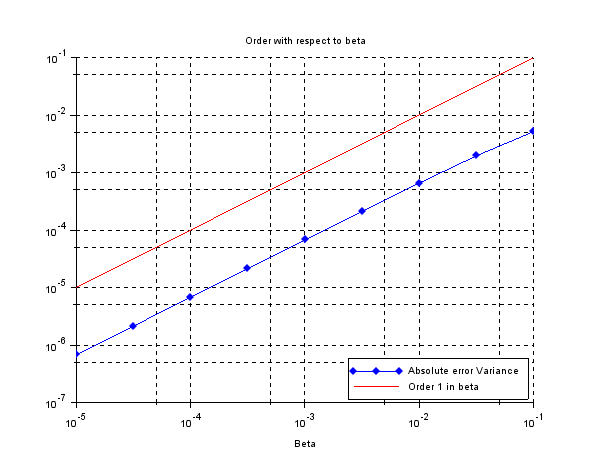}
\caption{Log-Log scale plot of the variance absolute error of $R_T^{(p=15)}$ at $T=1$ for different $\beta = 10^{[-5:0.5:-1]}$ compared with order 1. The parameters of the Vasicek interest model are $\alpha = 0.1$, $\sigma = 0.15$ and $R_0=110$.}  \label{fig:fig17}
\end{figure}

The absolute error of the variance is
\begin{equation}
\epsilon^{(p)}_{VAR} = \abs{\Var{R_T}-\Var{R^{(p)}_T} } \le  \abs{\Var{R_T}-\Var{RR_T} }+ \abs{\Var{R^{(p)}_T}-\Var{RR_T} }\;, 
\end{equation}\label{sec:eq32}
Due to spectral convergence of PCE-approximation, the second summand on the right hand side is irrelevant if compared with the second one. Therefore let us focus our attention on $\abs{\Var{R_T}-\Var{RR_T} }$, in particular let us compute the Taylor expansion,  with respect to $\beta$ and centered at $\beta = 0$ , of $\Var{R_T}$, resp. of $\Var{RR_T}$, so that the variance of \eqref{sec:eq23} can be approximated by
\begin{equation*}
\begin{split}
\Var{R_T} & =  \frac{\sigma^2}{2 \beta}\Big(1-e^{-2 \beta T}\Big)  \\
& = \frac{\sigma^2}{2 \beta} \left( 1- (1-2\beta T+\frac{4\beta^2 T^2}{2})+ \mathcal{O}(\beta^2)\right)  \\
& =  \sigma^2 T-\sigma^2 T^2  \beta +\mathcal{O}(\beta^2) \;,
\end{split}
\end{equation*}
while the Taylor expansion of \eqref{sec:eq31} reads as follow
\begin{equation*}
\begin{split}
\Var{RR_T} &=\sigma^2 \left(\sqrt{T}-\beta e^{-\beta T} \int_0^T \sqrt{s} e^{\beta s} ds  \right)^2  \\
& = \sigma^2  \left( \sqrt{T}- \frac{2}{3}\beta T^{3/2}+\mathcal{O}(\beta^2) \right)^2  \\
& = \sigma^2 \left(T-\frac{4}{3}\beta T^2\right)+\mathcal{O}(\beta^2) \;, 
\end{split}
\end{equation*}
therefore
\begin{equation}
\epsilon^{(p)}_{VAR}  =  \abs{\sigma^2 T-\sigma^2 T^2  \beta -\sigma^2 T+\frac{4}{3}\sigma^2  T^2 \beta +\mathcal{O}(\beta^2)}  \le \frac{1}{3} T^2 \beta +\mathcal{O}(\beta^2)\;. \label{sec:eq33}
\end{equation}
Hence let us compute $\epsilon^{(p)}_{VAR} $ for a  set of decreasing values of $\beta$
\begin{equation*}
\beta = 10^{[-5:0.5:-1]}\;,
\end{equation*}
while the other parameters are $\sigma = 15\%$, $\alpha = 0.1$, $R_0 =110$ and $T=1$.
The results are displayed in Figure {\ref{fig:fig17}}, and we note that they agree with the theoretical bound given by (\ref{sec:eq33}). \newline 

\newpage 

\begin{remark}.
By means of \eqref{sec:eq7} the error of the variance is meaningful for the convergence in mean square sense of $R^{(p)}_T$ to $R_T$ as $p\to +\infty$. Moreover such computations show that the PCE-approximation actually achieves spectral convergence, but to \eqref{sec:eq29}, which differs in {\it mean square norm} to the analytical solution $R_T$, see eq. \eqref{sec:eq24}, of $\abs{\Var{R_T}-\Var{RR_T} }$, hence motivating the behaviour of the error for variance in Fig. \ref{fig:fig16}
\end{remark}

\begin{remark}.
We note that $\sigma$ does not influence the computations shown above, therefore the same procedure and related results are also valid for the other values of volatility, whose associated analyses are therefore skipped.
\end{remark}

\begin{figure}[!h]
\centering
\includegraphics[scale=0.35]{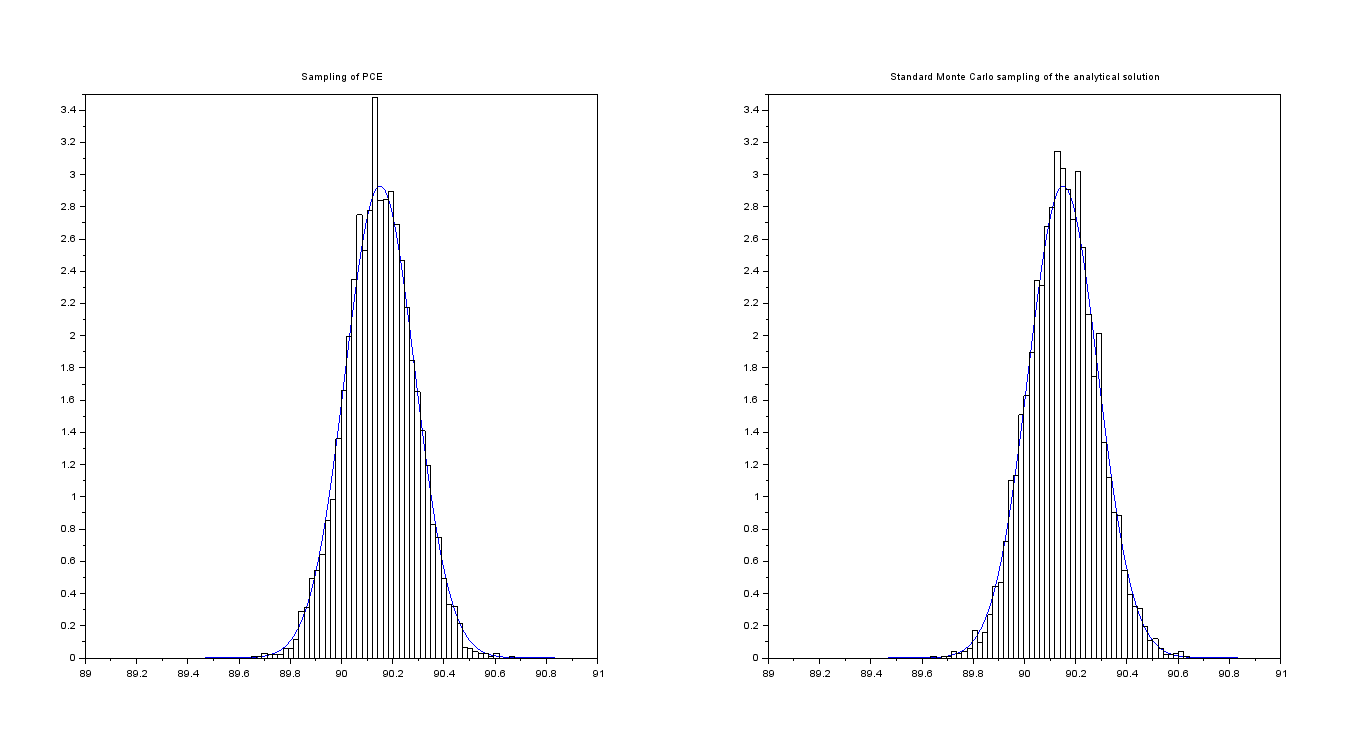}
\caption{Left plot: Probability density function of the Vasicek interest rate model, with parameters  $\alpha = 0.1$, $\beta = 0.2$, $\sigma = 0.15$ and $R_0 =110$, (blue curve) and histogram of a Monte Carlo sampling (size = $5000$) of the $R_T^{(p)}$ for $p=15$. With the same parameters, in the right plot we compare the analytical probability density function of the Vasicek model, with the one obtained using the Monte Carlo sampling of size $5000$, for $R_T$, being $T=1$. }  \label{fig:fig18}
\end{figure}

Coming back to the values in Table \ref{tab:tab11}, let us compute a Monte Carlo sampling of size $5000$ of $R^{(p)}_T$, with $p=15$, whose histogram, see fig. \ref{fig:fig18}, is compared with the probability density function of the normal random variable $R_T$.

Moreover in the same fig. \ref{fig:fig18} we show a sampling obtained exploiting the standard Monte Carlo technique.

\begin{figure}[!h]
\centering
\includegraphics[scale=0.5]{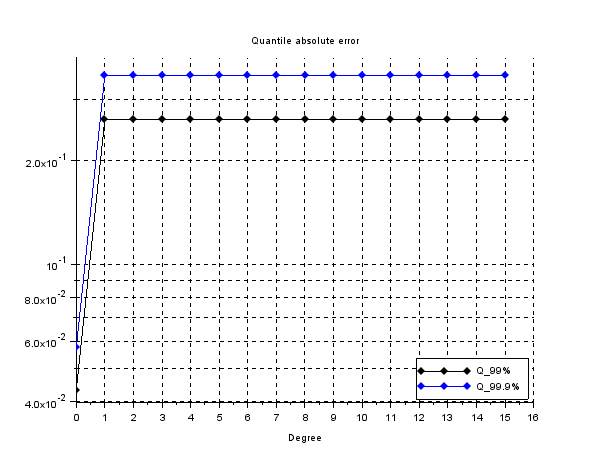}
\caption{Absolute errors of the two quantiles $\hat{Q}_{99\%}$  and $\hat{Q}_{99.9\%}$ of the PCE approximation of the Vasicek interest rate model at time $T=1$. The parameters are set as $\alpha = 0.1$, $\beta = 0.2$, $\sigma = 0.15$ and $R_0=110$.} \label{fig:fig19}
\end{figure}

The errors of quantiles for $\gamma  = 99\%$, resp. for $\gamma=99.9\%$, are shown in Fig. \ref{fig:fig19} and in Table \ref{tab:tab13}. These values are estimated by means of the sample quantile, see Sec. \ref{sec:Nvas} for further details.  \newline

\begin{table}[!h]
\centering
\begin{tabular}{cccc}
 Degree of PCE &  $\epsilon_{99\%}$ & $\epsilon_{99.9\%}$   \tabularnewline
\hline
 0 & 4.3141e-02 & 5.7307e-02 \tabularnewline 
 1 & 2.6330e-01 & 3.5319e-01 \tabularnewline 
 2 & 2.6330e-01 & 3.5319e-01 \tabularnewline 
 3 & 2.6330e-01 & 3.5319e-01 \tabularnewline 
 4 & 2.6330e-01 & 3.5319e-01 \tabularnewline 
 5 & 2.6330e-01 & 3.5319e-01 \tabularnewline 
 6 & 2.6330e-01 & 3.5319e-01 \tabularnewline 
 7 & 2.6330e-01 & 3.5319e-01 \tabularnewline 
 8 & 2.6330e-01 & 3.5319e-01 \tabularnewline 
 9 & 2.6330e-01 & 3.5319e-01 \tabularnewline 
 10 & 2.6330e-01 & 3.5319e-01 \tabularnewline 
 11 & 2.6330e-01 & 3.5319e-01 \tabularnewline 
 12 & 2.6330e-01 & 3.5319e-01 \tabularnewline 
 13 & 2.6330e-01 & 3.5319e-01 \tabularnewline 
 14 & 2.6330e-01 & 3.5319e-01 \tabularnewline 
 15 & 2.6330e-01 & 3.5319e-01 \tabularnewline 
\end{tabular}
\caption{Absolute errors of the two quantiles $\hat{Q}_{99\%}$  and $\hat{Q}_{99.9\%}$ of the PCE approximation of the Vasicek Interest rate model at time $T=1$. The parameters are set as $\alpha = 0.1$, $\beta = 0.2$, $\sigma = 0.15$ and $R_0=110$.}  \label{tab:tab13}
\end{table}

\newpage
The quantiles computed by means of a standard Monte Carlo sampling of the analytical solution, whose size is equal to the one used for PCE, give the following standard error
\begin{align*}
&SE_{Q_{99\%}^{MC}} =0.0004395             \\ &SE_{Q_{99.9\%}^{MC}} =   0.0013269 
\end{align*}

\subsection{$\boldsymbol{\sigma = 25\%}$}

In this section the value of the volatility is set to $\sigma = 25\%$, while the other parameters are chosen as in Table \ref{tab:tab11}. \newline

\begin{figure}[!h]
\centering
\includegraphics[scale=0.35]{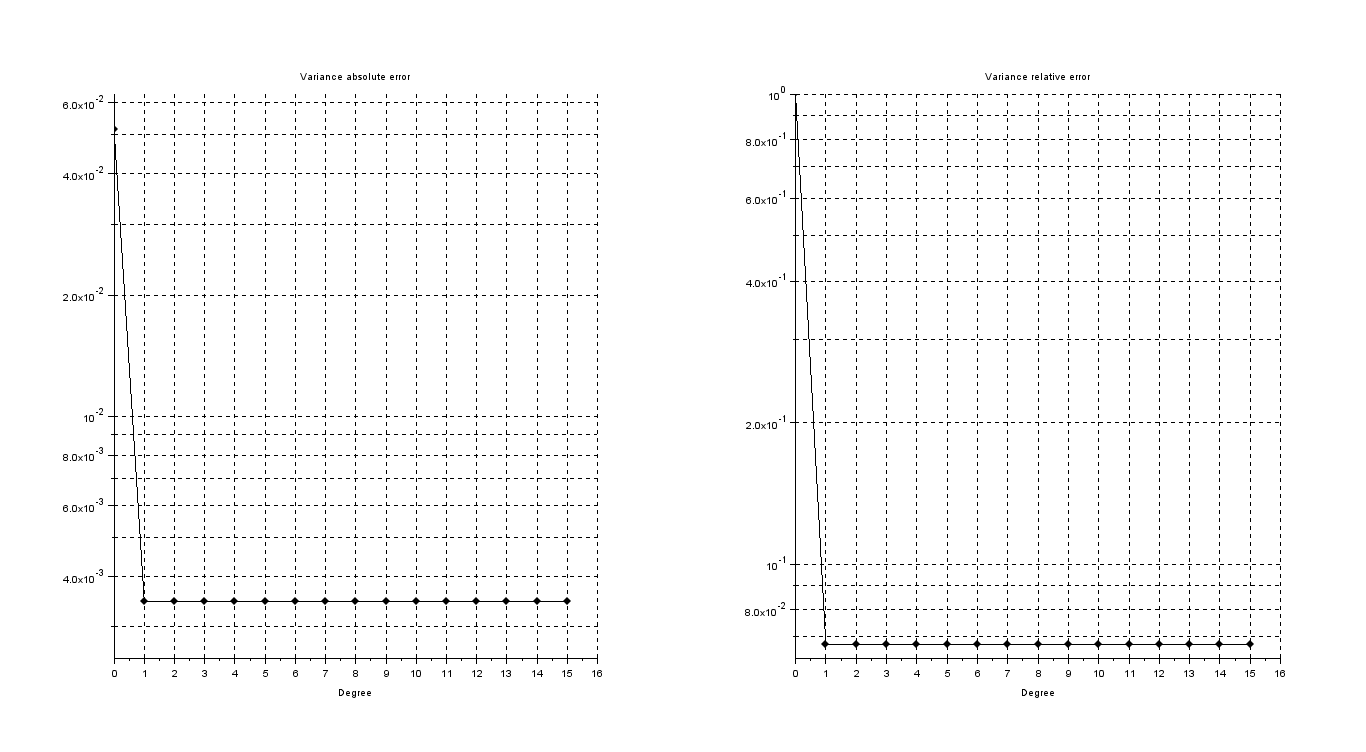}
\caption{Semilogy scale plot of the absolute error of the variance of $R_T^{(p)}$ evaluated at time $T=1$ (left) and the absolute  relative error of the variance (right) of $R_T^{(p)}$ at time $T=1$.  In both cases the parameters of the Vasicek interest rate model are $\alpha = 0.1$, $\beta = 0.2$, $\sigma = 0.25$.}  \label{fig:fig20}
\end{figure}

First let us display in Figure \ref{fig:fig20} the absolute and relative error of the variance, while in \ref{tab:tab14} are collected both the absolute and absolute value of relative error for mean and variance concerning the Vasicek interest rate model.

\begin{table}[!h]
\centering
\begin{tabular}{ccccc}
 Degree of PCE &  Average Error & Variance Error & Average relative error & Variance relative error   \tabularnewline
\hline

 0 & 8.2289e-08 & 5.1512e-02 & 9.1279e-10 & 1.0000e+00 \tabularnewline 
 1 & 8.2346e-08 & 3.4693e-03 & 9.1342e-10 & 6.7349e-02 \tabularnewline 
 2 & 8.2320e-08 & 3.4693e-03 & 9.1314e-10 & 6.7349e-02 \tabularnewline 
 3 & 8.2375e-08 & 3.4693e-03 & 9.1374e-10 & 6.7349e-02 \tabularnewline 
 4 & 8.2274e-08 & 3.4693e-03 & 9.1262e-10 & 6.7349e-02 \tabularnewline 
 5 & 8.2351e-08 & 3.4693e-03 & 9.1348e-10 & 6.7349e-02 \tabularnewline 
 6 & 8.2342e-08 & 3.4693e-03 & 9.1338e-10 & 6.7349e-02 \tabularnewline 
 7 & 8.2285e-08 & 3.4693e-03 & 9.1275e-10 & 6.7349e-02 \tabularnewline 
 8 & 8.2322e-08 & 3.4693e-03 & 9.1316e-10 & 6.7349e-02 \tabularnewline 
 9 & 8.2348e-08 & 3.4693e-03 & 9.1345e-10 & 6.7349e-02 \tabularnewline 
 10 & 8.2306e-08 & 3.4693e-03 & 9.1298e-10 & 6.7349e-02 \tabularnewline 
 11 & 8.2321e-08 & 3.4693e-03 & 9.1314e-10 & 6.7349e-02 \tabularnewline 
 12 & 8.2323e-08 & 3.4693e-03 & 9.1317e-10 & 6.7349e-02 \tabularnewline 
 13 & 8.2356e-08 & 3.4693e-03 & 9.1353e-10 & 6.7349e-02 \tabularnewline 
 14 & 8.2346e-08 & 3.4693e-03 & 9.1343e-10 & 6.7349e-02 \tabularnewline 
 15 & 8.2359e-08 & 3.4693e-03 & 9.1357e-10 & 6.7349e-02 \tabularnewline 

\end{tabular}
\caption{Absolute error of the average and the variance of $R_T^{(p)}$ at time $T=1$, with respect to Vasicek parameters $\alpha = 0.1$, $\beta = 0.2$, $\sigma = 25 \%$.} \label{tab:tab14}
\end{table}

As for $\sigma = 15\%$ the Table \ref{tab:tab14} shows the stationary behavior of the absolute and relative errors of mean and variance. Such features can be motivated as in the  $\sigma = 25\%$ case, indeed both the same computations and bounds hold again.

\begin{figure}[!h]
\centering
\includegraphics[scale=0.35]{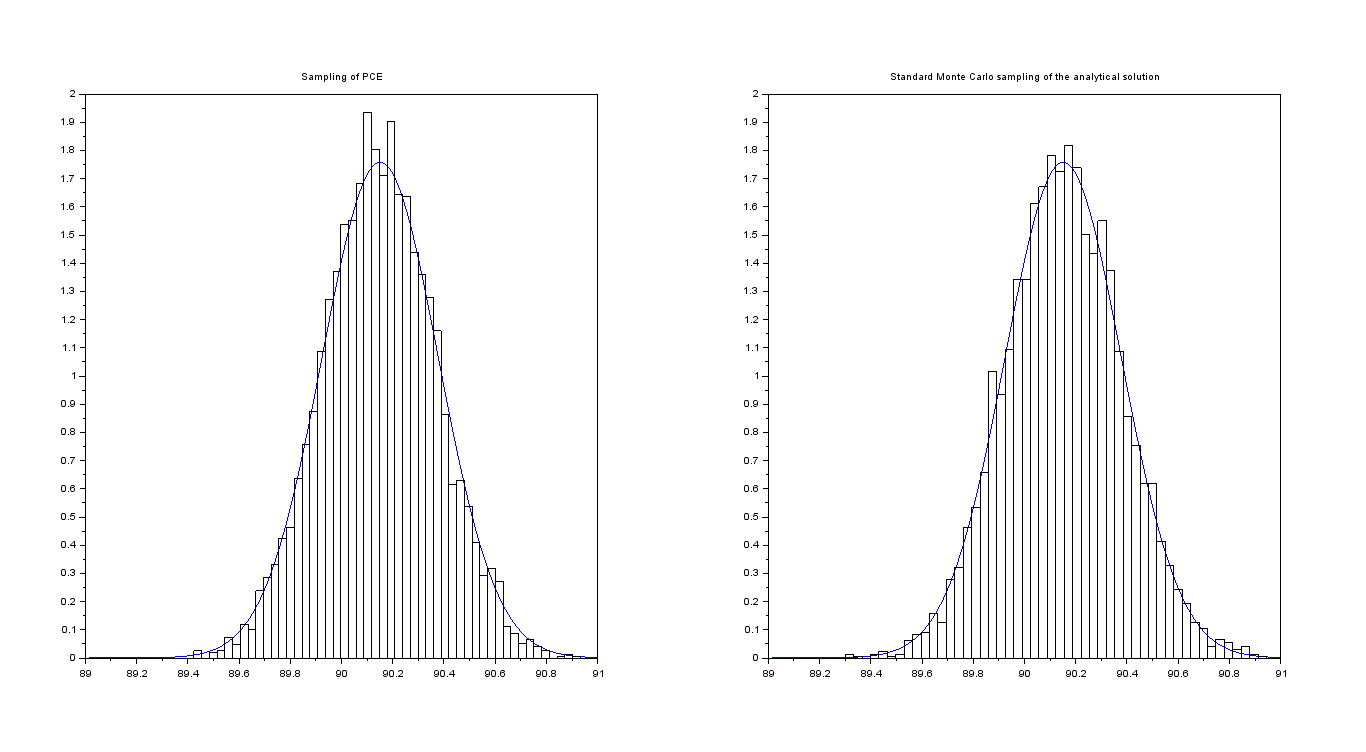}
\caption{Left plot: Probability density function of the Vasicek interest rate model, with parameters  $\alpha = 0.1$, $\beta = 0.2$, $\sigma = 0.25$ and $R_0 =110$, (blue curve) and histogram of a Monte Carlo sampling (size = $5000$) of the $R_T^{(p)}$ for $p=15$. With the same parameters, in the right plot we compare the analytical probability density function of the Vasicek model, with the one obtained using the Monte Carlo sampling of size $5000$, for $R_T$, being $T=1$. }  \label{fig:fig21}
\end{figure}

Coming back to the values in Table \ref{tab:tab11}, let us compute a Monte Carlo sampling of size $5000$ of $R^{(p)}_T$, with $p=15$, whose histogram, see fig. \ref{fig:fig21}, is compared with the probability density function of the normal random variable $R_T$.
Moreover in the same fig. \ref{fig:fig21} we show a sampling obtained exploiting the standard Monte Carlo technique.

\begin{figure}[!h]
\centering
\includegraphics[scale=0.45]{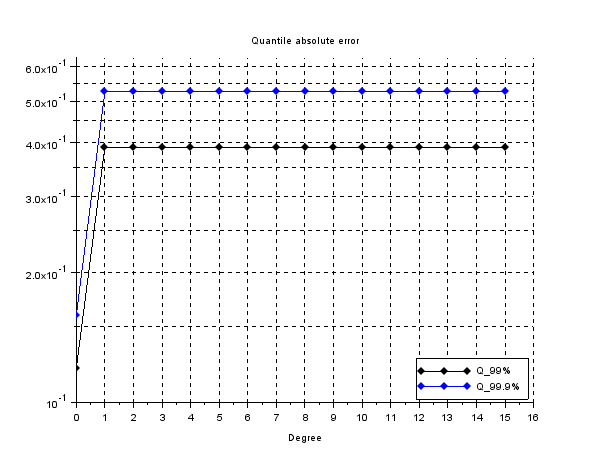}
\caption{Absolute errors of the two quantiles $\hat{Q}_{99\%}$  and $\hat{Q}_{99.9\%}$ of the PCE approximation of the Vasicek interest rate model at time $T=1$. The parameters are set as $\alpha = 0.1$, $\beta = 0.2$, $\sigma = 0.25$ and $R_0=110$.}  \label{fig:fig22}
\end{figure}

The errors of quantiles for $\gamma  = 99\%$, resp. for $\gamma=99.9\%$, are shown in Fig. \ref{fig:fig22} and in Table \ref{tab:tab15}.   \newline

\begin{table}[!h]
\centering
\begin{tabular}{cccc}
 Degree of PCE &  $\epsilon_{99\%}$ & $\epsilon_{99.9\%}$   \tabularnewline
\hline
 0 & 1.1984e-01 & 1.5919e-01 \tabularnewline 
 1 & 3.9090e-01 & 5.2498e-01 \tabularnewline 
 2 & 3.9090e-01 & 5.2498e-01 \tabularnewline 
 3 & 3.9090e-01 & 5.2498e-01 \tabularnewline 
 4 & 3.9090e-01 & 5.2498e-01 \tabularnewline 
 5 & 3.9090e-01 & 5.2498e-01 \tabularnewline 
 6 & 3.9090e-01 & 5.2498e-01 \tabularnewline 
 7 & 3.9090e-01 & 5.2498e-01 \tabularnewline 
 8 & 3.9090e-01 & 5.2498e-01 \tabularnewline 
 9 & 3.9090e-01 & 5.2498e-01 \tabularnewline 
 10 & 3.9090e-01 & 5.2498e-01 \tabularnewline 
 11 & 3.9090e-01 & 5.2498e-01 \tabularnewline 
 12 & 3.9090e-01 & 5.2498e-01 \tabularnewline 
 13 & 3.9090e-01 & 5.2498e-01 \tabularnewline 
 14 & 3.9090e-01 & 5.2498e-01 \tabularnewline 
 15 & 3.9090e-01 & 5.2498e-01 \tabularnewline 

\end{tabular}
\caption{Absolute errors of the two quantiles $\hat{Q}_{99\%}$  and $\hat{Q}_{99.9\%}$ of the PCE approximation of the Vasicek Interest rate model at time $T=1$. The parameters are set as $\alpha = 0.1$, $\beta = 0.2$, $\sigma = 0.25$ and $R_0=110$.} \label{tab:tab15}
\end{table}

The quantile computed by means of a standard Monte Carlo sampling of the analytical solution, whose size is equal to the one used for PCE, gives the following standard error
\begin{align*}
SE_{Q_{99\%}^{MC}} =0.0008740     \; ,            \quad \quad SE_{Q_{99.9\%}^{MC}} =     0.0022574 \;.   
\end{align*}

\newpage

\subsection{$\boldsymbol{\sigma =30\%}$}

In this section the value of the volatility is set to $\sigma = 30\%$, while the other parameters are chosen as in Table \ref{tab:tab11}.

\begin{figure}[!h]
\centering
\includegraphics[scale=0.35]{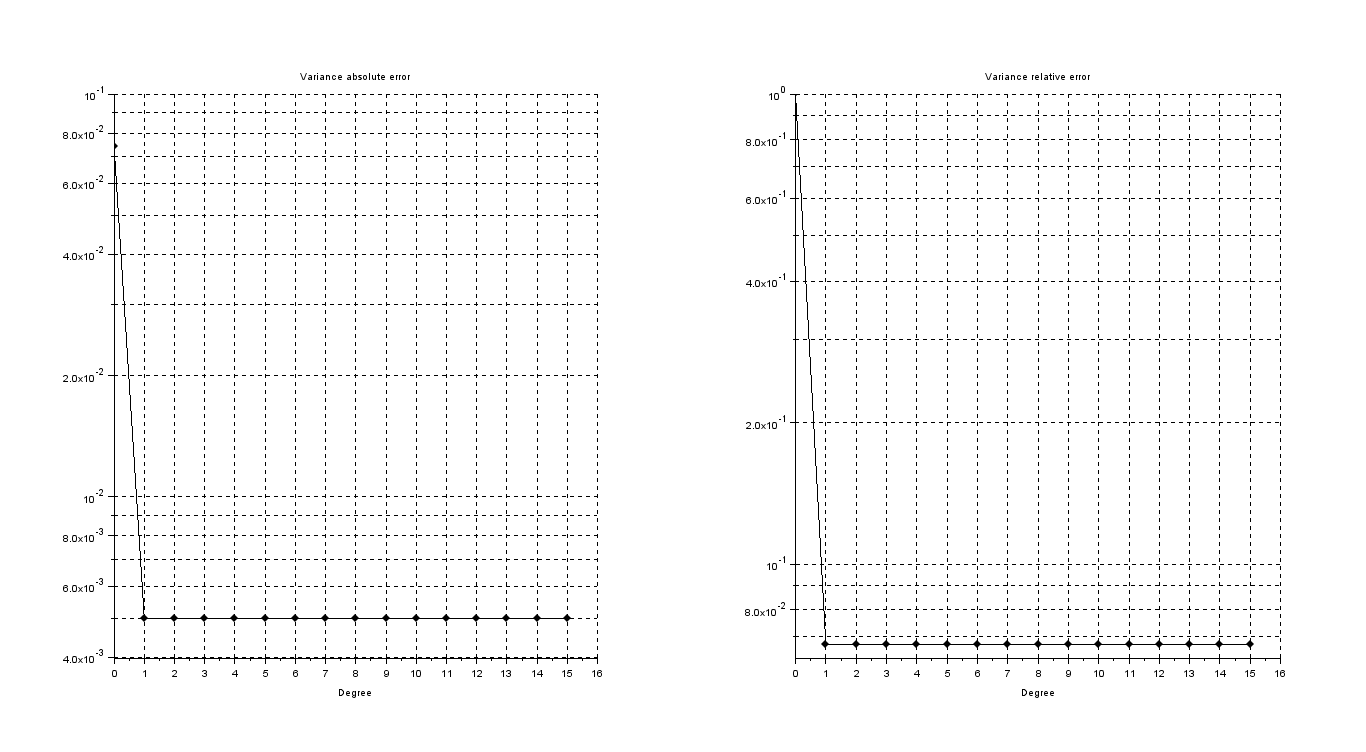}
\caption{Semilogy scale plot of the absolute error of the variance of $R_T^{(p)}$ evaluated at time $T=1$ (left) and the absolute  relative error of the variance (right) of $R_T^{(p)}$ at time $T=1$.  In both cases the parameters of the Vasicek interest rate model are $\alpha = 0.1$, $\beta = 0.2$, $\sigma = 30 \%$.}  \label{fig:fig23}
\end{figure}

First let us display in Figure \ref{fig:fig23} the absolute and relative error of the variance, while in Table \ref{tab:tab16} are collected both the absolute and absolute value of relative error for mean and variance concerning the Vasicek interest rate model.

\begin{table}[!h]
\centering
\begin{tabular}{ccccc}
 Degree of PCE &  Average Error & Variance Error & Average relative error & Variance relative error   \tabularnewline
\hline

 0 & 8.2289e-08 & 7.4178e-02 & 9.1279e-10 & 1.0000e+00 \tabularnewline 
 1 & 8.2400e-08 & 4.9958e-03 & 9.1402e-10 & 6.7349e-02 \tabularnewline 
 2 & 8.2292e-08 & 4.9958e-03 & 9.1282e-10 & 6.7349e-02 \tabularnewline 
 3 & 8.2276e-08 & 4.9958e-03 & 9.1264e-10 & 6.7349e-02 \tabularnewline 
 4 & 8.2334e-08 & 4.9958e-03 & 9.1328e-10 & 6.7349e-02 \tabularnewline 
 5 & 8.2374e-08 & 4.9958e-03 & 9.1374e-10 & 6.7349e-02 \tabularnewline 
 6 & 8.2287e-08 & 4.9958e-03 & 9.1277e-10 & 6.7349e-02 \tabularnewline 
 7 & 8.2332e-08 & 4.9958e-03 & 9.1326e-10 & 6.7349e-02 \tabularnewline 
 8 & 8.2349e-08 & 4.9958e-03 & 9.1345e-10 & 6.7349e-02 \tabularnewline 
 9 & 8.2326e-08 & 4.9958e-03 & 9.1320e-10 & 6.7349e-02 \tabularnewline 
 10 & 8.2346e-08 & 4.9958e-03 & 9.1342e-10 & 6.7349e-02 \tabularnewline 
 11 & 8.2296e-08 & 4.9958e-03 & 9.1286e-10 & 6.7349e-02 \tabularnewline 
 12 & 8.2324e-08 & 4.9958e-03 & 9.1318e-10 & 6.7349e-02 \tabularnewline 
 13 & 8.2349e-08 & 4.9958e-03 & 9.1346e-10 & 6.7349e-02 \tabularnewline 
 14 & 8.2309e-08 & 4.9958e-03 & 9.1301e-10 & 6.7349e-02 \tabularnewline 
 15 & 8.2345e-08 & 4.9958e-03 & 9.1342e-10 & 6.7349e-02 \tabularnewline

\end{tabular}
\caption{Absolute error of the average and the variance of $R_T^{(p)}$ at time $T=1$, with respect to the Vasicek parameters $\alpha = 0.1$, $\beta = 0.2$, $\sigma = 30 \%$.} \label{tab:tab16}
\end{table}
\newpage 
As for $\sigma = 15\%$ and $\sigma = 25\%$, the Table \ref{tab:tab16} shows the stationary behavior of the absolute and relative errors of mean and variance, which can be motivated for the same reasoning for $\sigma = 15\%$, and we recover  the same computations and bounds obtained for the previous values of $\sigma$.

\begin{figure}[!h]
\centering
\includegraphics[scale=0.35]{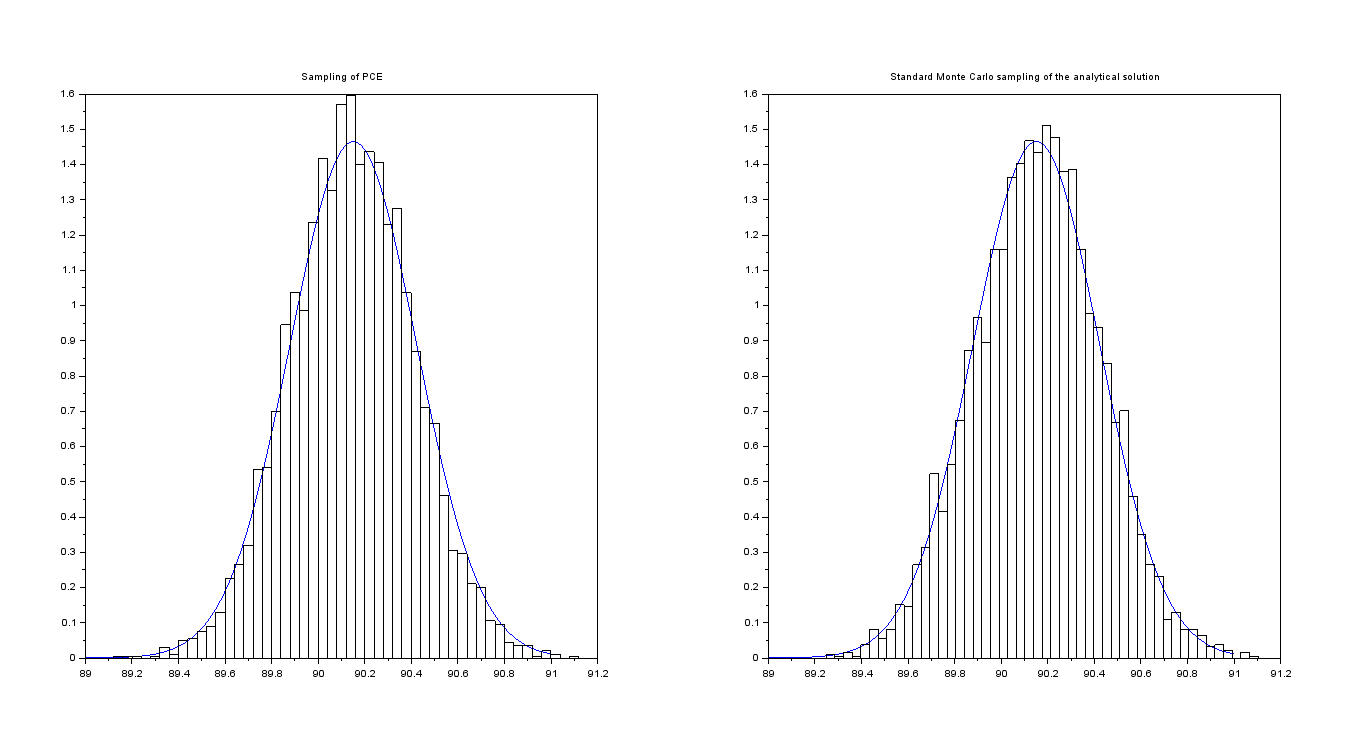}
\caption{  Left plot: Probability density function of the Vasicek interest rate model, with parameters  $\alpha = 0.1$, $\beta = 0.2$, $\sigma = 0.30$ and $R_0 =110$, (blue curve) and histogram of a Monte Carlo sampling (size = $5000$) of the $R_T^{(p)}$ for $p=15$. With the same parameters, in the right plot, we compare the analytical probability density function \eqref{sec:eq24}, with the one obtained using the Monte Carlo sampling of size $5000$, for $R_T$, being $T=1$. }  \label{fig:fig24}
\end{figure}

Coming back to the values in Table \ref{tab:tab11}, let us compute a Monte Carlo sampling of size $5000$ of $R^{(p)}_T$, with $p=15$, whose histogram, see fig. \ref{fig:fig24}, is compared with the probability density function of the normal random variable $R_T$.
Moreover in the same fig. \ref{fig:fig24} we show a standard Monte Carlo sampling. \newline

\begin{figure}[!h]
\centering
\includegraphics[scale=0.5]{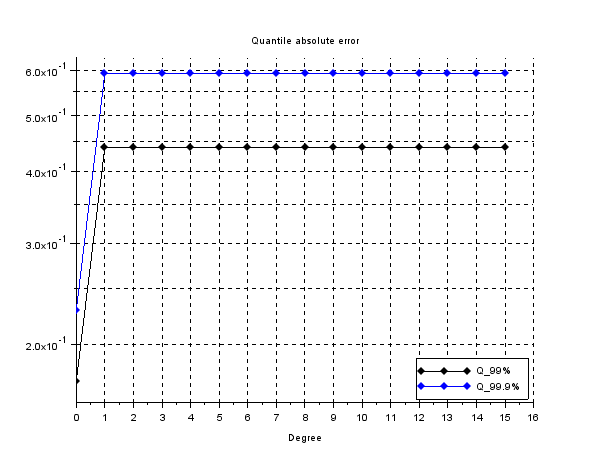}
\caption{Absolute errors of the two quantiles $\hat{Q}_{99\%}$  and $\hat{Q}_{99.9\%}$ of the PCE approximation of the Vasicek interest rate model at time $T=1$. The parameters are set as $\alpha = 0.1$, $\beta = 0.2$, $\sigma = 0.30$ and $R_0=110$.} \label{fig:fig25}
\end{figure}

\newpage 
The errors of quantiles for $\gamma  = 99\%$, resp. for $\gamma=99.9\%$, are shown in Fig. \ref{fig:fig25} and in Table \ref{tab:tab17}.  \newline

\begin{table}[!h]
\centering
\begin{tabular}{cccc}
 Degree of PCE &  $\epsilon_{99\%}$ & $\epsilon_{99.9\%}$   \tabularnewline
\hline
 0 & 1.7256e-01 & 2.2923e-01 \tabularnewline 
 1 & 4.4032e-01 & 5.9178e-01 \tabularnewline 
 2 & 4.4032e-01 & 5.9178e-01 \tabularnewline 
 3 & 4.4032e-01 & 5.9178e-01 \tabularnewline 
 4 & 4.4032e-01 & 5.9178e-01 \tabularnewline 
 5 & 4.4032e-01 & 5.9178e-01 \tabularnewline 
 6 & 4.4032e-01 & 5.9178e-01 \tabularnewline 
 7 & 4.4032e-01 & 5.9178e-01 \tabularnewline 
 8 & 4.4032e-01 & 5.9178e-01 \tabularnewline 
 9 & 4.4032e-01 & 5.9178e-01 \tabularnewline 
 10 & 4.4032e-01 & 5.9178e-01 \tabularnewline 
 11 & 4.4032e-01 & 5.9178e-01 \tabularnewline 
 12 & 4.4032e-01 & 5.9178e-01 \tabularnewline 
 13 & 4.4032e-01 & 5.9178e-01 \tabularnewline 
 14 & 4.4032e-01 & 5.9178e-01 \tabularnewline 
 15 & 4.4032e-01 & 5.9178e-01 \tabularnewline 
\end{tabular}
\caption{Absolute errors of the two quantiles $\hat{Q}_{99\%}$  and $\hat{Q}_{99.9\%}$ of the PCE approximation of the Vasicek Interest rate model at time $T=1$. The parameters are set as $\alpha = 0.1$, $\beta = 0.2$, $\sigma = 0.30$ and $R_0=110$.}  \label{tab:tab17}
\end{table}

The quantiles computed by means of a standard Monte Carlo sampling of the analytical solution, whose size is equal to the one used for PCE, give the following standard error
\begin{align*}
&SE_{Q_{99\%}^{MC}} =0.0010324           \\ &SE_{Q_{99.9\%}^{MC}} =       0.0025387  
\end{align*}



\section{CIR interest rate model} \label{sec:secCIR}
In what follows we consider the PCE method for the CIR interest rate model,  introduced in 1985 by John C. Cox, Jonathan E. Ingersoll and Stephen A. Ross, see \cite{Cox}, and  defined by the following SDE

\begin{equation}
dR_t = \left(\alpha-\beta R_t\right) dt+\sigma \sqrt{R_t} dW_t \;,\label{sec:eq34}
\end{equation}
where $\alpha,\beta$ and $\sigma$ are positive, real-valued constants, and $W_t$ is a standard Wiener process.
Let us recall that eq. \eqref{sec:eq34} generalizes the Vasicek model seen in Sec. \ref{sec:secVAS}, in fact, even if the drift term $\left(\alpha-\beta R_t\right)$ is common between the twos, the standard deviation factor is different and, in particular, it equals $\sigma \sqrt{R_t}$ for the CIR model ensuring to avoid
negative interest rates if $\alpha$ and $\beta$ are positive. Moreover if $2 \alpha \beta \geq \sigma^2$, we cannot have $R_t$ to be equal to zero.
It is worth to mention that, unlike in the case of the Vasicek model, there is not an analytical solution to eq. \eqref{sec:eq34}. The CIR process is  ergodic and possesses a stationary distribution, see, e.g.,  \cite[Section 3.2.3]{Brigo} and \cite[Section 23.5]{Hull} for further details.

The statistics of the CIR model for fixed $T>0$, read as follows
\begin{align*}
\Expect{R_T} &= R(0) e^{-\beta T}+\frac{\alpha}{\beta}(1-e^{-\beta T})\;, \\ 
\Var{R_T} &= \frac{\sigma^2}{\beta}R(0)(e^{-\beta t}-e^{-2 \beta t})+\frac{\alpha \sigma^2}{2\beta^2}\left(1- 2e^{-\beta T}+e^{-2 \beta T} \right)\:.
\end{align*}

\subsection{PCE approximation}

In order to apply the PCE method to the process $R_t$ in \eqref{sec:eq34} evaluated at a specific time $T>0$, let us consider the stochastic process $\{Y_t\}_{t \ge0}$
\begin{equation}
Y_t = \sqrt{R_t} \label{sec:eq35}\;,
\end{equation}
then, applying the It\=o-D\"oebling Lemma, $Y_t$  satisfies the following SDE 
\begin{equation}
dY_t = \frac{1}{2Y_t}\left(\alpha-\beta Y^2_t -\frac{1}{4}\sigma^2 \right) dt+ \frac{1}{2} \sigma dW_t \;, \label{sec:eq36}
\end{equation}
therefore we have canceled  the interactions between the state of the original process $R_t$  and the increments of the
Wiener process. The transformation defined by \eqref{sec:eq35}, see \cite[Chapter 2]{Iacus} and \cite[Chapter 4]{Jackel} for details, allows to reduce simulations instability as well as it simplifies the application of Doss approach, developed in Sec. \ref{sec:secDoss}. Indeed the volatility is constant thus its derivative is null. 

Actually there is a subtleties concerning the drift $\mu(x) = \frac{1}{x} \left(\alpha -\beta x^2-\frac{1}{4}\sigma^2 \right)$, it is Lipschitz on intervals $(a,+\infty)$, where $a>0$. Since the solution of eq. \eqref{sec:eq36} is strictly positive, providing  $2\alpha \beta \geq \sigma^2$,  it is quite reasonable to apply the Doss theory in such setting. \newline

As seen studying the gBm model, see Sec. \ref{sec:secGBM} and the Vasicek model, see Sec. \ref{sec:secVAS}, the solution to (\ref{sec:eq35}) can be expressed in terms of a functional $H$ 
\begin{equation}
Y_T = H(D_T,W_T)\;, \label{sec:eq37}
\end{equation}
for every fixed  time $T$, which is often referred as {\it maturity time}.
In particular the scalar function $H(x,y)$ is determined by solving the ODE \eqref{sec:eq14}, namely
\begin{equation*}
H(x,y) = \frac{\sigma}{2} y+x\:.
\end{equation*}
Then the process $\{D_t\}_{t \ge 0}$ is path-wise  determined by solving the following ODE
\begin{equation} \label{sec:eq38}
\begin{cases}
\dot{D}{(\omega)} = \dfrac{1}{2H(D(\omega),W_t(\omega))}\left(\alpha-\beta \big(H(D(\omega),W_t(\omega))\big)^2 -\frac{1}{4}\sigma^2 \right)   \hspace{.5 cm} t>0 \\
D(\omega) = R_0  \hspace{8.25 cm} t=0
\end{cases}\;,
\end{equation}
for almost all $\omega \in \Omega$. Notice that the dependence on $t$ of the function $D(\omega)$, for each fixed sample event $\omega$, is omitted in order to shorten the notation. Moreover  $\dot{D}{(\omega)}$ represents its derivative with respect to time.
As in Sec. \ref{sec:secGBM} and Sec. \ref{sec:secVAS},  the Wiener process is
\begin{equation}
W_t \sim \mathcal{N}(0,t) \label{sec:eq39} \;,  t \in [0,T],
\end{equation}
which allows to set $W_t = \sqrt{2t} \xi$, and we still consider the {\it basic random variable} s $\xi\sim\mathcal{N}(0,1/2)$, as made in the previous {\it gBm and Vasicek sections}, to fulfill the restrictions imposed by the software we have used to numerically implement the PCE method.
In particular \eqref{sec:eq38} is numerically solved, for a fixed sample event $\omega$, by means of an adaptive Runge-Kutta method of order 4 (RK4), where the absolute tolerance, resp. relative tolerance, is set as $1e$-$7$, resp. $1e$-$5$, see, e.g., \cite[Section II.1, Scetion II.4]{Hairer} for further details. \newline

We can therefore define the solution of \eqref{sec:eq38} at time $T$ as $D_T = \mathcal{M}_1(\xi,\Theta)$, where $\Theta \in\ R^3$ collects all the parameters that characterized the dynamics of the CIR interest rate model \eqref{sec:eq34}. Then using both \eqref{sec:eq37} and \eqref{sec:eq35}, we get
\begin{equation}
 \mathcal{M}(\xi,\Theta) := R_T =\left( \frac{\sigma}{2}  W_T+D_T \right)^2 =  \left( \frac{\sigma}{2} \sqrt{2T}\xi+\mathcal{M}_1(\xi,\Theta) \right)^2  \;. \label{sec:eq40}
\end{equation}
We shorten the notation omitting the explicit dependence of the CIR model on the parameters vector $\Theta = (\alpha,\beta,\sigma) \in \R^3$, being $\alpha,\beta,\sigma$ positive constants. Applying  the truncated, at degree $p$, PCE to $R_T$, we obtain
\begin{equation*}
R_T^{(p)} = \sum_{i=0}^p c_i \Psi_i(\xi)\;,
\end{equation*}
where the coefficients are detecting using (\ref{sec:eq10}), evaluating $R_T = \mathcal{M}(\xi)$ on a set of realizations of the basic random variable, namely we compute \eqref{sec:eq40} at $\{\xi_j\}_{j=1}^N$, taking $N=p$ and referring to these values  as
\begin{equation}
\{R_{T,j}\}_{j=1}^N = \{\mathcal{M}(\xi_j)\}_{j=1}^N \;, \label{sec:eq41}
\end{equation}
where such realizations are determined in two steps:
\begin{itemize}
\item each Gaussian quadrature nodes, belonging to $\{\xi_j\}_{j=1}^N$, can be interpreted as the image of a suitable $\omega_j \in \Omega$ through $\xi$. Therefore by integrating \eqref{sec:eq38} for each path in $\{\omega_1,\dots,\omega_N\} \subset \Omega$ we achieve a set of realization of $D_T$, let us say $\{\mathcal{M}_1(\xi_j)\}_{j=1}^N$.
 It is worth to mentioned that  we are integrating $N$ independent ODEs.
\item By means of \eqref{sec:eq40} we get the required set $\{R_{T,j}\}_{j=1}^N$, where the second summand is simply $\{{\frac{\sigma}{2} \sqrt{2 T}\xi_j}\}_{j=1}^N$.
\end{itemize}

\subsection{Numerical application: overview of the computations} \label{sec:secnum}

Even if the CIR model does not admit close solution, see, e.g. \cite[Section 4.4.4]{Shreve}, it possesses a stationary distribution, namely, by following \cite[Section 3.2.3]{Brigo}, $R_T$ features a non-central chi-squared distribution. In particular let us denote by $f_{R_T}$ the analytical probability density function of $R_T$, for a positive end time $T$, then
\begin{equation}
f_{R_T}(x) = f_{\chi^2(q,\lambda_T)/c_T} = c_T f_{\chi^2(q,\lambda_T)}(c_Tx),\quad x \in [0,+\infty) \; , \label{sec:eq42}
\end{equation}
where the subscript denotes the random variable considered. Moreover the constants are defined as
\begin{align}
c_T &= \frac{4 \beta}{\sigma^2 (1-\exp(-\beta T))} \; \label{sec:eq43},\\
q &= \frac{4 \alpha}{\sigma^2} \; ,\label{sec:eq44} \\
\lambda_T &=  c_T R_0 \exp(-\beta T)\;\label{sec:eq45} ,
\end{align}
where the last two constants are, respectively, the degree of freedom and the non-centrality parameter, see, e.g., \cite[Section 3.2.3]{Brigo}, for further details.
In order to achieve a natural number $q=3$ let us set $\alpha = \frac{3}{4}\sigma^2$, where $\sigma$ is considered as a known data. \newline

Therefore to compute the  probability density function and the analytic quantiles we set $\alpha$ in order to satisfy \eqref{sec:eq44}, where $q=3$, moreover the PCE-approximation is computed for a set of volatility, namely
\begin{equation*}
\sigma = \{15 \%, 25 \%, 30 \%\}
\end{equation*}
while the other parameters are shown in Table \ref{tab:tab18}. \newline

\begin{table}[!h]
\centering
\begin{tabular}{c|ccccc}
\centering Parameters & \centering $\beta$ &  \centering$ R_0 $& \centering T \tabularnewline
\hline
\centering Values &\centering 0.2  & \centering 110 & \centering2 \tabularnewline
\end{tabular}
\caption{Parameters involved in the CIR interest model.} \label{tab:tab18}
\end{table}

Then for each value of $\sigma$ and for an increasing set of degree $ p \in \{0,1, \ldots,15\}$ the absolute errors of the mean, resp. of the variance, are computed as follows
\begin{align*}
\epsilon^{(p)}_{MEAN} & = \abs{\Expect{R_T} -\Expect{R^{(p)}_T}}\;,\\
\epsilon^{(p)}_{VAR} & = \abs{\Var{R_T}-\Var{R^{(p)}_T} }\;,
\end{align*}
furthermore the absolute value of the relative errors are considered, where, as we have assumed discussing the gBm and the Vasicek model, due to the logarithmic scale, we report their absolute values, to highlight its order, rather that its numerical value. These errors are given by
\begin{align*}
RE^{(p)}_{MEAN} &= \frac{\epsilon^{(p)}_{MEAN}}{\Expect{R_T}}\;,\\
RE^{(p)}_{VAR} &= \frac{\epsilon^{(p)}_{VAR}}{\Var{R_T}}\:.
\end{align*}

Lastly we compute the two quantiles $\hat{Q}_\gamma$, where $\gamma = 99\%$ and $\gamma=99.9\%$, of the PCE approximation $R_T^{(p)}$, where $ p =0,1,\ldots,15$.
Proceeding as in Sec. \ref{sec:secGBM}, we consider the $\gamma$-th sample quantile, namely the $(K+1)$-th realization of the sampling of $R_T^{(p)}$, sorted in ascending order, such that $K \le [\gamma M]$, where $M$ is the size of the sampling, and $[\cdot]$ denotes the integer part of the real number within the brackets. In what follows we always employ a  Latin Hypercube Sampling (LHS) technique of $R_T^{(p)}$, see \cite{Tang, Owen} and also \cite{Iman} for further references, of size $M=5000$. In particular we  have $K_{99\%} = 4951$, while  $K_{99.9\%} = 4996$, and we compute the absolute error of $\hat{Q}_{\gamma}$ versus the analytical values ${Q}_\gamma$. Therefore we evaluate
$
\epsilon_{\gamma} = \abs{\hat{Q}_\gamma-{Q}_\gamma}
$, 
where $Q_\gamma$ is the quantile of a normal random variable of mean $\Expect{R_T}$ and variance $\Var{R_T}$ and, indicating its cumulative density function by $F_{R_T*}$, we have $
Q_\gamma = F_{R_T}^{-1}(\gamma)
$. \newline

We compare these quantiles with the ones estimated using a standard Monte Carlo sampling, whose size is $M=5000$, of the analytical solution $R_T$. In particular we compute the standard error for $L=200$ independent estimates  $\left\{Q_\gamma^{MC}(l)\right\}_{l=1}^L$ of the quantile. Thus
%
\begin{equation*}
SE_{Q_{99\%}^{MC}} = \frac{\hat{\sigma}}{\sqrt{L}}
\end{equation*}
where $\hat{\sigma}$ estimates the standard deviation of $L$ independent estimates of  $Q_\gamma$, thus
\begin{equation*}
\hat{\sigma}^2 = \frac{1}{L-1}\sum_{l=1}^L \left(Q_\gamma^{MC}(l)-\overline{Q}_\gamma^{MC}\right)^2\;,
\end{equation*}
while $\overline{Q}_\gamma^{MC}$ represents the arithmetic mean of $\left\{Q_\gamma^{MC}(l)\right\}_{l=1}^L$.

 \begin{remark}.
As for the gBm and the Vasicek case, the choice of the basic estimator for quantiles, i.e. the sample quantile, 
is due at focusing our attention to the efficacy of the method, instead that on  the accuracy of the estimates as well as providing a fair comparison between the data achieved. More accurate techniques are, e.g., those presented in \cite{Chen}, the \emph{L-estimator}, and also the \emph{Harrel Davis} (HD) estimators, see, e.g., \cite{Mausser} for further details.
\end{remark}

\subsection{$\boldsymbol{\sigma = 15\%}$}

In this section the value of the volatility is set to $\sigma = 15\%$, therefore as discussed in Sec \ref{sec:secnum}, $\alpha$ satisfies \eqref{sec:eq44}, for $q=3$. Its numerical value is
\begin{equation*}
\alpha = 0.005625\;,
\end{equation*}
while the other parameters are chosen as in Table \ref{tab:tab18}. \newline 

\begin{figure}[!h]
\centering
\includegraphics[scale=0.34]{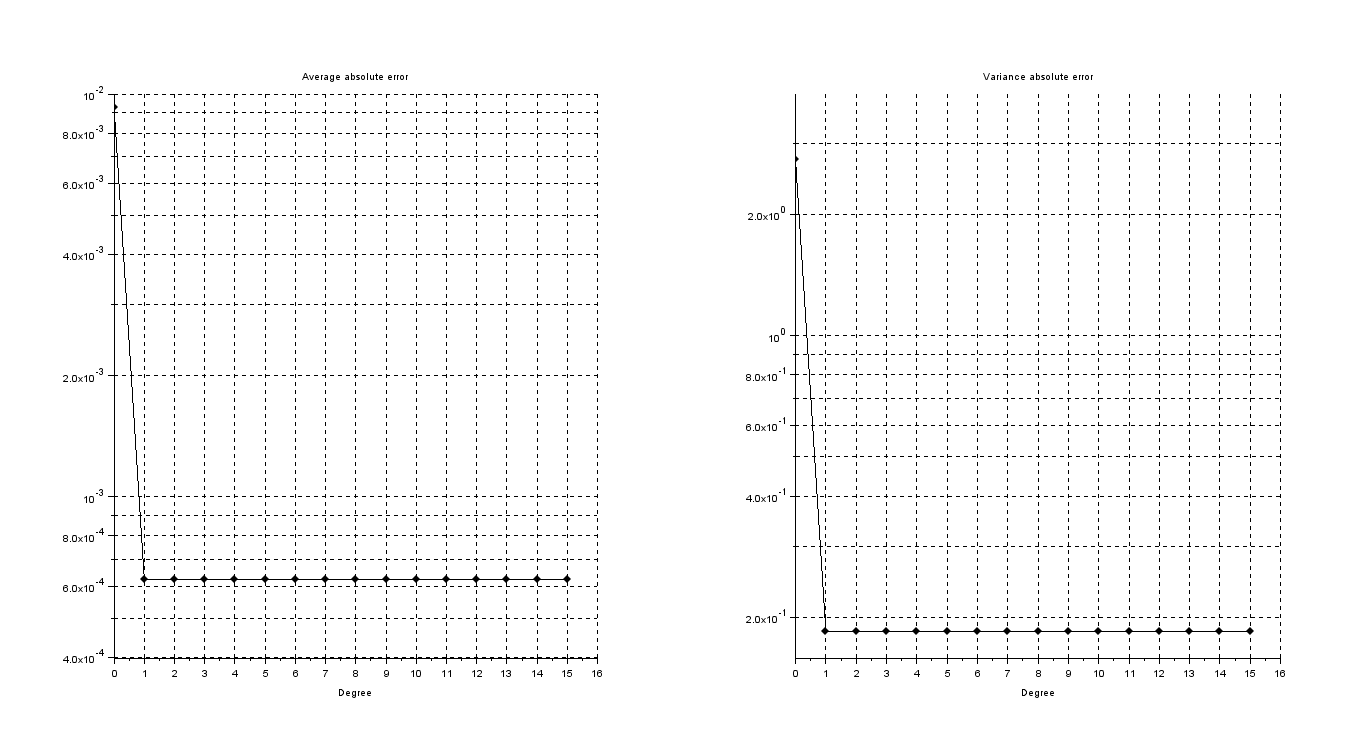}
\caption{Semilogy scale plot of the absolute error of the mean (left) and the variance (right) computed via PCE for the CIR at time $T=2$, whose parameters are $\alpha = 0.005625$, $\beta = 0.2$, $\sigma = 15 \%$ and starting value $R_0= 110$, for a set of degrees $p = \{0,1,2,\dots,15\}$.}  \label{fig:fig26}
\end{figure}

\begin{figure}[!h]
\centering
\includegraphics[scale=0.34]{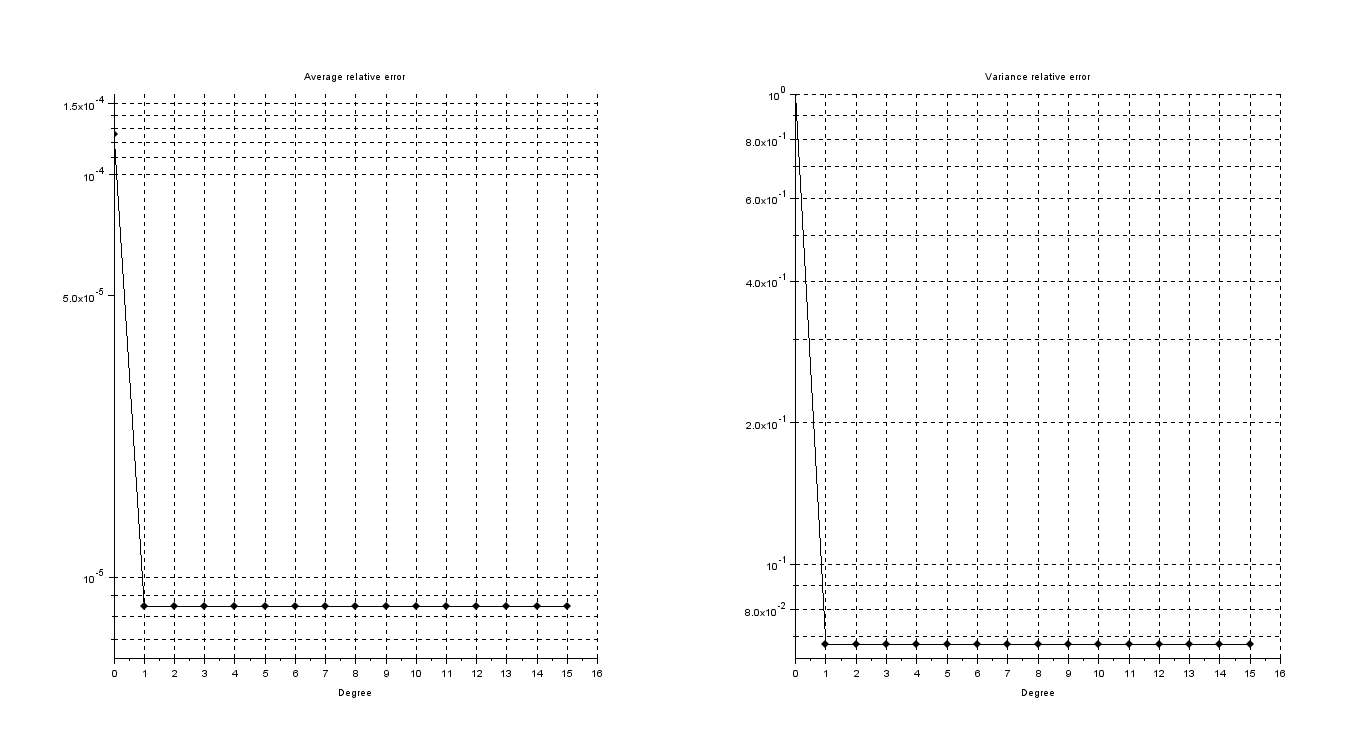}
\caption{Semilogy scale plot of the absolute value of the relative error of the mean (left) and the variance (right) computed via PCE for the CIR at time $T=2$, whose parameters are $\alpha = 0.005625$, $\beta = 0.2$, $\sigma = 15 \%$ and starting value $R_0= 110$, for a set of degrees $p = \{0,1,2,\dots,15\}$.}  \label{fig:fig27}
\end{figure}

First let us display in Figure \ref{fig:fig26}, Figure \ref{fig:fig27} and Table \ref{tab:tab19}, the absolute and relative error of the average, resp. variance, for the CIR interest rate model.

\begin{table}[!h]
\centering
\begin{tabular}{ccccc}
 Degree of PCE &  Average Error & Variance Error & Average relative error & Variance relative error   \tabularnewline
\hline
 0 & 9.2721e-03 & 2.7353e+00 & 1.2572e-04 & 1.0000e+00 \tabularnewline 
 1 & 6.2470e-04 & 1.8449e-01 & 8.4691e-06 & 6.7449e-02 \tabularnewline 
 2 & 6.2470e-04 & 1.8434e-01 & 8.4691e-06 & 6.7395e-02 \tabularnewline 
 3 & 6.2470e-04 & 1.8434e-01 & 8.4691e-06 & 6.7395e-02 \tabularnewline 
 4 & 6.2470e-04 & 1.8434e-01 & 8.4691e-06 & 6.7395e-02 \tabularnewline 
 5 & 6.2470e-04 & 1.8434e-01 & 8.4691e-06 & 6.7395e-02 \tabularnewline 
 6 & 6.2470e-04 & 1.8434e-01 & 8.4691e-06 & 6.7395e-02 \tabularnewline 
 7 & 6.2470e-04 & 1.8434e-01 & 8.4691e-06 & 6.7395e-02 \tabularnewline 
 8 & 6.2470e-04 & 1.8434e-01 & 8.4691e-06 & 6.7395e-02 \tabularnewline 
 9 & 6.2470e-04 & 1.8434e-01 & 8.4691e-06 & 6.7395e-02 \tabularnewline 
 10 & 6.2470e-04 & 1.8434e-01 & 8.4691e-06 & 6.7395e-02 \tabularnewline 
 11 & 6.2470e-04 & 1.8434e-01 & 8.4691e-06 & 6.7395e-02 \tabularnewline 
 12 & 6.2470e-04 & 1.8434e-01 & 8.4691e-06 & 6.7395e-02 \tabularnewline 
 13 & 6.2470e-04 & 1.8434e-01 & 8.4691e-06 & 6.7395e-02 \tabularnewline 
 14 & 6.2470e-04 & 1.8434e-01 & 8.4691e-06 & 6.7395e-02 \tabularnewline 
 15 & 6.2470e-04 & 1.8434e-01 & 8.4691e-06 & 6.7395e-02 \tabularnewline 
\end{tabular}
\caption{Absolute error of the average and the variance of $R_T^{(p)}$ at time $T=2$, with respect to the CIR parameters $\alpha = 0.005625$, $\beta = 0.2$, $\sigma = 15 \%$.} \label{tab:tab19}
\end{table}

\newpage

The absolute and relative errors of mean and variance are stationary. Therefore there is an error that corrupts the spectral convergence of PCE-approximation. In particular the numerical values suggest that it is not due to numerical error made in order to compute $\{R_{T,j}\}_{j=1}^N$,  but it is due to the approximation  of the solution (\ref{sec:eq34}) obtained by means of \eqref{sec:eq40}.

\begin{figure}[!h]
\centering
\includegraphics[scale=0.34]{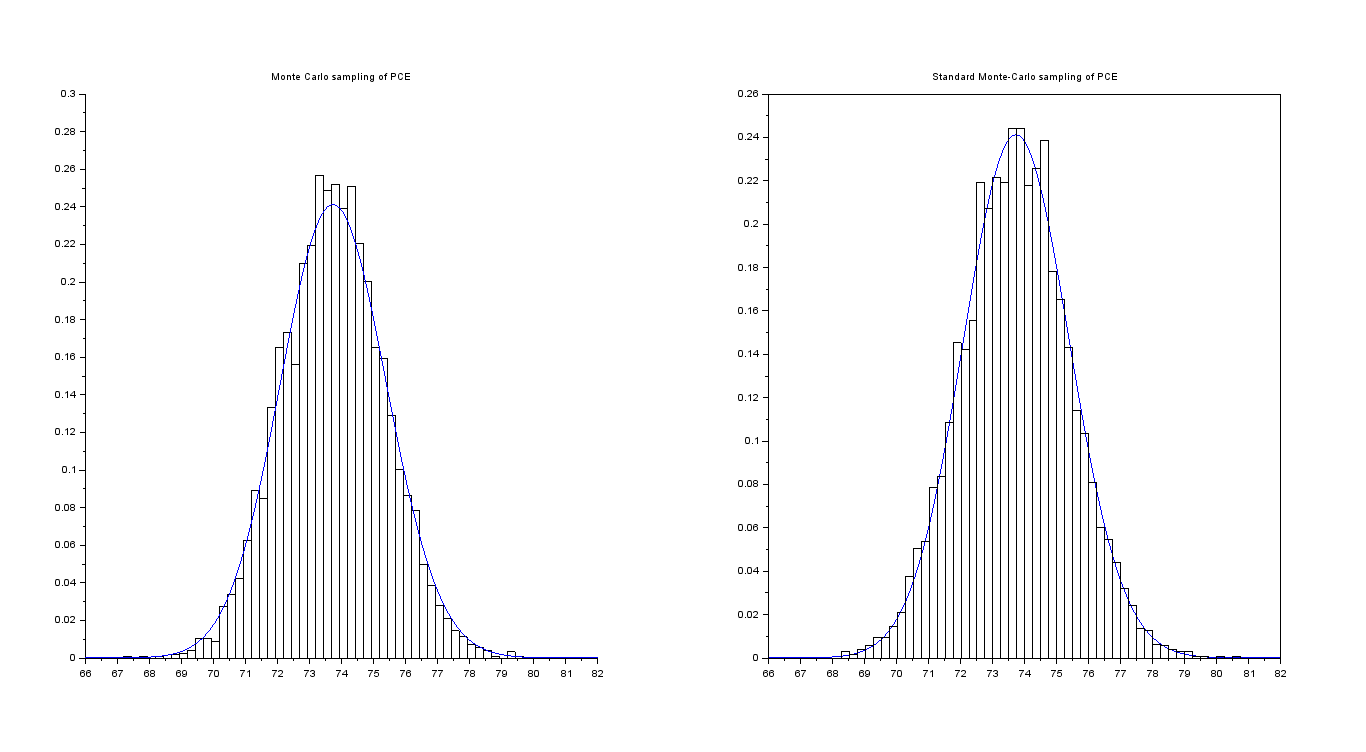}
\caption{Left plot: Probability density function of the CIR interest rate model, with parameters  $\alpha = 0.005625$, $\beta = 0.2$, $\sigma = 0.15$ and $R_0 =110$, (blue curve) and histogram of a Monte Carlo sampling (size = $5000$) of the $R_T^{(p)}$ for $p=15$. With the same parameters, in the right plot we compare the analytical probability density function of the  CIR model, with the one obtained using the Monte Carlo sampling of size $5000$, for $R_T$, being $T=2$. }  \label{fig:fig28}
\end{figure}

Coming back to the values in Table \ref{tab:tab18}, let us compute a sampling of size $5000$ of $R^{(p)}_T$, with $p=15$, whose histogram, see fig. \ref{fig:fig28}, is compared with the probability density function of the normal random variable $R_T$.
Moreover in the same fig. \ref{fig:fig28} we show a sampling obtained exploiting the standard Monte Carlo technique.

\newpage

\begin{figure}[!h]
\centering
\includegraphics[scale=0.5]{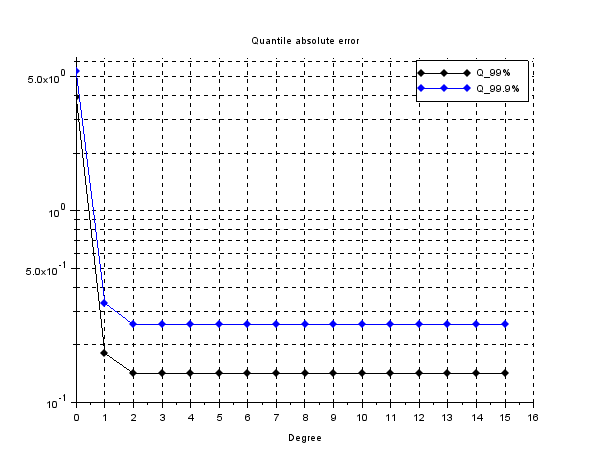}
\caption{Absolute errors of the two quantiles $\hat{Q}_{99\%}$  and $\hat{Q}_{99.9\%}$ of the PCE approximation of the CIR interest rate model at time $T=2$. The parameters are set as $\alpha = 0.005625$, $\beta = 0.2$, $\sigma = 0.15$ and $R_0=110$.} \label{fig:fig29}
\end{figure}

The errors of quantiles for $\gamma = 99\%$, resp. for $\gamma =99.9\%$, are shown in Fig. \ref{fig:fig29} and in Table \ref{tab:tab20}. \newline

\begin{table}[!h]
\centering
\begin{tabular}{cccc}
 Degree of PCE &  $\epsilon_{99\%}$ & $\epsilon_{99.9\%}$   \tabularnewline
\hline
 0 & 3.9106e+00 & 5.3240e+00 \tabularnewline 
 1 & 1.8048e-01 & 3.3009e-01 \tabularnewline 
 2 & 1.4218e-01 & 2.5449e-01 \tabularnewline 
 3 & 1.4218e-01 & 2.5449e-01 \tabularnewline 
 4 & 1.4218e-01 & 2.5449e-01 \tabularnewline 
 5 & 1.4218e-01 & 2.5449e-01 \tabularnewline 
 6 & 1.4218e-01 & 2.5449e-01 \tabularnewline 
 7 & 1.4218e-01 & 2.5449e-01 \tabularnewline 
 8 & 1.4218e-01 & 2.5449e-01 \tabularnewline 
 9 & 1.4218e-01 & 2.5449e-01 \tabularnewline 
 10 & 1.4218e-01 & 2.5449e-01 \tabularnewline 
 11 & 1.4218e-01 & 2.5449e-01 \tabularnewline 
 12 & 1.4218e-01 & 2.5449e-01 \tabularnewline 
 13 & 1.4218e-01 & 2.5449e-01 \tabularnewline 
 14 & 1.4218e-01 & 2.5449e-01 \tabularnewline 
 15 & 1.4218e-01 & 2.5449e-01 \tabularnewline 
\end{tabular}
\caption{Absolute errors of the two quantiles $\hat{Q}_{99\%}$  and $\hat{Q}_{99.9\%}$ of the PCE approximation of the CIR Interest rate model at time $T=2$. The parameters are set as $\alpha = 0.005625$, $\beta = 0.2$, $\sigma = 0.15$ and $R_0=110$.}  \label{tab:tab20}
\end{table}

The quantiles computed by means of a standard Monte Carlo sampling of the analytical solution, whose size is equal to the one used for PCE, give the following standard error
\begin{align*}
&SE_{Q_{99\%}^{MC}} = 0.0201547              \\ &SE_{Q_{99.9\%}^{MC}} =   0.0534291   
\end{align*}

\subsection{$\boldsymbol{\sigma = 25\%}$}

\begin{figure}[!h]
\centering
\includegraphics[scale=0.31]{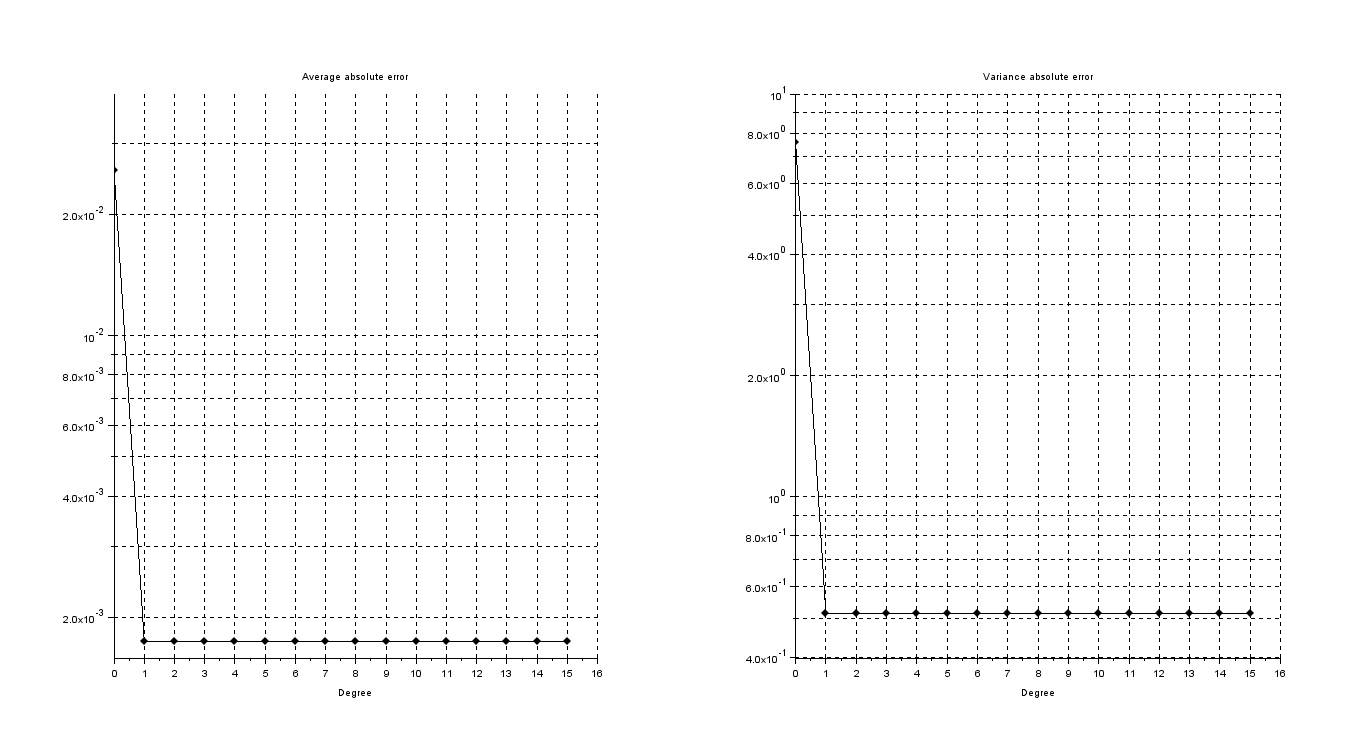}
\caption{Semilogy scale plot of the absolute error of the mean (left) and the variance (right) computed via PCE for the CIR at time $T=2$, whose parameters are $\alpha =  0.046875$, $\beta = 0.2$, $\sigma = 25 \%$ and starting value $R_0= 110$, for a set of degrees $p = \{0,1,2,\dots,15\}$.}  \label{fig:fig30}
\end{figure}

\begin{figure}[!h]
\centering
\includegraphics[scale=0.31]{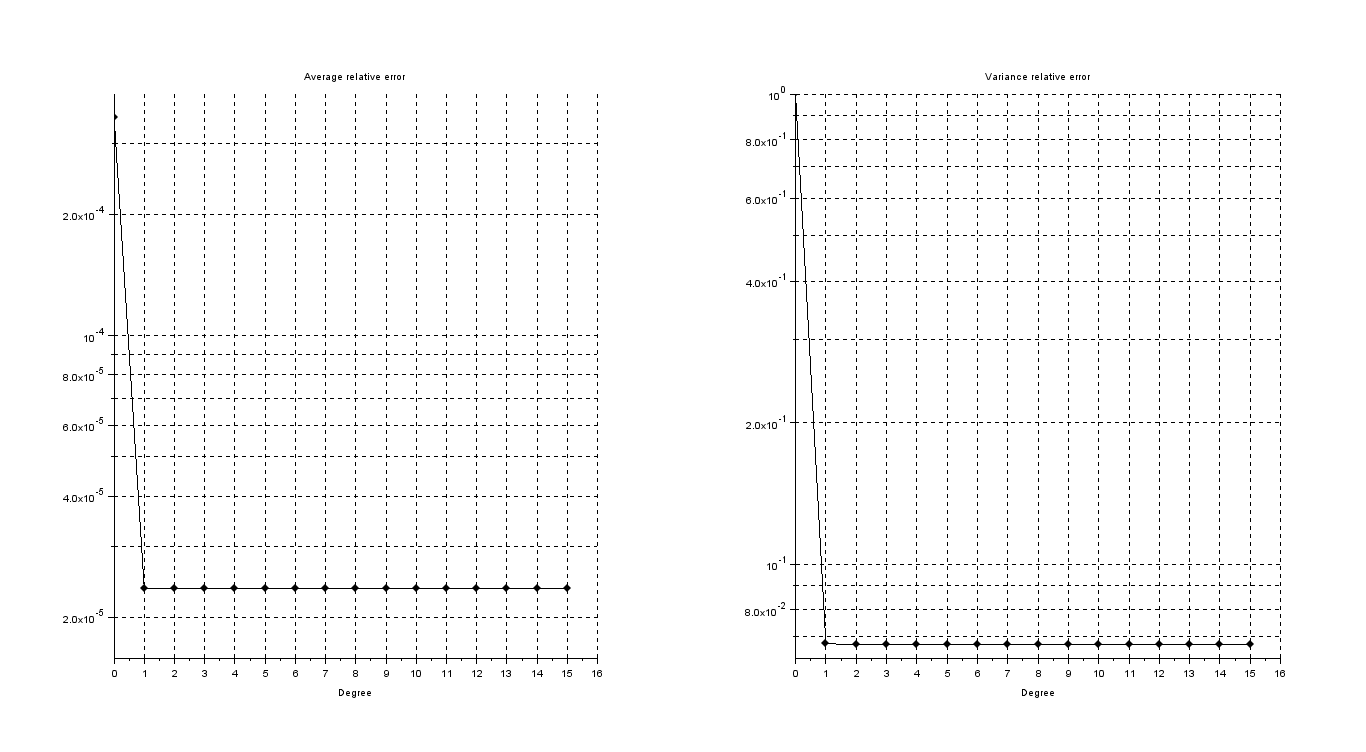}
\caption{Semilogy scale plot of the absolute value of the relative error of the mean (left) and the variance (right) computed via PCE for the CIR at time $T=2$, whose parameters are $\alpha =   0.046875$, $\beta = 0.2$, $\sigma = 25 \%$ and starting value $R_0= 110$, for a set of degrees $p = \{0,1,2,\dots,15\}$.}  \label{fig:fig31}
\end{figure}

In this section the value of the volatility is set to $\sigma = 25\%$, consequently by means of  \eqref{sec:eq44}, where $q=3$, we get
\begin{equation*}
\alpha =0.046875\;.
\end{equation*}
The other parameters are chosen as in Table \ref{tab:tab18}.
First let us display in Figure \ref{fig:fig30}, Figure \ref{fig:fig31} and Table \ref{tab:tab21} the absolute and relative error of the average, resp. variance, for the CIR interest rate model.

\begin{table}[!h]
\centering
\begin{tabular}{ccccc}
 Degree of PCE &  Average Error & Variance Error & Average relative error & Variance relative error   \tabularnewline
\hline
 0 & 2.5756e-02 & 7.6005e+00 & 3.4906e-04 & 1.0000e+00 \tabularnewline 
 1 & 1.7373e-03 & 5.1400e-01 & 2.3537e-05 & 6.7627e-02 \tabularnewline 
 2 & 1.7373e-03 & 5.1284e-01 & 2.3537e-05 & 6.7475e-02 \tabularnewline 
 3 & 1.7373e-03 & 5.1284e-01 & 2.3537e-05 & 6.7475e-02 \tabularnewline 
 4 & 1.7373e-03 & 5.1284e-01 & 2.3537e-05 & 6.7475e-02 \tabularnewline 
 5 & 1.7373e-03 & 5.1284e-01 & 2.3537e-05 & 6.7475e-02 \tabularnewline 
 6 & 1.7373e-03 & 5.1284e-01 & 2.3537e-05 & 6.7475e-02 \tabularnewline 
 7 & 1.7373e-03 & 5.1284e-01 & 2.3537e-05 & 6.7475e-02 \tabularnewline 
 8 & 1.7373e-03 & 5.1284e-01 & 2.3537e-05 & 6.7475e-02 \tabularnewline 
 9 & 1.7373e-03 & 5.1284e-01 & 2.3537e-05 & 6.7475e-02 \tabularnewline 
 10 & 1.7373e-03 & 5.1284e-01 & 2.3537e-05 & 6.7475e-02 \tabularnewline 
 11 & 1.7373e-03 & 5.1284e-01 & 2.3537e-05 & 6.7475e-02 \tabularnewline 
 12 & 1.7373e-03 & 5.1284e-01 & 2.3537e-05 & 6.7475e-02 \tabularnewline 
 13 & 1.7373e-03 & 5.1284e-01 & 2.3537e-05 & 6.7475e-02 \tabularnewline 
 14 & 1.7373e-03 & 5.1284e-01 & 2.3537e-05 & 6.7475e-02 \tabularnewline 
 15 & 1.7373e-03 & 5.1284e-01 & 2.3537e-05 & 6.7475e-02 \tabularnewline 

\end{tabular}
\caption{Absolute error of the average and the variance of $R_T^{(p)}$ at time $T=2$, with respect to the CIR parameters $\alpha =   0.046875$, $\beta = 0.2$, $\sigma = 25 \%$.} \label{tab:tab21}
\end{table}

The absolute and relative errors of mean and variance are stationary, see Table \ref{tab:tab21}. Therefore there is an error that corrupts the spectral convergence of PCE-approximation, and the obtained  values suggest that it is not due to numerical errors related to the computation of $\{R_{T,j}\}_{j=1}^N$, but, instead, it is due to the choice of approximating the solution (\ref{sec:eq34}) by means of \eqref{sec:eq40}.

\begin{figure}[!h]
\centering
\includegraphics[scale=0.31]{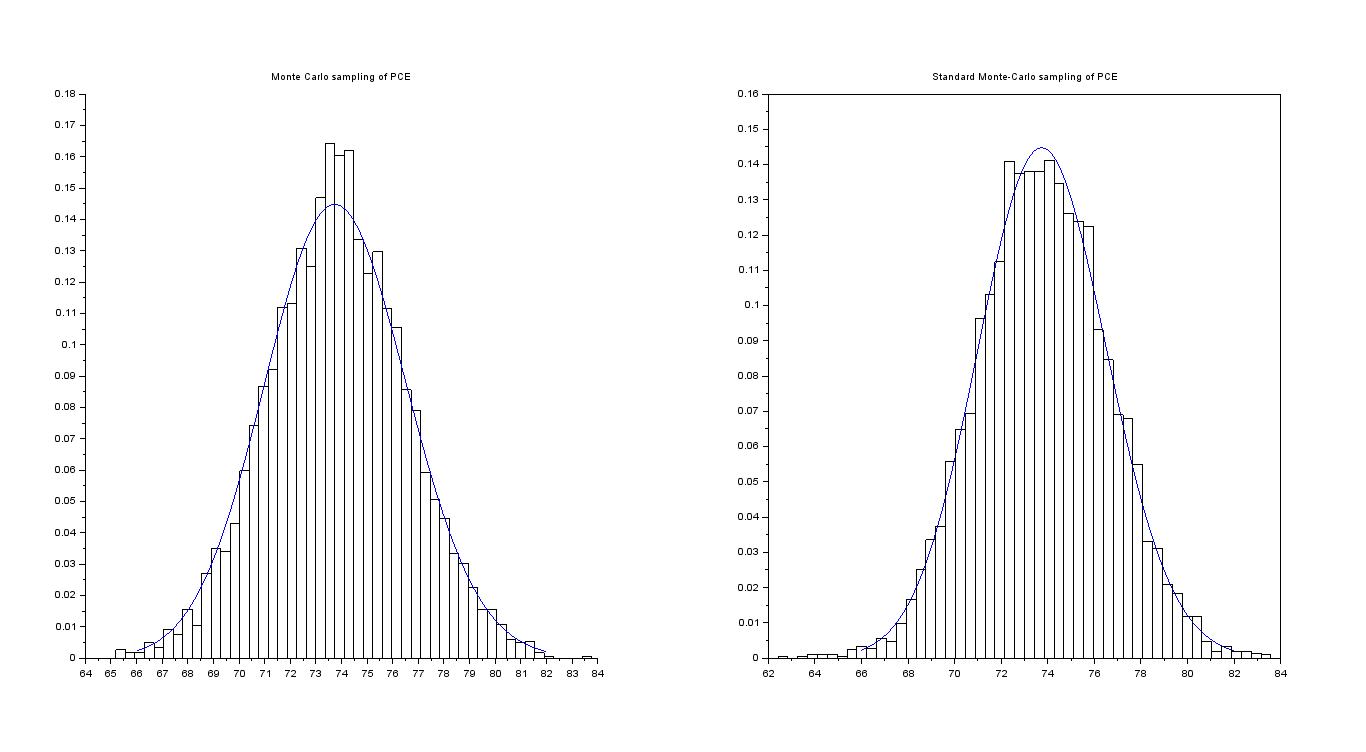}
\caption{\small Left plot: Probability density function of the CIR interest rate model, with parameters  $\alpha =   0.046875$, $\beta = 0.2$, $\sigma = 0.25$ and $R_0 =110$, (blue curve) and histogram of a Monte Carlo sampling (size = $5000$) of the $R_T^{(p)}$ for $p=15$. With the same parameters, in the right plot, we compare the analytical probability density function of the CIR model and the standard Monte Carlo sampling of size $5000$, for $R_T$, being $T=2$. }  \label{fig:fig32}
\end{figure}

Coming back to the values in Table \ref{tab:tab21}, let us compute a sampling of size $5000$ of $R^{(p)}_T$, with $p=15$, whose histogram, see fig. \ref{fig:fig32}, is compared with the probability density function of the normal random variable $R_T$.
Moreover in the same fig. \ref{fig:fig32} we show a sampling obtained exploiting the standard Monte Carlo technique.

\newpage

\begin{figure}[!h]
\centering
\includegraphics[scale=0.5]{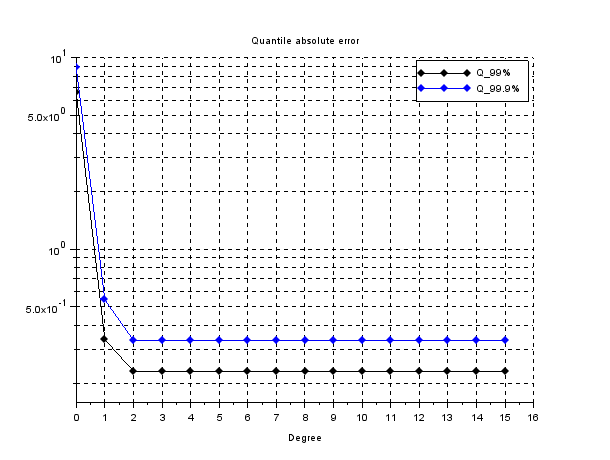}
\caption{Absolute errors of the two quantiles $\hat{Q}_{99\%}$  and $\hat{Q}_{99.9\%}$ of the PCE approximation of the CIR interest rate model at time $T=2$. The parameters are set as $\alpha =    0.046875  $, $\beta = 0.2$, $\sigma = 0.25$ and $R_0=110$.} \label{fig:fig33}
\end{figure}

The errors of quantiles for $\gamma = 99\%$, resp. for $\gamma =99.9\%$, are shown in Fig. \ref{fig:fig33} and in Table \ref{tab:tab22}. \newline

\begin{table}[!h]
\centering
\begin{tabular}{cccc}
 Degree of PCE &  $\epsilon_{99\%}$ & $\epsilon_{99.9\%}$   \tabularnewline
\hline
 0 & 6.5644e+00 & 8.8778e+00 \tabularnewline 
 1 & 3.3747e-01 & 5.4443e-01 \tabularnewline 
 2 & 2.3108e-01 & 3.3443e-01 \tabularnewline 
 3 & 2.3108e-01 & 3.3443e-01 \tabularnewline 
 4 & 2.3108e-01 & 3.3443e-01 \tabularnewline 
 5 & 2.3108e-01 & 3.3443e-01 \tabularnewline 
 6 & 2.3108e-01 & 3.3443e-01 \tabularnewline 
 7 & 2.3108e-01 & 3.3443e-01 \tabularnewline 
 8 & 2.3108e-01 & 3.3443e-01 \tabularnewline 
 9 & 2.3108e-01 & 3.3443e-01 \tabularnewline 
 10 & 2.3108e-01 & 3.3443e-01 \tabularnewline 
 11 & 2.3108e-01 & 3.3443e-01 \tabularnewline 
 12 & 2.3108e-01 & 3.3443e-01 \tabularnewline 
 13 & 2.3108e-01 & 3.3443e-01 \tabularnewline 
 14 & 2.3108e-01 & 3.3443e-01 \tabularnewline 
 15 & 2.3108e-01 & 3.3443e-01 \tabularnewline 
\end{tabular}
\caption{Absolute errors of the two quantiles $\hat{Q}_{99\%}$  and $\hat{Q}_{99.9\%}$ of the PCE approximation of the CIR Interest rate model at time $T=2$. The parameters are set as $\alpha =   0.046875$, $\beta = 0.2$, $\sigma = 0.25$ and $R_0=110$.}  \label{tab:tab22}
\end{table}

The quantiles computed by means of a standard Monte Carlo sampling of the analytical solution, whose size  equals  the one used for PCE, gives 
$ SE_{Q_{99\%}^{MC}} =0.0353593$   and      $SE_{Q_{99.9\%}^{MC}} =   0.0913327$.        


\newpage

\subsection{$\boldsymbol{\sigma = 30\%}$}

\begin{figure}[!h]
\centering
\includegraphics[scale=0.31]{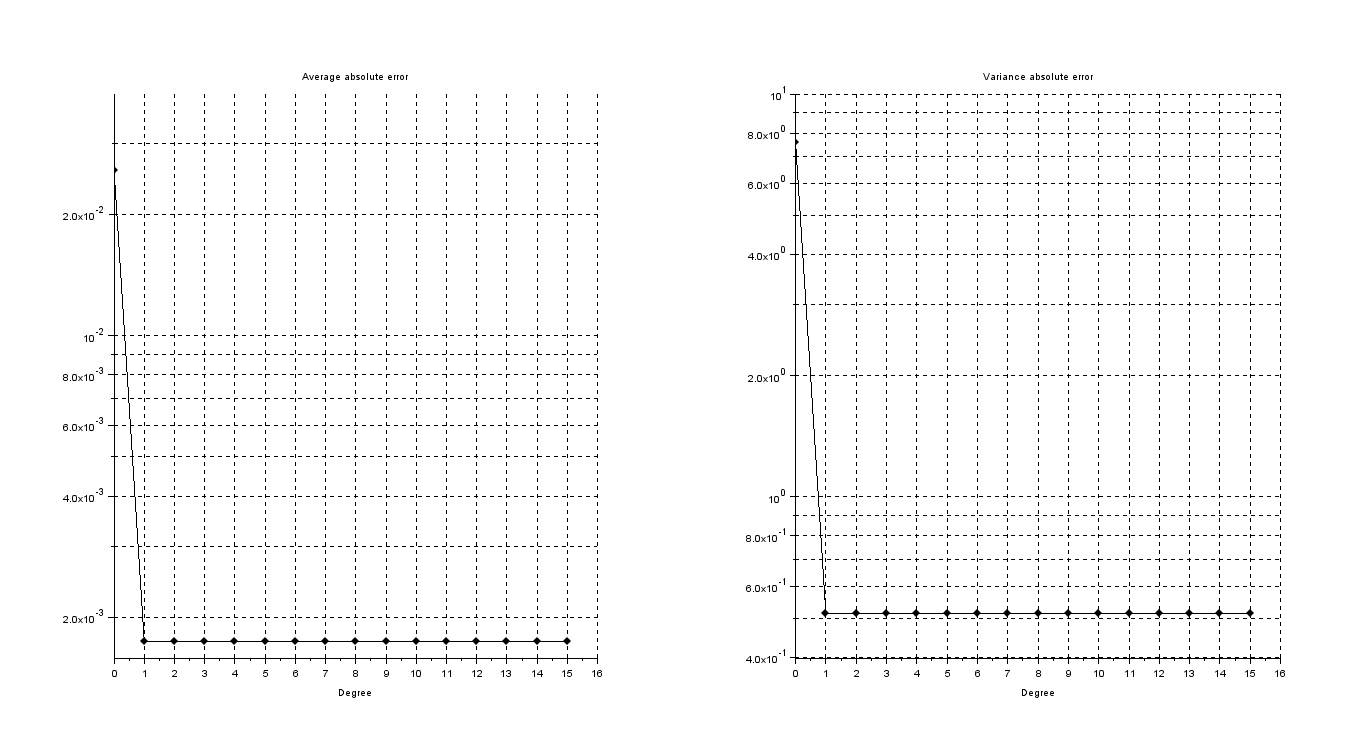}
\caption{Semilogy scale plot of the absolute error of the mean (left) and the variance (right) computed via PCE for the CIR at time $T=2$, whose parameters are $\alpha =  0.0675 $, $\beta = 0.2$, $\sigma = 30 \%$ and starting value $R_0= 110$, for a set of degrees $p = \{0,1,2,\dots,15\}$.}  \label{fig:fig34}
\end{figure}

\begin{figure}[!h]
\centering
\includegraphics[scale=0.31]{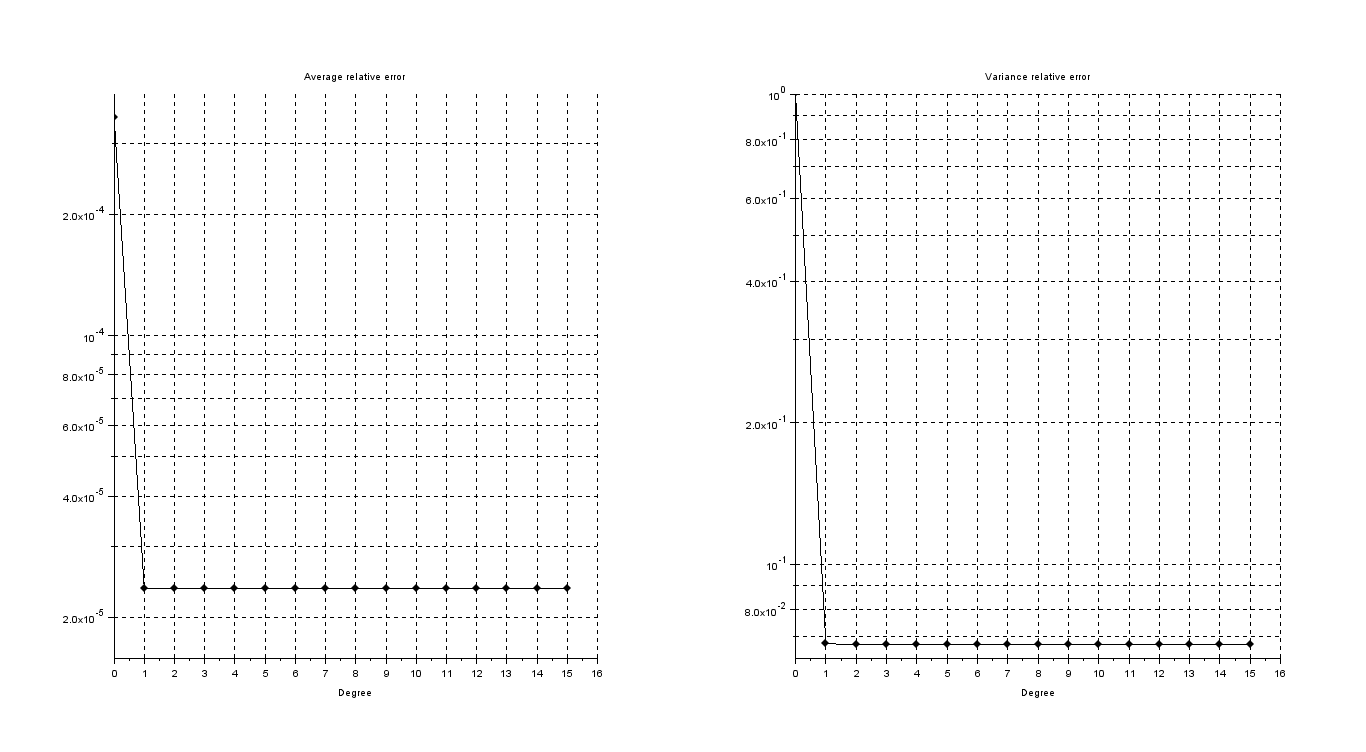}
\caption{Semilogy scale plot of the absolute value of the relative error of the mean (left) and the variance (right) computed via PCE for the CIR at time $T=2$, whose parameters are $\alpha = 0.0675 $, $\beta = 0.2$, $\sigma = 30 \%$ and starting value $R_0= 110$, for a set of degrees $p = \{0,1,2,\dots,15\}$.}  \label{fig:fig35}
\end{figure}

In this section the value of the volatility is set to $\sigma = 30\%$, consequently by means of  \eqref{sec:eq44}, where $q=3$, we get
\begin{equation*}
\alpha =0.0675\;,
\end{equation*}
while the other parameters are chosen as in Table \ref{tab:tab18}. First let us display in Figure \ref{fig:fig34}, Figure \ref{fig:fig35} and Table \ref{tab:tab23}, the absolute and relative error of the average, resp. variance, for the CIR interest rate model.

\begin{table}[!h]
\centering
\begin{tabular}{ccccc}
 Degree of PCE &  Average Error & Variance Error & Average relative error & Variance relative error   \tabularnewline
\hline
 0 & 3.7089e-02 & 1.0947e+01 & 5.0250e-04 & 1.0000e+00 \tabularnewline 
 1 & 2.5035e-03 & 7.4166e-01 & 3.3903e-05 & 6.7748e-02 \tabularnewline 
 2 & 2.5035e-03 & 7.3927e-01 & 3.3903e-05 & 6.7530e-02 \tabularnewline 
 3 & 2.5035e-03 & 7.3927e-01 & 3.3903e-05 & 6.7530e-02 \tabularnewline 
 4 & 2.5035e-03 & 7.3927e-01 & 3.3903e-05 & 6.7530e-02 \tabularnewline 
 5 & 2.5035e-03 & 7.3927e-01 & 3.3903e-05 & 6.7530e-02 \tabularnewline 
 6 & 2.5035e-03 & 7.3927e-01 & 3.3903e-05 & 6.7530e-02 \tabularnewline 
 7 & 2.5035e-03 & 7.3927e-01 & 3.3903e-05 & 6.7530e-02 \tabularnewline 
 8 & 2.5035e-03 & 7.3927e-01 & 3.3903e-05 & 6.7530e-02 \tabularnewline 
 9 & 2.5035e-03 & 7.3927e-01 & 3.3903e-05 & 6.7530e-02 \tabularnewline 
 10 & 2.5035e-03 & 7.3927e-01 & 3.3903e-05 & 6.7530e-02 \tabularnewline 
 11 & 2.5035e-03 & 7.3927e-01 & 3.3903e-05 & 6.7530e-02 \tabularnewline 
 12 & 2.5035e-03 & 7.3927e-01 & 3.3903e-05 & 6.7530e-02 \tabularnewline 
 13 & 2.5035e-03 & 7.3927e-01 & 3.3903e-05 & 6.7530e-02 \tabularnewline 
 14 & 2.5035e-03 & 7.3927e-01 & 3.3903e-05 & 6.7530e-02 \tabularnewline 
 15 & 2.5035e-03 & 7.3927e-01 & 3.3903e-05 & 6.7530e-02 \tabularnewline

\end{tabular}
\caption{Absolute error of the average and the variance of $R_T^{(p)}$ at time $T=2$, with respect to the CIR parameters $\alpha = 0.0675 $, $\beta = 0.2$, $\sigma = 30 \%$.} \label{tab:tab23}
\end{table}

The absolute and relative errors of mean and variance are stationary. Therefore there is an error that corrupts the spectral convergence of PCE-approximation. In particular the values suggest that it is not due to numerical error coming from the computations of  $\{R_{T,j}\}_{j=1}^N$, but, instead, it is due to the approximation of the solution (\ref{sec:eq34}) by means of \eqref{sec:eq40}.

\begin{figure}[!h]
\centering
\includegraphics[scale=0.31]{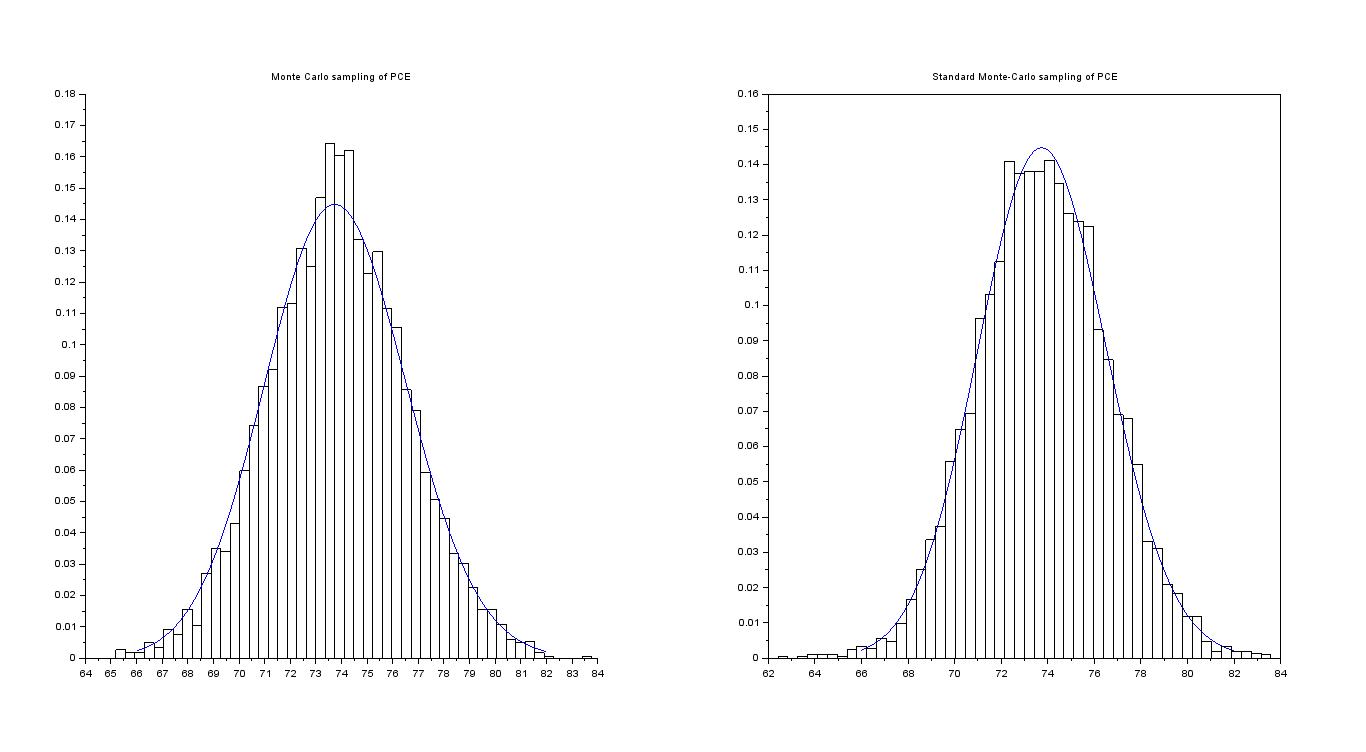}
\caption{\small Left plot: Probability density function of the CIR interest rate model, with parameters  $\alpha =  0.0675 $, $\beta = 0.2$, $\sigma = 0.30$ and $R_0 =110$, (blue curve) and histogram of a Monte Carlo sampling (size = $5000$) of the $R_T^{(p)}$ for $p=15$. With the same parameters, in the right plot we compare the analytical probability density function of the  CIR model, with the one obtained using the Monte Carlo sampling of size $5000$, for $R_T$, being $T=2$. }  \label{fig:fig36}
\end{figure}

Coming back to the values in Table \ref{tab:tab18}, let us compute a Monte Carlo sampling of size $5000$ of $R^{(p)}_T$, with $p=15$, whose histogram, see fig. \ref{fig:fig36}, is compared with the probability density function of the normal random variable $R_T$.
Moreover in the same fig. \ref{fig:fig36} we show a standard Monte Carlo sampling.

\newpage

\begin{figure}[!h]
\centering
\includegraphics[scale=0.5]{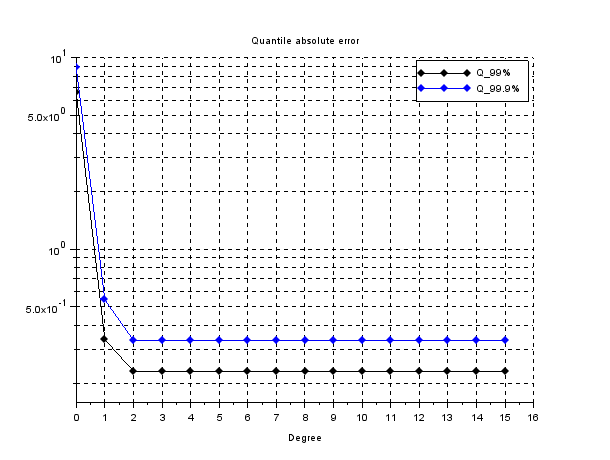}
\caption{Absolute errors of the two quantiles $\hat{Q}_{99\%}$  and $\hat{Q}_{99.9\%}$ of the PCE approximation of the CIR interest rate model at time $T=2$. The parameters are set as $\alpha =  0.0675 $, $\beta = 0.2$, $\sigma = 0.30$ and $R_0=110$.} \label{fig:fig37}
\end{figure}

The errors of quantiles for $\gamma = 99\%$, resp. for $\gamma =99.9\%$, are shown in Fig. \ref{fig:fig37} and in Table \ref{tab:tab24}. \newline

\begin{table}[!h]
\centering
\begin{tabular}{cccc}
 Degree of PCE &  $\epsilon_{99\%}$ & $\epsilon_{99.9\%}$   \tabularnewline
\hline
 0 & 7.9085e+00 & 1.0687e+01 \tabularnewline 
 1 & 4.3007e-01 & 6.8095e-01 \tabularnewline 
 2 & 2.7688e-01 & 3.7857e-01 \tabularnewline 
 3 & 2.7688e-01 & 3.7857e-01 \tabularnewline 
 4 & 2.7688e-01 & 3.7857e-01 \tabularnewline 
 5 & 2.7688e-01 & 3.7857e-01 \tabularnewline 
 6 & 2.7688e-01 & 3.7857e-01 \tabularnewline 
 7 & 2.7688e-01 & 3.7857e-01 \tabularnewline 
 8 & 2.7688e-01 & 3.7857e-01 \tabularnewline 
 9 & 2.7688e-01 & 3.7857e-01 \tabularnewline 
 10 & 2.7688e-01 & 3.7857e-01 \tabularnewline 
 11 & 2.7688e-01 & 3.7857e-01 \tabularnewline 
 12 & 2.7688e-01 & 3.7857e-01 \tabularnewline 
 13 & 2.7688e-01 & 3.7857e-01 \tabularnewline 
 14 & 2.7688e-01 & 3.7857e-01 \tabularnewline 
 15 & 2.7688e-01 & 3.7857e-01 \tabularnewline 
\end{tabular}
\caption{Absolute errors of the two quantiles $\hat{Q}_{99\%}$  and $\hat{Q}_{99.9\%}$ of the PCE approximation of the CIR Interest rate model at time $T=2$. The parameters are set as $\alpha =  0.0675 $, $\beta = 0.2$, $\sigma = 0.30$ and $R_0=110$.}  \label{tab:tab24}
\end{table}

The quantiles computed by means of a standard Monte Carlo sampling of the analytical solution, whose size is equal to the one used for PCE, give the following standard error
\begin{align*}
&SE_{Q_{99\%}^{MC}} = 0.0416182               \\ &SE_{Q_{99.9\%}^{MC}} =  0.1128376  
\end{align*}

\begin{remark}. These computations, for each dynamics considered, point out that the volatility $\sigma$ influences the accuracy of PCE computations. Indeed the errors both of statistics and quantile increase proportionally to $\sigma$, since the randomness of the Brwonian motion becomes more relevant. \newline
\end{remark}

\begin{remark}
In each case considered the lack of spectral convergence arises. Nevertheless
such a point is not connected with the model nor with the related parameters, in fact it is a general
consequence due to the approximations implied by the concrete application of the PCE-machinery used to
compute the solution at a certain positive time T.
Indeed the mean square convergence is ruled by the coefficients, see \eqref{sec:eq7}, therefore from a numerical point of view they are closed to machine precision, therefore after a certain degree they comes irrelevant. This motivates once more such behavior.
\end{remark}

\section{Polynomial Chaos Expansion compared with MC and QMC} \label{sec:Comparison}

In this section we compare the PCE method with the standard Monte Carlo (MC) method as well as with the {\it quasi-}Monte Carlo (QMC) one. In particular we consider the accuracy in approximating the statistics of each model studied, namely the gBm equity model, the Vasicek and the CIR model, evaluated at time $T$, as well as the computational time costs to detect their mean and variance, as comparison criteria. \newline


It is worth to mention that both the MC and the  QMC methods are usually applied  to approximate analytical solutions, hence, from our {\it interest models point of view} we can consider them only for the gBm and the Vasicek case, while, for the CIR model, we regard them in relation with the Euler–Maruyama scheme.

The usual Monte Carlo approximations require a set of $M$ independent realizations $\left \{X_{T,i}\right\}_{i=1}^M$ of the process $X_T$. Then the mean and variance are determined as
\begin{align}
\Expect{X_T } &\approx \mu_{MC} = \frac{1}{M} \sum_{i=1}^M X_{T,i} \label{sec:eq46}\;,\\
\Var{X_T} &\approx \sigma_{MC}^2 = \frac{1}{M-1} \sum_{i=1}^M \left(X_{T,i}-{\mu_{MC}}\right)^2 \:,\label{sec:eq47}
\end{align}
then, exploiting the {\it Law of large numbers}, we have that both (\ref{sec:eq46}) and (\ref{sec:eq47}), converges to the correspondent {\it true value}, hence their standard errors 
\begin{align*}
SE_{\mu_{MC}} &= \frac{\sigma_{MC}}{\sqrt{n}} \;,\\
SE_{\sigma^2_{MC}} &= \sigma^2_{MC} \sqrt{\frac{2}{n-1}}\;,
\end{align*}
are more informative than the single execution error.
The QMC method uses low-discrepancy sequence $\{\phi_j\}_{j=1}^M$  generated by Sobol algorithm, that maximizes the uniformity of the sample points, see, e.g.,  \cite[Chapter 5]{Glasserman}, then it uses such values to approximate the statistics we are interested in, or, generally speaking, to compute integrals of the following type
\begin{equation}
\int_{D}g(x) f(x)dx \approx \frac{1}{M}\sum_{i=0}^M g(\phi_i) \;,\label{sec:eq48}
\end{equation}
where $f(x)dx$ is the measure induced by the considered random variable which characterizes the related image space. By setting $g(x) = x$, resp.  $g(x) = (x-\mu)^2$, we can approximate the mean of the random variable $X_T$, resp. its variance. Moreover, in QMC computations, the parameters for accuracy are given by the absolute errors  
\begin{align*}
\epsilon_{\mu_{QMC}} &= \abs{\Expect{X_T} - \mu_{QMC}}\;,\\
\epsilon_{\sigma^2_{QMC}} &= \abs{\Var{X_T} - \sigma^2_{QMC}}\:.
\end{align*} 
We would like to underline that for both MC and QMC, we consider only the effective time to compute the statistics, without taking into account the amount of time spent for the  detection of the grid of samplings which is necessary to simulate the process. Moreover we highlight that in each of the case we consider, namely the gBm, resp. the Vasicek, resp. the CIR model, all the computations concerning the PCE method have been performed following the schemes proposed in sections \ref{sec:secGBM}, resp. \ref{sec:secVAS}, resp. \ref{sec:secCIR}, with related accuracy estimated by means of the absolute error of the average as well as of the variance.

\subsection{Geometric Brownian Motion}\label{Sec:GBMcomparisons}

\begin{figure}[!h]
\includegraphics[scale=0.33]{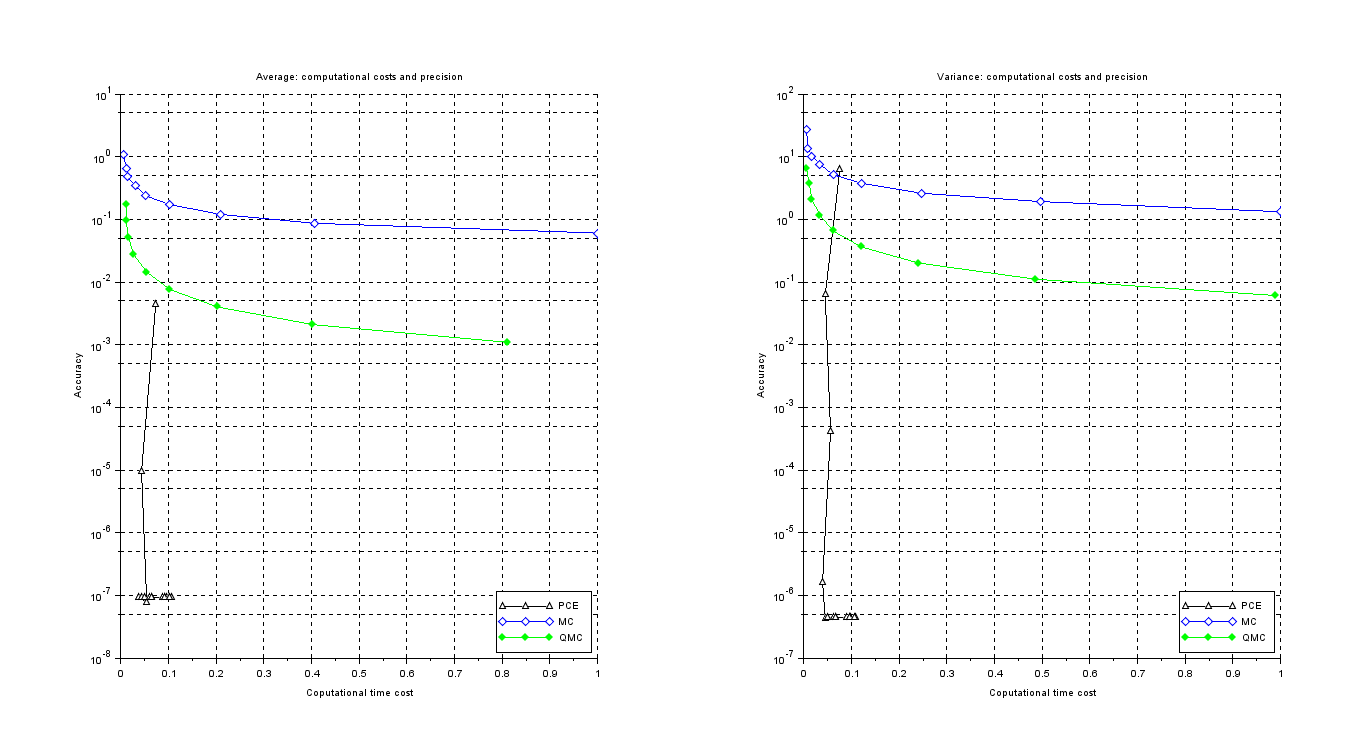}
\caption{Semilogy scale plot which compares the accuracy and computational time costs for detecting the average (left) and variance (right) of the gBm by means of PCE, MC and QMC approaches. The parameters of the gBm equity model are $r=3\%,\sigma = 15\%$, $T=1$ and $S_0=100$.}  \label{fig:fig38}
\end{figure}

\begin{figure}[!h]
\includegraphics[scale=0.33]{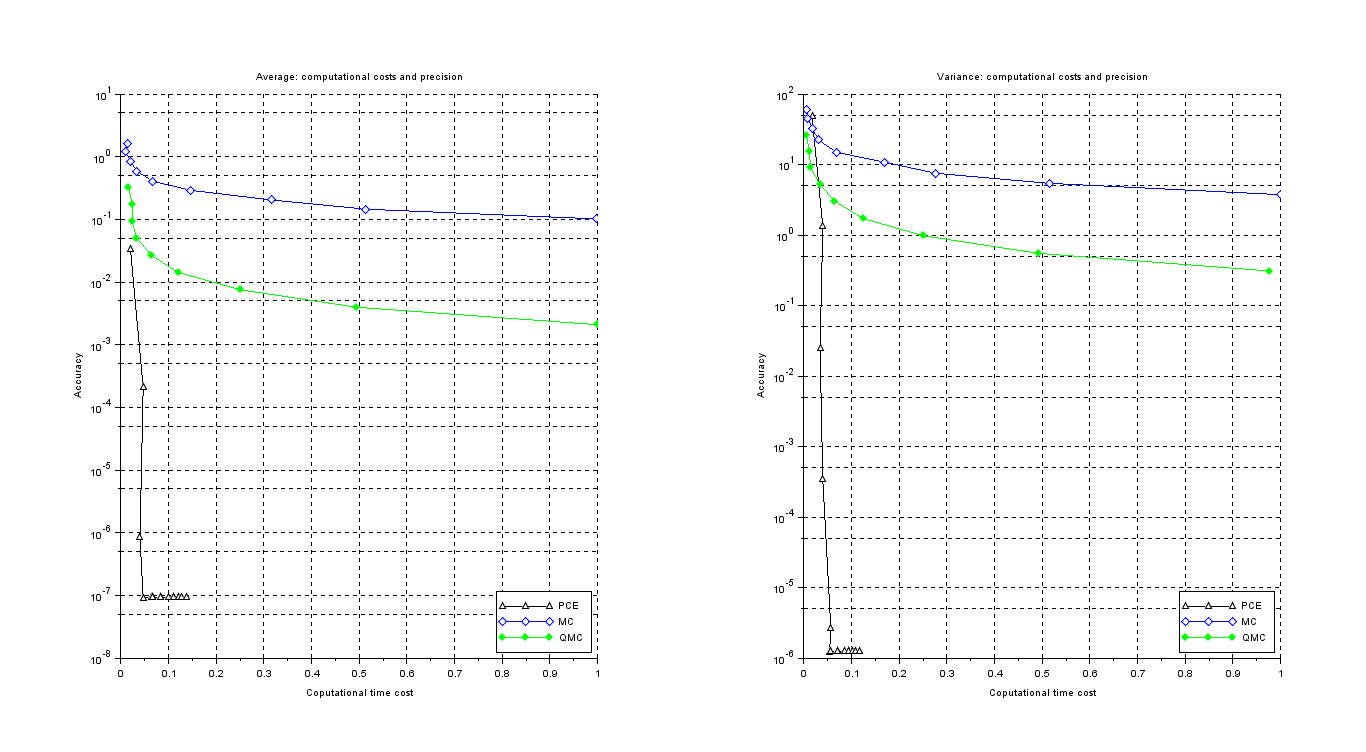}
\caption{Semilogy scale plot which compares the accuracy and computational time costs for detecting the average (left) and variance (right) of the gBm by means of PCE, MC and QMC approaches. The parameters of the gBm equity model are $r=3\%,\sigma = 25\%$, $T=1$ and $S_0=100$.}  \label{fig:fig39}
\end{figure}

\begin{figure}[!h]
\includegraphics[scale=0.33]{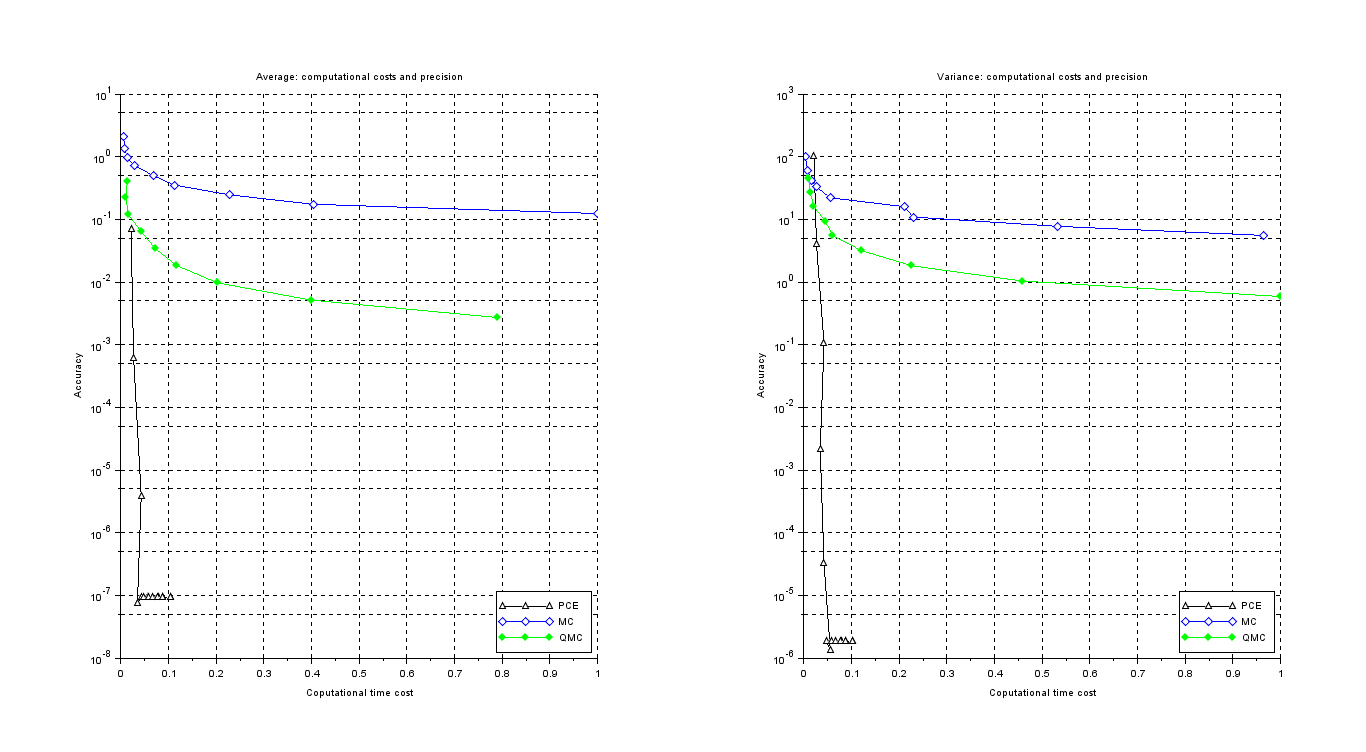}
\caption{Semilogy scale plot which compares the accuracy and computational time costs for detecting the average (left) and variance (right) of the gBm by means of PCE, MC and QMC approaches. The parameters of the gBm equity model are $r=3\%,\sigma = 30\%$, $T=1$ and $S_0=100$.}  \label{fig:fig40}
\end{figure}

Let us consider the SDE in eq. (\ref{sec:eq16}), where the parameters are taken as follows $T=1, r=3\%$ while $\sigma = \{ 15\%,25\%,30\%\}$ and $S_0=100$. The MC and QMC methods approximate the statistics of the solution to the gBm model exploiting its analytical solution, see eq. (\ref{sec:eq22}), for an increasing size $M$ of samplings
$
M=\left\{2^{8},2^{9},\dots,2^{16}\right\}\;,
$
and we then compute, for each value of $M$, the standard errors for the MC, resp. the  absolute error for the QMC, are computed. 
In order to compare the accuracy and computational time costs of the three aforementioned methods, let us consider the plots in Fig. \ref{fig:fig38}, Fig. \ref{fig:fig39} and Fig. \ref{fig:fig40}, correspoinding to $\sigma = \{15\%,25\%,30\%\}$.

Where the x-axis represents the computational time costs of the methods, while the error of the statistics, namely $SE_{\mu_{MC}}$, $SE_{\sigma^2_{MC}}$, $\epsilon_{\mu_{QMC}}$, $\epsilon_{\sigma^2_{QMC}}$, $\epsilon^{(p)}_{MEAN}$ and $\epsilon^{(p)}_{VAR}$,  are displayed along the y-axis. 
Moreover the x-axis is normalized with respect to the highest data among MC, QMC and PCE values, actually on the x-axis we report related  relative computational time costs is shown, which are more effective than the absolute ones. We also note that each point of the plot represents the computations for different number of realization points, namely $M$ for MC and QMC,  $p$ for the PCE method. \newline

\newpage
Fig. \ref{fig:fig38}, Fig. \ref{fig:fig39} and Fig. \ref{fig:fig40} point out the high accuracy as well as the very low computational effort required  by PCE  to get the solution. 

Let us display the values of computational time cost for the PCE, see Table \ref{tab:tab25}, resp. for the MC and the QMC methods, see Table \ref{tab:tab26},  as well as the error made using both the MC and the QMC methods, see Table \ref{tab:tab27}. \newline

\begin{table}[!h]
\centering
\begin{tabular}{cccc}
\centering Degree of PCE $p$ & \centering Time costs $\sigma = 15\%$ & \centering Time costs $\sigma = 25\%$  & \centering Time costs $\sigma = 30\%$  \tabularnewline
\hline
1 &     0.0000 &0.0000 &0.0000 \tabularnewline  
2 &     0.0000 &0.0160 &0.0000 \tabularnewline 
3 &     0.0000 &0.0000 &0.0000 \tabularnewline 
4 &     0.0160 &0.0000 &0.0160 \tabularnewline 
5 &     0.0150 &0.0150 &0.0160 \tabularnewline 
6 &     0.0150 &0.0160 &0.0320 \tabularnewline 
7 &     0.0320 &0.0150 &0.0310 \tabularnewline
8 &     0.0310 &0.0150 &0.0310 \tabularnewline 
9 &     0.0310 &0.0160 &0.0470 \tabularnewline 
10 &    0.0150 &0.0310 &0.0320 \tabularnewline 
11 &    0.0310 &0.0320 &0.0470 \tabularnewline 
12 &    0.0310 &0.0310 &0.0310 \tabularnewline 
13 &    0.0310 &0.0310 &0.0320 \tabularnewline  
14 &    0.0460 &0.0310 &0.0310 \tabularnewline 
15 &    0.0470 &0.0460 &0.0470 \tabularnewline 
\end{tabular}
\caption{Time cost of computations for PCE, approximating gBm at $T=1$ for $r=3\%$, \newline  $\sigma = \{15\%,25\%,30\%\}$ and $S_0 = 100$.} \label{tab:tab25}
\end{table}

The computational time costs required for the detection of PCE's coefficients, see eq. (\ref{sec:eq10}), and its statistics,  see equations (\ref{sec:eq8}) and (\ref{sec:eq9}), are irrelevant if compared with time required to evaluate the process, eq. (\ref{sec:eq21}), at $\{\xi_j\}_{j=1}^N$, which are defined by Gaussian quadrature formula as stated in (\ref{sec:eq10}),
this motivates the data in Table \ref{tab:tab25}. 

\begin{table}[!h]
\centering
\begin{tabular}{c|cc|cc|cc|cc|cc|cc}
$\ $ &\multicolumn{4}{c}{$\sigma = 15\%$}  & \multicolumn{4}{|c}{$\sigma = 25\%$} & \multicolumn{4}{|c}{$\sigma = 30\%$}  \tabularnewline
\hline
$\ $ &\multicolumn{2}{c}{Average}  & \multicolumn{2}{|c|}{Variance} &\multicolumn{2}{c}{Average}  & \multicolumn{2}{|c|}{Variance}  &\multicolumn{2}{c}{Average}  & \multicolumn{2}{|c}{Variance}  \tabularnewline
\hline
\centering $M$ & \centering MC & \centering QMC  &  MC &  QMC &  MC &  QMC &  MC &  QMC &  MC &  QMC &  MC &  QMC \tabularnewline
\hline
        256 &     0.0000 &      0.0000 &     0.0000 &     0.0000 &     0.0000 &      0.0000 &     0.0000 &     0.0150 &     0.0000 &      0.0000 &     0.0160 &     0.0160 \\
        512 &     0.0150 &      0.0000 &     0.0000 &     0.0150 &     0.0000 &      0.0000 &     0.0000 &     0.0000 &     0.0000 &      0.0000 &     0.0000 &     0.0150 \\
       1024 &     0.0000 &      0.0000 &     0.0160 &     0.0000 &     0.0000 &      0.0160 &     0.0150 &     0.0000 &     0.0160 &      0.0000 &     0.0000 &     0.0160 \\
       2048 &     0.0150 &      0.0000 &     0.0000 &     0.0150 &     0.0000 &      0.0160 &     0.0160 &     0.0150 &     0.0150 &      0.0000 &     0.0160 &     0.0160 \\
       4096 &     0.0160 &      0.0160 &     0.0150 &     0.0310 &     0.0310 &      0.0160 &     0.0160 &     0.0310 &     0.0150 &      0.0310 &     0.0160 &     0.0150 \\
       8192 &     0.0470 &      0.0310 &     0.0470 &     0.0470 &     0.0310 &      0.0310 &     0.0320 &     0.0470 &     0.0310 &      0.0470 &     0.0310 &     0.0310 \\
      16384 &     0.0620 &      0.0780 &     0.0780 &     0.0780 &     0.0780 &      0.0620 &     0.0780 &     0.0940 &     0.0620 &      0.0780 &     0.0780 &     0.0780 \\
      32768 &     0.1400 &      0.1250 &     0.1720 &     0.1710 &     0.1400 &      0.1400 &     0.1720 &     0.1560 &     0.1240 &      0.1250 &     0.1720 &     0.1560 \\
      65536 &     0.2810 &      0.2650 &     0.3270 &     0.3280 &     0.2800 &      0.2650 &     0.3280 &     0.3740 &     0.2960 &      0.2810 &     0.3120 &     0.3270 \\

\end{tabular}
\caption{Computational time costs achieved by MC and QMC methods to detect the mean and variance of the process $S_T$, at $T=1$, with parameters  $r=3\%,\sigma = \{15\%,25\%,30\%\}$ and $S_0=100$.} \label{tab:tab26}
\end{table}

\begin{table}[!h]
\centering
\begin{tabular}{c|c|cc|cc}
$\ $ &$\ $ &\multicolumn{2}{c}{Average}  & \multicolumn{2}{|c}{Variance}  \tabularnewline
\hline
$\ $&\centering Size of sampling  $M$ & \centering $SE_{\mu_{MC}}$   & \centering  $\epsilon_{\mu_{QMC}}$  & \centering $SE_{\sigma^2_{MC}}$ & \centering  $\epsilon_{\sigma^2_{QMC}}$ \tabularnewline
\hline
\multirow{8}{*}{$\sigma = 15 \%$} &
      256 &8.7186e-01 & 1.7295e-01 & 1.7234e+01 & 6.5079e+00 \tabularnewline 
 $\ $ &      512 &6.9261e-01 & 9.4193e-02 & 1.5365e+01 & 3.6819e+00 \tabularnewline 
  $\ $ &    1024 &4.9055e-01 & 5.0744e-02 & 1.0896e+01 & 2.0760e+00 \tabularnewline 
  $\ $ &    2048 &3.4608e-01 & 2.7110e-02 & 7.6670e+00 & 1.1656e+00 \tabularnewline 
$\ $ &      4096 &2.4726e-01 & 1.4388e-02 & 5.5344e+00 & 6.5155e-01 \tabularnewline 
$\ $ &      8192 &1.6965e-01 & 7.5954e-03 & 3.6842e+00 & 3.6254e-01 \tabularnewline 
$\ $ &     16384 &1.2111e-01 & 3.9921e-03 & 2.6552e+00 & 2.0082e-01 \tabularnewline 
$\ $ &     32768 &8.5602e-02 & 2.0905e-03 & 1.8759e+00 & 1.1076e-01 \tabularnewline 
$\ $ &     65536 &6.1000e-02 & 1.0913e-03 & 1.3472e+00 & 6.0845e-02 \tabularnewline 
\hline
\multirow{8}{*}{$\sigma = 25 \%$} &
       256 &1.4618e+00 & 3.1675e-01 & 4.8443e+01 & 2.5888e+01 \tabularnewline 
$\ $ &       512 &1.1236e+00 & 1.7307e-01 & 4.0438e+01 & 1.5299e+01 \tabularnewline 
$\ $ &      1024 &8.1200e-01 & 9.3620e-02 & 2.9853e+01 & 8.9624e+00 \tabularnewline 
$\ $ &      2048 &5.7963e-01 & 5.0254e-02 & 2.1507e+01 & 5.2056e+00 \tabularnewline 
$\ $ &      4096 &4.1155e-01 & 2.6811e-02 & 1.5332e+01 & 2.9990e+00 \tabularnewline 
$\ $ &      8192 &2.8569e-01 & 1.4233e-02 & 1.0448e+01 & 1.7149e+00 \tabularnewline 
$\ $ &     16384 &2.0385e-01 & 7.5252e-03 & 7.5225e+00 & 9.7390e-01 \tabularnewline 
$\ $ &     32768 &1.4416e-01 & 3.9648e-03 & 5.3204e+00 & 5.4964e-01 \tabularnewline 
$\ $ &     65536 &1.0226e-01 & 2.0828e-03 & 3.7858e+00 & 3.0844e-01 \tabularnewline 
\hline
\multirow{8}{*}{$\sigma = 30 \%$} &
       256 &1.9748e+00 & 3.9966e-01 & 8.8419e+01 & 4.4469e+01 \tabularnewline 
$\ $ &       512 &1.4752e+00 & 2.1905e-01 & 6.9708e+01 & 2.6743e+01 \tabularnewline 
$\ $ &      1024 &9.5791e-01 & 1.1890e-01 & 4.1545e+01 & 1.5910e+01 \tabularnewline 
$\ $ &      2048 &6.8569e-01 & 6.4066e-02 & 3.0099e+01 & 9.3693e+00 \tabularnewline 
$\ $ &      4096 &4.9136e-01 & 3.4316e-02 & 2.1855e+01 & 5.4664e+00 \tabularnewline 
$\ $ &      8192 &3.4695e-01 & 1.8292e-02 & 1.5409e+01 & 3.1625e+00 \tabularnewline 
$\ $ &     16384 &2.4803e-01 & 9.7119e-03 & 1.1136e+01 & 1.8158e+00 \tabularnewline 
$\ $ &     32768 &1.7542e-01 & 5.1389e-03 & 7.8780e+00 & 1.0354e+00 \tabularnewline 
$\ $ &     65536 &1.2311e-01 & 2.7113e-03 & 5.4873e+00 & 5.8685e-01 \tabularnewline 
\end{tabular}
\caption{Errors of average and variance computed by MC and QMC methods for the gBm evaluated at time $T=1$, with parameters  $r=3\%,\sigma = \{15\%,25\%,30\%\}$ and $S_0=100$.} \label{tab:tab27}
\end{table}

\newpage

\subsection{Vasicek interest rate model}\label{Sec:Vasicekapproximation}
Proceeding as in Sec. \ref{Sec:GBMcomparisons}, we consider the SDE (\ref{sec:eq23})  characterizing the Vasicek model for the following set of values for its parameters $\alpha=0.1, \beta = 2 \cdot 10^{-5}$, $\sigma = \{15\%,25\%,30\%\}$, and we apply both the
MC and QMC method to approximate the analytical solution (\ref{sec:eq24}) to eq. (\ref{sec:eq23}), and to compute  the statistics of $R_T$ at time  $T=1$ for an increasing size of samplings $
M=\left\{2^{8},\dots,2^{16}\right\}\;.
$
We use the same type of plot exploited in the previous section, in order to  show the accuracy, the computational time costs and the errors of the PCE, resp. the MC , resp. the QMC, method.

\begin{figure}[!h]
\includegraphics[scale=0.33]{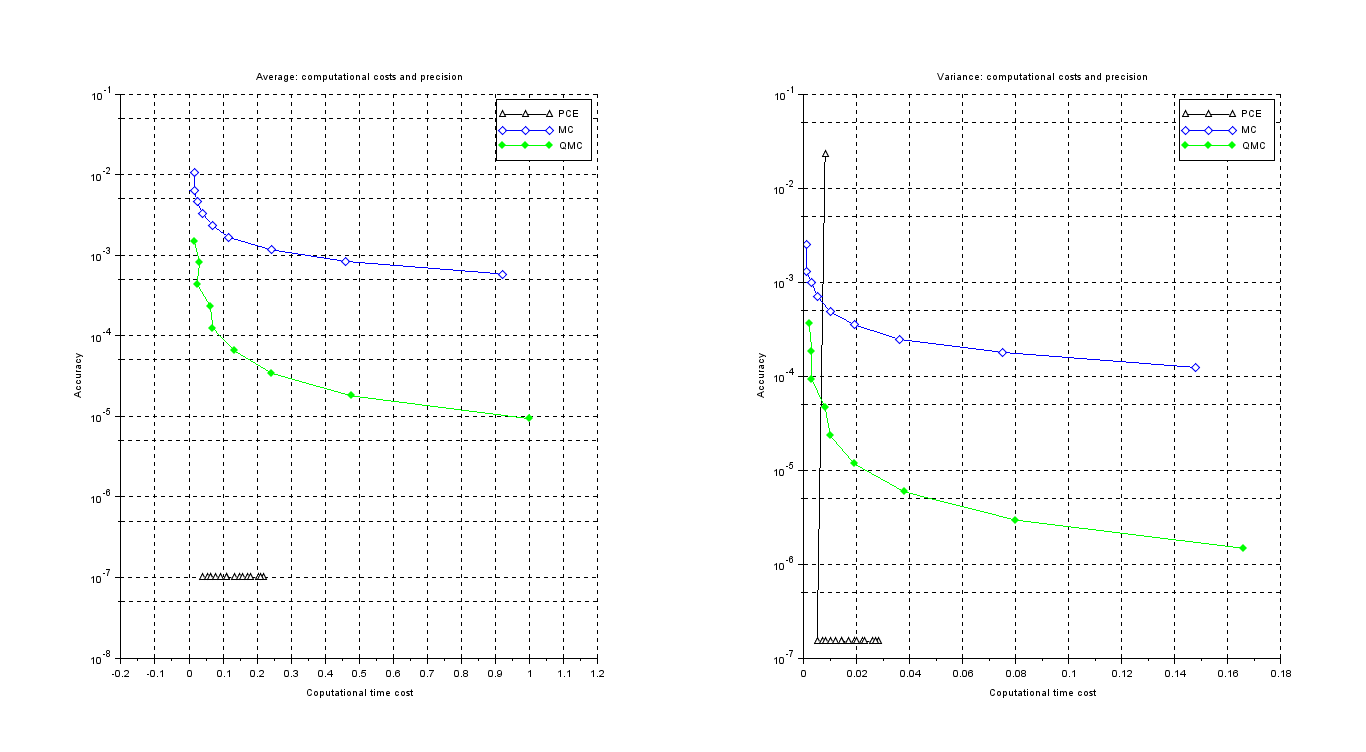}
\caption{Semilogy scale plot which compares the accuracy and computational time costs for detecting the average (left) and variance (right) of the Vasicek interest rate model  by means of PCE, MC and QMC approaches. The parameters are $\alpha=0.1, \beta = 2\cdot 10^{-5},\sigma = 15\%$, $T=1$ and $R_0=110$.}  \label{fig:fig41}
\end{figure}

\begin{figure}[!h]
\includegraphics[scale=0.33]{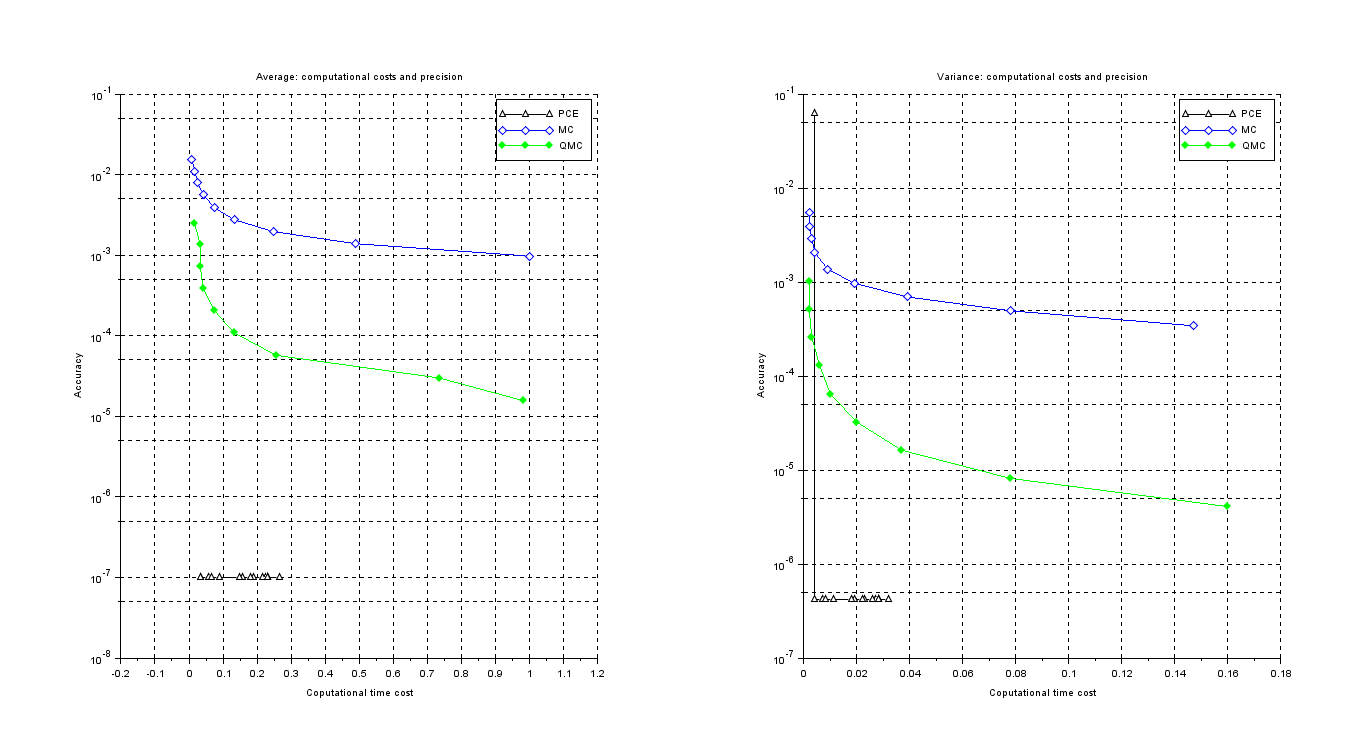}
\caption{Semilogy scale plot which compares the accuracy and computational time costs for detecting the average (left) and variance (right) of the Vasicek interest rate model by means of PCE, MC and QMC approaches. The parameters are $\alpha=0.1, \beta = 2\cdot 10^{-5}, \sigma = 25\%$, $T=1$ and $R_0=110$.}  \label{fig:fig42}
\end{figure}

\begin{figure}[!h]
\includegraphics[scale=0.33]{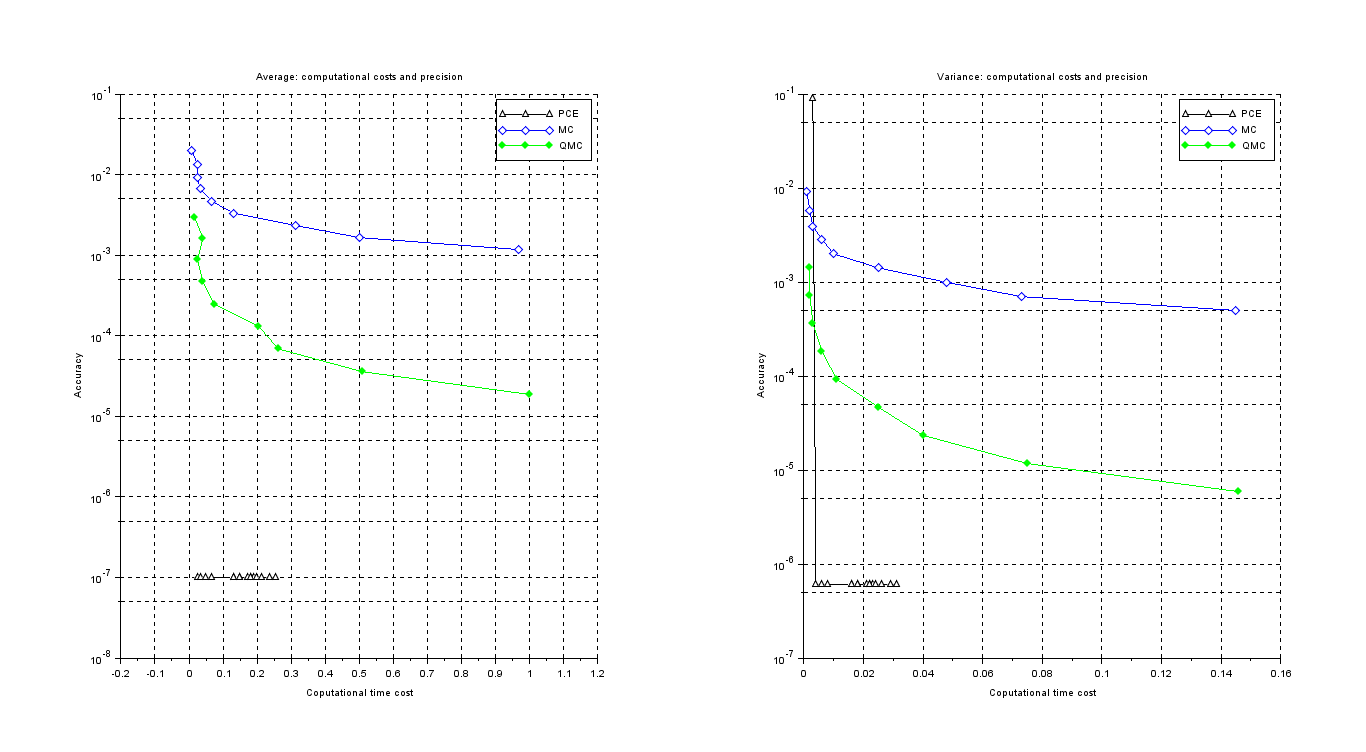}
\caption{Semilogy scale plot which compares the accuracy and computational time costs for detecting the average (left) and variance (right) of the Vasicek interest rate model  by means of PCE, MC and QMC approaches. The parameters are $\alpha=0.1, \beta = 2\cdot 10^{-5},\sigma = 30\%$, $T=1$ and $R_0=110$.}  \label{fig:fig43}
\end{figure}

In particular Fig. \ref{fig:fig41}, Fig. \ref{fig:fig42} and Fig. \ref{fig:fig43} point out the high accuracy of PCE and the corresponding low computational effort to get these results: the key points are the low number of simulation required to get the solution of  eq. (\ref{sec:eq23}). \newline

\newpage

In order to underline the efficiency and accuracy of the PCE method, let us show  the related computational time costs, see Table \ref{tab:tab28}, compared to those of the MC and QMC methods, see Table \ref{tab:tab30}, also we highlight the numerical values concerning the absolute errors achieved using PCE method, see Table \ref{tab:tab29}, as well as MC and QMC approaches, see Table \ref{tab:tab31}.

As observed in the gBm-case the computational time costs required for the detection of PCE's coefficients, equation (\ref{sec:eq10}), and its statistics,  equations (\ref{sec:eq8}) and (\ref{sec:eq9}), are irrelevant if compared with time required to get the evaluation of the process in eq. (\ref{sec:eq28}) at quadrature nodes $\{\xi_j\}_{j=1}^N$. 
\begin{table}[!h]
\centering
\begin{tabular}{cccc}
\centering Degree of PCE $p$ & \centering Time costs $\sigma = 15\%$ & \centering Time costs $\sigma = 25\%$  & \centering Time costs $\sigma = 30\%$  \tabularnewline
\hline
 0 &     0.0030 &     0.0030 &     0.0030 \\
 1 &     0.0050 &     0.0040 &     0.0050 \\
 2 &     0.0080 &     0.0060 &     0.0080 \\
 3 &     0.0100 &     0.0080 &     0.0100 \\
 4 &     0.0140 &     0.0100 &     0.0140 \\
 5 &     0.0140 &     0.0110 &     0.0140 \\
 6 &     0.0160 &     0.0150 &     0.0160 \\
 7 &     0.0180 &     0.0150 &     0.0180 \\
 8 &     0.0200 &     0.0160 &     0.0200 \\
 9 &     0.0240 &     0.0180 &     0.0240 \\
 10 &     0.0270 &     0.0190 &     0.0270 \\
 11 &     0.0220 &     0.0210 &     0.0220 \\
 12 &     0.0220 &     0.0240 &     0.0220 \\
 13 &     0.0260 &     0.0260 &     0.0260 \\
 14 &     0.0260 &     0.0270 &     0.0260 \\
 15 &     0.0310 &     0.0310 &     0.0310 \\

\end{tabular}
\caption{Time cost of computations for PCE, approximating the Vasicek interest rate model at $T=1$ for $\alpha=0.1, \beta = 2\cdot 10^{-5},\sigma = \{15\%,25\%,30\%\}$ and $R_0 = 110$.} \label{tab:tab28}
\end{table}

\begin{table}[!h]
\centering
\begin{tabular}{c|cc|cc|cc}
$\ $ &\multicolumn{2}{c|}{$\sigma =15\%$}  & \multicolumn{2}{c|}{$\sigma =25\%$}  &\multicolumn{2}{c}{$\sigma =30\%$}  \\
\hline
\centering Degree of PCE $p$ & $\epsilon_{MEAN}^{(p)}$& $\epsilon_{VAR}^{(p)}$  & $\epsilon_{MEAN}^{(p)}$ & $\epsilon_{VAR}^{(p)}$& $\epsilon_{MEAN}^{(p)}$ & $\epsilon_{VAR}^{(p)}$\\
\hline
 0 & 1.0006e-07 & 2.2500e-02 & 1.0006e-07 & 6.2499e-02 & 1.0006e-07 & 8.9998e-02 \\
 1 & 1.0006e-07 & 1.4999e-07 & 1.0006e-07 & 4.1664e-07 & 1.0006e-07 & 5.9996e-07 \\
 2 & 1.0006e-07 & 1.4999e-07 & 1.0006e-07 & 4.1664e-07 & 1.0006e-07 & 5.9997e-07 \\
 3 & 1.0006e-07 & 1.4999e-07 & 1.0006e-07 & 4.1664e-07 & 1.0006e-07 & 5.9996e-07 \\
 4 & 1.0006e-07 & 1.4999e-07 & 1.0006e-07 & 4.1664e-07 & 1.0006e-07 & 5.9997e-07 \\
 5 & 1.0006e-07 & 1.4999e-07 & 1.0006e-07 & 4.1664e-07 & 1.0006e-07 & 5.9996e-07 \\
 6 & 1.0006e-07 & 1.4999e-07 & 1.0006e-07 & 4.1664e-07 & 1.0006e-07 & 5.9997e-07 \\
 7 & 1.0006e-07 & 1.4999e-07 & 1.0006e-07 & 4.1664e-07 & 1.0006e-07 & 5.9996e-07 \\
 8 & 1.0006e-07 & 1.4999e-07 & 1.0006e-07 & 4.1664e-07 & 1.0006e-07 & 5.9997e-07 \\
 9 & 1.0006e-07 & 1.4999e-07 & 1.0006e-07 & 4.1664e-07 & 1.0006e-07 & 5.9996e-07 \\
 10 & 1.0006e-07 & 1.4999e-07 & 1.0006e-07 & 4.1664e-07 & 1.0006e-07 & 5.9997e-07 \\
 11 & 1.0006e-07 & 1.4999e-07 & 1.0006e-07 & 4.1664e-07 & 1.0006e-07 & 5.9996e-07 \\
 12 & 1.0006e-07 & 1.4999e-07 & 1.0006e-07 & 4.1664e-07 & 1.0006e-07 & 5.9997e-07 \\
 13 & 1.0006e-07 & 1.4999e-07 & 1.0006e-07 & 4.1664e-07 & 1.0006e-07 & 5.9997e-07 \\
 14 & 1.0006e-07 & 1.4999e-07 & 1.0006e-07 & 4.1664e-07 & 1.0006e-07 & 5.9997e-07 \\
 15 & 1.0006e-07 & 1.4999e-07 & 1.0006e-07 & 4.1664e-07 & 1.0006e-07 & 5.9997e-07 \\
\end{tabular}
\caption{Absolut errors of PCE-approximation of the  Vasicek interest rate model at $T=1$ for $\alpha=0.1$, \newline$ \beta = 2\cdot 10^{-5},\sigma = \{15\%,25\%,30\%\}$ and $R_0 = 110$.} \label{tab:tab29}
\end{table}

\begin{table}[!h]
\centering
\begin{tabular}{c|cc|cc|cc|cc|cc|cc}
$\ $ &\multicolumn{4}{c}{$\sigma = 15\%$}  & \multicolumn{4}{|c}{$\sigma = 25\%$} & \multicolumn{4}{|c}{$\sigma = 30\%$}  \tabularnewline
\hline
$\ $ &\multicolumn{2}{c}{Average}  & \multicolumn{2}{|c|}{Variance} &\multicolumn{2}{c}{Average}  & \multicolumn{2}{|c|}{Variance}  &\multicolumn{2}{c}{Average}  & \multicolumn{2}{|c}{Variance}  \tabularnewline
\hline
\centering $M$ & \centering MC & \centering QMC  &  MC &  QMC &  MC &  QMC &  MC &  QMC &  MC &  QMC &  MC &  QMC \tabularnewline
\hline

        256 &     0.0020 &      0.0020 &     0.0020 &     0.0020 &     0.0020 &      0.0020 &     0.0020 &     0.0030 &     0.0020 &      0.0020 &     0.0010 &     0.0020 \\
        512 &     0.0020 &      0.0030 &     0.0010 &     0.0020 &     0.0020 &      0.0020 &     0.0010 &     0.0030 &     0.0020 &      0.0030 &     0.0030 &     0.0020 \\
       1024 &     0.0030 &      0.0030 &     0.0020 &     0.0030 &     0.0040 &      0.0050 &     0.0030 &     0.0030 &     0.0030 &      0.0030 &     0.0030 &     0.0040 \\
       2048 &     0.0050 &      0.0060 &     0.0050 &     0.0060 &     0.0070 &      0.0050 &     0.0050 &     0.0060 &     0.0040 &      0.0050 &     0.0050 &     0.0050 \\
       4096 &     0.0080 &      0.0110 &     0.0100 &     0.0130 &     0.0110 &      0.0100 &     0.0110 &     0.0100 &     0.0080 &      0.0080 &     0.0100 &     0.0090 \\
       8192 &     0.0160 &      0.0170 &     0.0210 &     0.0210 &     0.0170 &      0.0170 &     0.0200 &     0.0230 &     0.0180 &      0.0170 &     0.0220 &     0.0220 \\
      16384 &     0.0290 &      0.0350 &     0.0350 &     0.0400 &     0.0340 &      0.0340 &     0.0380 &     0.0430 &     0.0330 &      0.0310 &     0.0400 &     0.0390 \\
      32768 &     0.0580 &      0.0640 &     0.0740 &     0.0820 &     0.0640 &      0.0700 &     0.0790 &     0.0890 &     0.0630 &      0.0660 &     0.0770 &     0.0820 \\
      65536 &     0.1220 &      0.1170 &     0.1570 &     0.1430 &     0.1260 &      0.1300 &     0.1560 &     0.1580 &     0.1170 &      0.1380 &     0.1410 &     0.1560 \\

\end{tabular}
\caption{Computational time costs achieved by MC and QMC methods to detect the mean and variance of the process $R_T$,at $T=1$ for $\alpha=0.1, \beta = 2\cdot 10^{-5},\sigma = \{15\%,25\%,30\%\}$ and $R_0 = 110$.} \label{tab:tab30}
\end{table}

\begin{table}[!h]
\centering
\begin{tabular}{c|c|cc|cc}
$\ $ &$\ $ &\multicolumn{2}{c}{Average}  & \multicolumn{2}{|c}{Variance}  \tabularnewline
\hline
$\ $&\centering Size of sampling  $M$ & \centering $SE_{\mu_{MC}}$   & \centering  $\epsilon_{\mu_{QMC}}$  & \centering $SE_{\sigma^2_{MC}}$ & \centering  $\epsilon_{\sigma^2_{QMC}}$ \tabularnewline
\hline
\multirow{8}{*}{$\sigma = 15 \%$} &
        256 & 9.6455e-03 & 1.4768e-03 & 2.1093e-03 & 3.6049e-04 \\ 
$\ $ &        512 & 6.9057e-03 & 8.0728e-04 & 1.5275e-03 & 1.8175e-04 \\ 
$\ $ &       1024 & 4.6269e-03 & 4.3579e-04 & 9.6929e-04 & 9.1525e-05 \\ 
$\ $ &       2048 & 3.3442e-03 & 2.3302e-04 & 7.1593e-04 & 4.6045e-05 \\ 
$\ $ &       4096 & 2.3732e-03 & 1.2367e-04 & 5.0984e-04 & 2.3146e-05 \\ 
$\ $ &       8192 & 1.6639e-03 & 6.5245e-05 & 3.5439e-04 & 1.1627e-05 \\ 
$\ $ &      16384 & 1.1707e-03 & 3.4252e-05 & 2.4810e-04 & 5.8373e-06 \\ 
$\ $ &      32768 & 8.2966e-04 & 1.7907e-05 & 1.7622e-04 & 2.9291e-06 \\ 
$\ $ &      65536 & 5.8410e-04 & 9.3297e-06 & 1.2352e-04 & 1.4692e-06 \\
\hline 
\multirow{8}{*}{$\sigma = 25 \%$} &
     256 & 1.5392e-02 & 2.4614e-03 & 5.3712e-03 & 1.0014e-03 \\ 
$\ $ &        512 & 1.1543e-02 & 1.3455e-03 & 4.2681e-03 & 5.0487e-04 \\ 
$\ $ &       1024 & 7.9822e-03 & 7.2632e-04 & 2.8848e-03 & 2.5424e-04 \\ 
$\ $ &       2048 & 5.5877e-03 & 3.8837e-04 & 1.9987e-03 & 1.2790e-04 \\ 
$\ $ &       4096 & 3.9699e-03 & 2.0612e-04 & 1.4266e-03 & 6.4294e-05 \\ 
$\ $ &       8192 & 2.7871e-03 & 1.0874e-04 & 9.9438e-04 & 3.2297e-05 \\ 
$\ $ &      16384 & 1.9696e-03 & 5.7086e-05 & 7.0224e-04 & 1.6215e-05 \\ 
$\ $ &      32768 & 1.3777e-03 & 2.9845e-05 & 4.8592e-04 & 8.1365e-06 \\ 
$\ $ &      65536 & 9.7726e-04 & 1.5549e-05 & 3.4576e-04 & 4.0811e-06 \\ 
\hline
\multirow{8}{*}{$\sigma = 30 \%$} &
        256 & 1.8935e-02 & 2.9537e-03 & 8.1284e-03 & 1.4420e-03 \\ 
$\ $ &        512 & 1.3966e-02 & 1.6146e-03 & 6.2476e-03 & 7.2701e-04 \\ 
$\ $ &       1024 & 9.3332e-03 & 8.7158e-04 & 3.9441e-03 & 3.6610e-04 \\ 
$\ $ &       2048 & 6.5312e-03 & 4.6604e-04 & 2.7307e-03 & 1.8418e-04 \\ 
$\ $ &       4096 & 4.6692e-03 & 2.4735e-04 & 1.9735e-03 & 9.2584e-05 \\ 
$\ $ &       8192 & 3.3352e-03 & 1.3049e-04 & 1.4239e-03 & 4.6508e-05 \\ 
$\ $ &      16384 & 2.3304e-03 & 6.8504e-05 & 9.8311e-04 & 2.3349e-05 \\ 
$\ $ &      32768 & 1.6527e-03 & 3.5815e-05 & 6.9927e-04 & 1.1717e-05 \\ 
$\ $ &      65536 & 1.1749e-03 & 1.8659e-05 & 4.9977e-04 & 5.8768e-06 \\ 

\end{tabular}
\caption{Errors of the average and variance computed by MC and QMC for the Vasicek interest rate model $R_T$ at time $T=1$, with  parameters $\alpha=0.1$, $\beta = 2 \cdot 10^{-5}$, $,\sigma = \{15\%,25\%,30\%\}$ and and $R_0=110$.} \label{tab:tab31}
\end{table}

\clearpage

\subsection{CIR interest rate model}
Even if the CIR model does not have a solution in closed form, we can still apply both the MC and the QMC methods, but considering  the numerical solution of the SDE (\ref{sec:eq34}).

As described in \cite{Iacus}, the Euler-Maruyama scheme applied to the transformed process SDE (\ref{sec:eq36}) is actually the Milstein scheme applied to the original SDE (\ref{sec:eq34}). Hence let us integrate numerically the transformed SDE from $0$ up to $T=2$ using $200$ time step. 
The mean and the variance are estimated, via MC and QMC, using a set of samplings of increasing size 
$
M=\left\{2^8,\dots,2^{15}\right\}
$, setting the CIR model parameters as follows $ \beta = 0.002, \sigma = \{15\%,25\%,30\%\}$. Moreover exploiting results stated in Sec. \ref{sec:secnum}, $\alpha = (3/4)\cdot \sigma^2$, thus  we have $\alpha = \{0.005625,0.046875,0.0675\}$.
We note that, while the MC method applies straightforward, the QMC implementation requires some further discussion which is, in fact, close to the analysis made in section \ref{sec:NumPCE}, namely the numerical scheme allows to write the solution at $T=2$ as a functional of all the independent increments of the Wiener process, one for each time step.

Since the dimensionality of the problem equals the number of time steps, i.e. $200$, then a multidimensional low-discrepancy sequence is computed, where the size $M$ represents the number of points belonging to this sequence. Once the the SDE is solved  the Euler-Maruyama scheme prescribes that each independent increment has to be replaced by an entry of a point belonging to such a multidimensional low-discrepancy sequence.
Eventually, by applying (\ref{sec:eq48}) to the detected solution, we get the required statistics.

\begin{figure}[!h]
\includegraphics[scale=0.33]{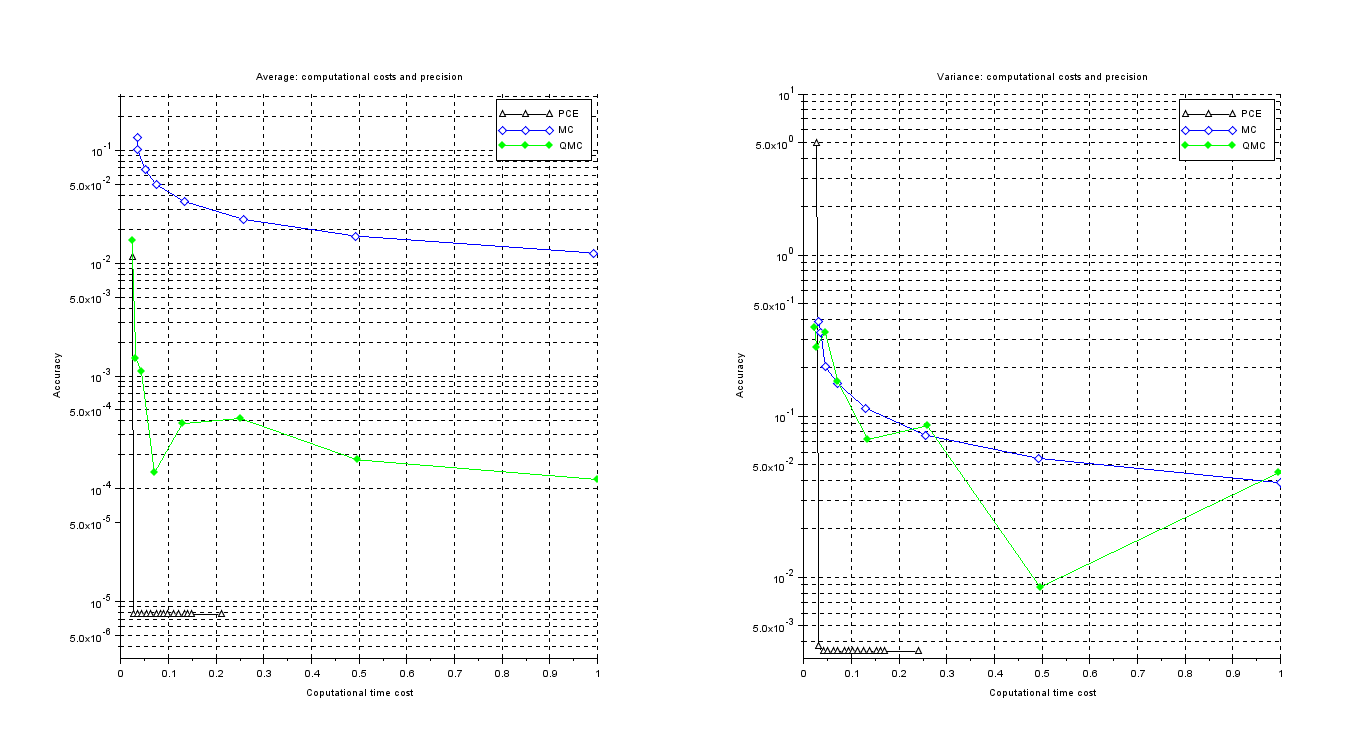}
\caption{Semilogy scale plot which compare the accuracy and computational time costs for detecting the average (left) and variance (right) of the CIR model at $T=2$ by means of PCE, MC and QMC approaches, with  parameters $\alpha=0.005625$, $\beta = 0.002$ and $\sigma = 15\%$.}  \label{fig:fig44}
\end{figure}

\begin{figure}[!h]
\includegraphics[scale=0.33]{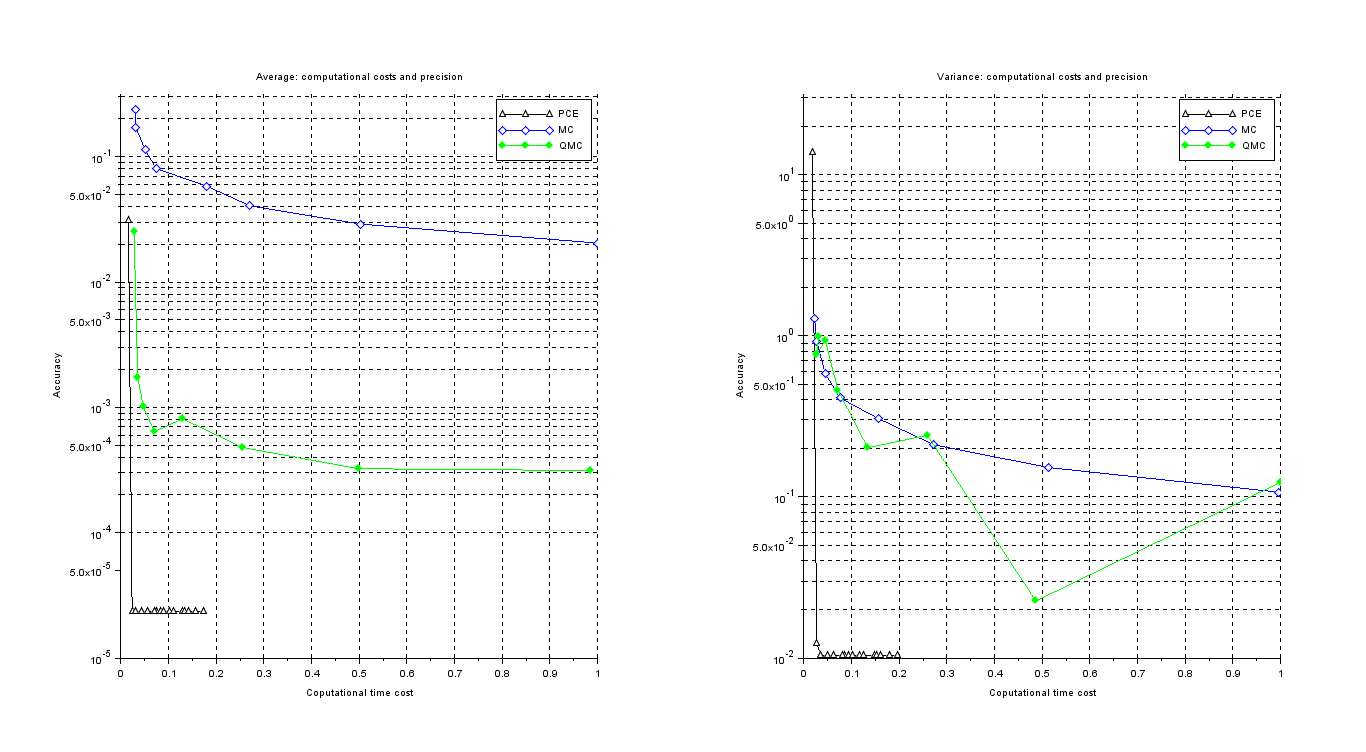}
\caption{Semilogy scale plot which compare the accuracy and computational time costs for detecting the average (left) and variance (right) of the CIR model at $T=2$ by means of PCE, MC and QMC approaches, with  parameters $\alpha=0.046875$, $\beta = 0.002$ and $\sigma = 25\%$.}  \label{fig:fig45}
\end{figure}

\begin{figure}[!h]
\includegraphics[scale=0.33]{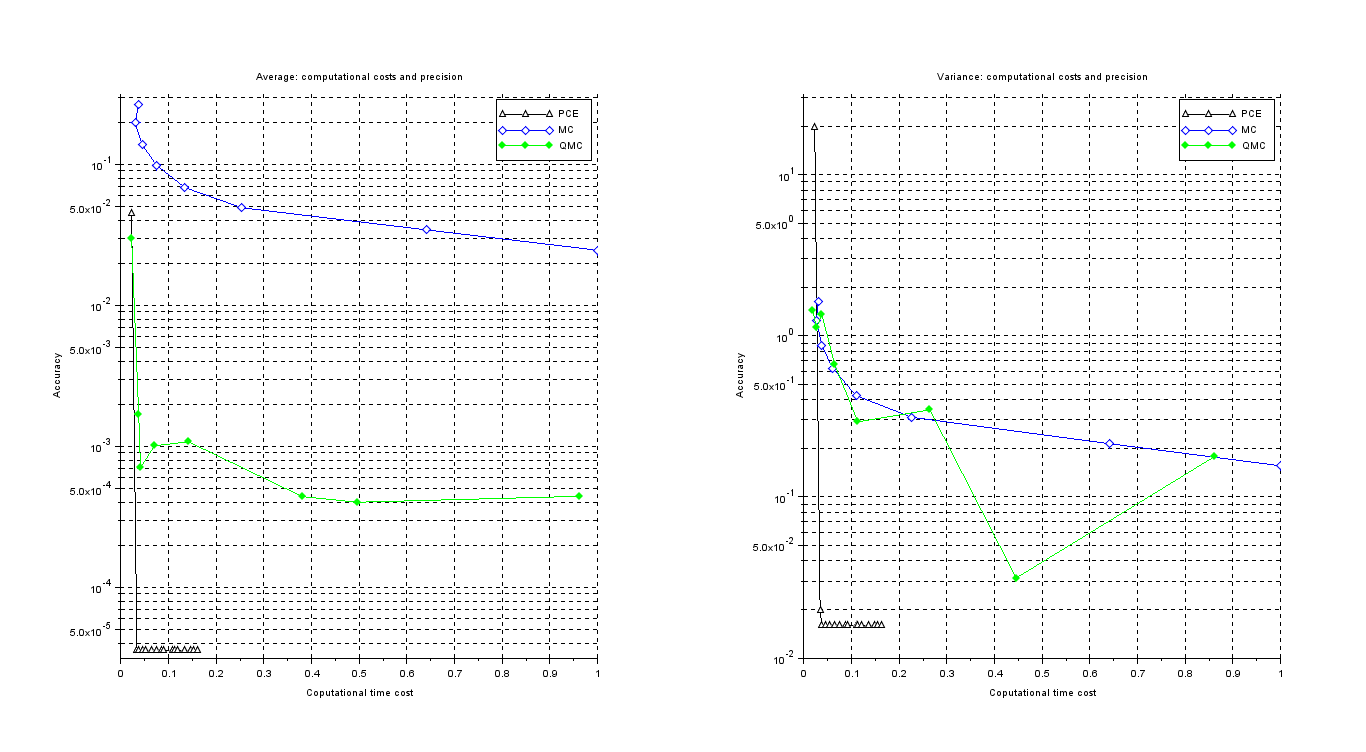}
\caption{Semilogy scale plot which compare the accuracy and computational time costs for detecting the average (left) and variance (right) of the CIR model at $T=2$ by means of PCE, MC and QMC approaches, with  parameters $\alpha=0.0675$, $\beta = 0.002$ and $\sigma = 30\%$.}  \label{fig:fig46}
\end{figure}

As we made in sections \ref{Sec:GBMcomparisons}  and \ref{Sec:Vasicekapproximation}, in Fig. \ref{fig:fig44}, Fig. \ref{fig:fig45} and Fig. \ref{fig:fig46} we compare the computational time costs and its related accuracy reached by the PCE method, resp. the MC and QMC methods. The data used to apply the PCE technique are computed following the approach developed in section \ref{sec:secCIR}. 
We recall that the plot is characterized by setting along the x-axis the computational time costs of the methods, while the error of the statistics $SE_{\mu_{MC}}$, $SE_{\sigma^2_{MC}}$, $\epsilon_{\mu_{QMC}}$ and $\epsilon_{\sigma^2_{QMC}}$,  and the PCE absolute errors, are placed along the y-axis.
Moreover the time costs are normalized with respect to the highest time value among MC, QMC and PCE, thus we display their   relative computational time costs, hence gaining more informations than merely using the related absolute values. \newline


Let us underline that Fig. \ref{fig:fig44}, Fig. \ref{fig:fig45} and \ref{fig:fig46} point out the high accuracy as well as the very low computational effort required  by the PCE method to get the solution. 
We can appreciate such an effectiveness property of the PCE approach looking at the following tables, where 
the values of computational time cost of PCE, see Table \ref{tab:tab32}, MC and QMC, see Table \ref{tab:tab34}, as well as the errors characterizing the  PCE, see Table \ref{tab:tab33}, MC and the QMC methods, see Table \ref{tab:tab35}.

\begin{table}[!h]
\centering
\begin{tabular}{cccc}
\centering Degree of PCE $p$ & \centering Time costs $\sigma = 15\%$ & \centering Time costs $\sigma = 25\%$  & \centering Time costs $\sigma = 30\%$  \tabularnewline
\hline
 0 &     0.0030 &     0.0050 &     0.0040 \\
 1 &     0.0070 &     0.0110 &     0.0070 \\
 2 &     0.0100 &     0.0090 &     0.0100 \\
 3 &     0.0110 &     0.0110 &     0.0190 \\
 4 &     0.0150 &     0.0210 &     0.0160 \\
 5 &     0.0190 &     0.0200 &     0.0180 \\
 6 &     0.0190 &     0.0210 &     0.0190 \\
 7 &     0.0230 &     0.0250 &     0.0220 \\
 8 &     0.0260 &     0.0290 &     0.0260 \\
 9 &     0.0270 &     0.0260 &     0.0270 \\
 10 &     0.0320 &     0.0290 &     0.0310 \\
 11 &     0.0310 &     0.0340 &     0.0340 \\
 12 &     0.0390 &     0.0380 &     0.0360 \\
 13 &     0.0360 &     0.0370 &     0.0410 \\
 14 &     0.0410 &     0.0410 &     0.0430 \\
 15 &     0.0430 &     0.0430 &     0.0440 \\

\end{tabular}
\caption{Computational time costs and absolute error of mean and variance for PCE-approximation of the CIR interest rate model evaluated at $T=2$, whose parameters are $\alpha = \{0.005625,0.046875,0.0675\}$, $\beta = 0.002$, $\sigma = \{15\%,25\%,30\%\}$ and $R_0=110$.} \label{tab:tab32}
\end{table}
\newpage

\begin{table}[!h]
\centering
\begin{tabular}{c|cc|cc|cc}
$\ $ &\multicolumn{2}{c|}{$\sigma =15\%$}  & \multicolumn{2}{c|}{$\sigma =25\%$}  &\multicolumn{2}{c}{$\sigma =30\%$}  \\
\hline
\centering Degree of PCE $p$ & $\epsilon_{MEAN}^{(p)}$& $\epsilon_{VAR}^{(p)}$  & $\epsilon_{MEAN}^{(p)}$ & $\epsilon_{VAR}^{(p)}$& $\epsilon_{MEAN}^{(p)}$ & $\epsilon_{VAR}^{(p)}$\\
\hline
 0 & 1.1227e-02 & 4.9211e+00 & 3.1187e-02 & 1.3674e+01 & 4.4910e-02 & 1.9694e+01 \\
 1 & 7.6695e-06 & 3.7012e-03 & 2.3551e-05 & 1.2358e-02 & 3.5874e-05 & 1.9850e-02 \\
 2 & 7.6696e-06 & 3.4494e-03 & 2.3552e-05 & 1.0415e-02 & 3.5876e-05 & 1.5822e-02 \\
 3 & 7.6695e-06 & 3.4494e-03 & 2.3552e-05 & 1.0415e-02 & 3.5876e-05 & 1.5822e-02 \\
 4 & 7.6696e-06 & 3.4494e-03 & 2.3552e-05 & 1.0415e-02 & 3.5876e-05 & 1.5822e-02 \\
 5 & 7.6695e-06 & 3.4494e-03 & 2.3552e-05 & 1.0415e-02 & 3.5876e-05 & 1.5822e-02 \\
 6 & 7.6696e-06 & 3.4494e-03 & 2.3552e-05 & 1.0415e-02 & 3.5876e-05 & 1.5822e-02 \\
 7 & 7.6695e-06 & 3.4494e-03 & 2.3552e-05 & 1.0415e-02 & 3.5876e-05 & 1.5822e-02 \\
 8 & 7.6696e-06 & 3.4494e-03 & 2.3552e-05 & 1.0415e-02 & 3.5876e-05 & 1.5822e-02 \\
 9 & 7.6695e-06 & 3.4494e-03 & 2.3552e-05 & 1.0415e-02 & 3.5876e-05 & 1.5822e-02 \\
 10 & 7.6696e-06 & 3.4494e-03 & 2.3552e-05 & 1.0415e-02 & 3.5876e-05 & 1.5822e-02 \\
 11 & 7.6695e-06 & 3.4494e-03 & 2.3552e-05 & 1.0415e-02 & 3.5876e-05 & 1.5822e-02 \\
 12 & 7.6696e-06 & 3.4494e-03 & 2.3552e-05 & 1.0415e-02 & 3.5876e-05 & 1.5822e-02 \\
 13 & 7.6696e-06 & 3.4494e-03 & 2.3552e-05 & 1.0415e-02 & 3.5876e-05 & 1.5822e-02 \\
 14 & 7.6696e-06 & 3.4494e-03 & 2.3552e-05 & 1.0415e-02 & 3.5876e-05 & 1.5822e-02 \\
 15 & 7.6695e-06 & 3.4494e-03 & 2.3552e-05 & 1.0415e-02 & 3.5876e-05 & 1.5822e-02 \\

\end{tabular}
\caption{Absolute errors of PCE-approximation of the CIR interest rate model at $T=2$ for \newline $\alpha = \{0.005625,0.046875,0.0675\},\; \beta = 0.002,\;\sigma = \{15\%,25\%,30\%\}$ and $R_0 = 110$.} \label{tab:tab33}
\end{table}

The computational time costs required for the detection of PCE's coefficients, equation (\ref{sec:eq10}), and its statistics,  equations (\ref{sec:eq8}) and (\ref{sec:eq9}), are irrelevant if compared with time required to get the evaluation of the process in eq. (\ref{sec:eq40}) at the quadrature nodes $\{\xi_j\}_{j=1}^N$, this motivates the time data in Table \ref{tab:tab32}

\begin{table}[!h]
\centering
\begin{tabular}{c|cc|cc|cc|cc|cc|cc}
$\ $ &\multicolumn{4}{c}{$\sigma = 15\%$}  & \multicolumn{4}{|c}{$\sigma = 25\%$} & \multicolumn{4}{|c}{$\sigma = 30\%$}  \tabularnewline
\hline
$\ $ &\multicolumn{2}{c}{Average}  & \multicolumn{2}{|c|}{Variance} &\multicolumn{2}{c}{Average}  & \multicolumn{2}{|c|}{Variance}  &\multicolumn{2}{c}{Average}  & \multicolumn{2}{|c}{Variance}  \tabularnewline
\hline
\centering $M$ & \centering MC & \centering QMC  &  MC &  QMC &  MC &  QMC &  MC &  QMC &  MC &  QMC &  MC &  QMC \tabularnewline
\hline

        256 &     0.0140 &     0.0070 &     0.0080 &     0.0050 &     0.0080 &     0.0110 &     0.0050 &     0.0050 &     0.0070 &     0.0090 &     0.0060 &     0.0060 \\
       512 &     0.0090 &     0.0090 &     0.0070 &     0.0060 &     0.0080 &     0.0110 &     0.0070 &     0.0070 &     0.0090 &     0.0080 &     0.0080 &     0.0060 \\
      1024 &     0.0110 &     0.0120 &     0.0110 &     0.0100 &     0.0120 &     0.0150 &     0.0100 &     0.0110 &     0.0120 &     0.0140 &     0.0100 &     0.0100 \\
      2048 &     0.0190 &     0.0200 &     0.0190 &     0.0170 &     0.0230 &     0.0240 &     0.0180 &     0.0210 &     0.0200 &     0.0200 &     0.0180 &     0.0170 \\
      4096 &     0.0330 &     0.0390 &     0.0340 &     0.0320 &     0.0370 &     0.0370 &     0.0320 &     0.0310 &     0.0380 &     0.0380 &     0.0360 &     0.0360 \\
      8192 &     0.0750 &     0.0670 &     0.0620 &     0.0590 &     0.0740 &     0.0790 &     0.0650 &     0.0580 &     0.0790 &     0.0680 &     0.0590 &     0.0610 \\
     16384 &     0.1410 &     0.1500 &     0.1190 &     0.1980 &     0.1330 &     0.1370 &     0.1260 &     0.1170 &     0.1570 &     0.1660 &     0.2010 &     0.1450 \\
     32768 &     0.2610 &     0.2690 &     0.2350 &     0.2390 &     0.2630 &     0.2810 &     0.2540 &     0.2580 &     0.2840 &     0.2690 &     0.2410 &     0.2430 \\

\end{tabular}
\caption{Computational cost of computation of statistics of the CIR interest rate model $R_{T}$ at $T=2$, determined  by MC, QMC, with $\alpha = \{0.005625,0.046875,0.0675\}$,  $ \beta = 0.002$, $\sigma = \{15\%,25\%,30\%\}$ and $R_0=110$.} \label{tab:tab34}
\end{table}

\begin{table}[!h]
\centering
\begin{tabular}{c|c|cc|cc}
$\ $ &$\ $ &\multicolumn{2}{c}{Average}  & \multicolumn{2}{|c}{Variance}  \tabularnewline
\hline
$\ $&\centering Size of sampling  $M$ & \centering $SE_{\mu_{MC}}$   & \centering  $\epsilon_{\mu_{QMC}}$  & \centering $SE_{\sigma^2_{MC}}$ & \centering  $\epsilon_{\sigma^2_{QMC}}$ \tabularnewline
\hline
\multirow{8}{*}{$\sigma = 15 \%$} &
        256 & 1.3515e-01 & 1.5744e-02 & 4.1411e-01 & 3.5681e-01 \tabularnewline 
    $\ $&    512 & 9.8423e-02 & 1.4373e-03 & 3.1029e-01 & 2.6892e-01 \tabularnewline 
   $\ $&    1024 & 6.9759e-02 & 1.1029e-03 & 2.2033e-01 & 3.3319e-01 \tabularnewline 
  $\ $&     2048 & 4.9068e-02 & 1.3868e-04 & 1.5413e-01 & 1.6354e-01 \tabularnewline 
 $\ $&      4096 & 3.4559e-02 & 3.7763e-04 & 1.0811e-01 & 7.1762e-02 \tabularnewline 
$\ $&       8192 & 2.4551e-02 & 4.2046e-04 & 7.7155e-02 & 8.7361e-02 \tabularnewline 
$\ $&      16384 & 1.7269e-02 & 1.8063e-04 & 5.3984e-02 & 8.6968e-03 \tabularnewline 
$\ $&      32768 & 1.2218e-02 & 1.2084e-04 & 3.8214e-02 & 4.4496e-02 \tabularnewline
\hline
\multirow{8}{*}{$\sigma = 25 \%$} & 
        256 & 2.3429e-01 & 2.5344e-02 & 1.2445e+00 & 9.9417e-01 \tabularnewline 
$\ $&        512 & 1.7451e-01 & 1.7405e-03 & 9.7546e-01 & 7.6735e-01 \tabularnewline 
$\ $&       1024 & 1.1406e-01 & 1.0069e-03 & 5.8900e-01 & 9.3635e-01 \tabularnewline 
$\ $&       2048 & 8.0223e-02 & 6.4006e-04 & 4.1199e-01 & 4.5978e-01 \tabularnewline 
$\ $&       4096 & 5.8942e-02 & 8.0988e-04 & 3.1449e-01 & 2.0169e-01 \tabularnewline 
  $\ $&     8192 & 4.0456e-02 & 4.7886e-04 & 2.0951e-01 & 2.4117e-01 \tabularnewline 
  $\ $&    16384 & 2.9153e-02 & 3.2369e-04 & 1.5385e-01 & 2.2580e-02 \tabularnewline 
  $\ $&    32768 & 2.0581e-02 & 3.1324e-04 & 1.0844e-01 & 1.2319e-01 \tabularnewline 
\hline
\multirow{8}{*}{$\sigma = 30 \%$} &
        256 & 2.6173e-01 & 2.9876e-02 & 1.5530e+00 & 1.4338e+00 \tabularnewline 
  $\ $&      512 & 2.0446e-01 & 1.6963e-03 & 1.3391e+00 & 1.1198e+00 \tabularnewline 
 $\ $&      1024 & 1.3572e-01 & 7.0999e-04 & 8.3396e-01 & 1.3563e+00 \tabularnewline 
$\ $&       2048 & 9.7037e-02 & 1.0130e-03 & 6.0278e-01 & 6.6614e-01 \tabularnewline 
$\ $&       4096 & 7.0453e-02 & 1.0797e-03 & 4.4931e-01 & 2.9220e-01 \tabularnewline 
$\ $&       8192 & 4.8704e-02 & 4.4112e-04 & 3.0364e-01 & 3.4621e-01 \tabularnewline 
$\ $&      16384 & 3.4759e-02 & 4.0243e-04 & 2.1871e-01 & 3.1364e-02 \tabularnewline 
$\ $&      32768 & 2.4443e-02 & 4.4340e-04 & 1.5295e-01 & 1.7711e-01 \tabularnewline

\end{tabular}
\caption{Errors of the average and variance computed by MC and QMC for the CIR interest rate model $R_T$ at time $T=2$, with  parameters $\alpha = \{0.005625,0.046875,0.0675\}$, $\beta = 0.002 $, $,\sigma = \{15\%,25\%,30\%\}$ and and $R_0=110$.} \label{tab:tab35}
\end{table}

\clearpage

\section{Conclusions}

In this paper we show how the Polynomial Chaos Expansion (PCE) method can be effectively used  to approximate the solution of a given SDE, exploiting only  some basic properties of the Brownian motion, which is assumed to drive the stochastic process we are interested in, and a suitable transformation of the original SDE we have taken into consideration. In particular we have applied the PCE technique to approximate the solutions of some of the most relevant interest rate models, namely the geometric Brownian motion  model, the Vasicek model and the Cox–Ingersoll–Ross (CIR) model.

The provided numerical examples deeply discuss the convergence properties of the PCE method in various scenarios of volatility. Indeed detailed analysis on error properties of the approximated statistics is given when analytical solution is available also providing  a comparison with respect to the related analytical probability density function. The latter is not the case for the CIR model, conversely to what happens for both the gBm and the Vasicek models. Hence, in the CIR case, the PCE approach is even more interesting also because it gives a general method to study those stochastic processes which are defined by SDEs which do not have solution in closed form. \newline
 \newline
We show by concrete examples how the PCE method overcomes the performances of the standard Monte Carlo (MC) method as well as those of the {\it quasi-}Monte Carlo (QMC) one. 
In particular a close comparison between PCE, MC and QMC techniques, points out the advantage in using the  PCE approach, especially in terms of computational costs. The latter result is due to the fact hat the PCE method requires just few realization of the random variable we want to approximate, therefore, even if the basic operations required by the PCE approach, are rather more time consuming,  than those implied by the MC or the QMC approach ( weighted average versus a quadrature formulas), the computational advantages in using the PCE technique are clear  as shown, e.g. ,  by the  CIR model analysis. \newline

Last but not least, it is worth to mention that
the PCE approach can be split  in two distinct section: the off-line computations and the real PCE  approximation of the random variable we want to study. 
Thus the basic random variable, the orthogonal polynomials and some data used in the Gaussian quadrature formulas, are computed once and for all, since they do not depend on the particular model, e.g. the gBm, the Vasicek or the CIR modelm as we made in our analysis. Hence a clear saving concerning the required computational efforts is achieved every time the model is sensible to time variations, as well as in the case when  a new calibration of the involved parameters is required.

\end{document}

%% file: macro.tex
\usepackage{amsmath}
\usepackage{amssymb}
\usepackage{bbm}
\usepackage{tabularx}
\usepackage{theorem}
\usepackage{color}
\usepackage{enumitem}

\usepackage{eurosym}   
\usepackage[a4paper,top=3cm,bottom=3cm]{geometry}
\usepackage{graphicx}
\usepackage{subfigure}
\usepackage{multirow}

\usepackage[T1]{fontenc}
\usepackage[utf8]{inputenc}
\usepackage[english]{babel}
\usepackage[all,cmtip,ps]{xy} 
\usepackage{pgfplots}
\usepackage{pgf,tikz}
\usetikzlibrary{arrows}
\usetikzlibrary{shapes,snakes}

\newtheorem{remark}{Remark}

\newcommand{\N}{{\mathbb N}}

\newcommand{\R}{{\mathbb R}}

\def\bmat{\left[ \begin{array}}
\def\emat{\end{array} \right]}

\newcommand{\PP}{\ensuremath{\mathbb{P}}}

\newcommand{\Prob}[1]{{\PP\left({#1}\right)}}
\newcommand{\EE}{\ensuremath{\mathbb{E}}}
\newcommand{\Expect}[1]{{\EE\left[{#1}\right]}}

\newcommand{\VVar}{\ensuremath{\mathsf{Var}}}
\newcommand{\Var}[1]{{\VVar\left[{#1}\right]}}

\newcommand{\norm}[1]{\left\lVert#1\right\rVert}
\newcommand{\abs}[1]{\left\lvert#1\right\rvert}
\newcommand{\normp}[1]{{\left\lVert#1\right\rVert}_{\PP}}
\newcommand{\XI}{\boldsymbol{\xi}}
\newcommand{\sprod}[2]{{\left<#1,#2\right>}_{\PP}}

\newcommand{\intPxi}[2]{\int_\Omega #1\left(\xi(\omega)\right)#2\left(\xi(\omega)\right)d\PP(\omega)}
\newcommand{\intPxiR}[2]{\int_\Omega #1\left(x\right)#2\left(x\right)f_\xi(x)dx}

{\left\lbrace\begin{array}{@{}l@{}}}%
{\end{array}\right.}

\usepackage{pgfplots}

\def\addlegendimage{\csname pgfplots@addlegendimage\endcsname}

\pgfkeys{/pgfplots/number in legend/.style={%
        /pgfplots/legend image code/.code={%
            \node at (0.295,-0.0225){#1};
        },%
    },
}